\documentclass[1p,review]{elsarticle}
\journal{Progress in Particle and Nuclear Physics}

\usepackage{amssymb}
\usepackage[pdfencoding=auto,colorlinks]{hyperref}

\graphicspath{{figures/}}

\usepackage{lineno}
\modulolinenumbers[5]

\makeatletter
\newcommand*\rel@kern[1]{\kern#1\dimexpr\macc@kerna}
\newcommand*\widebar[1]{%
  \begingroup
  \def\mathaccent##1##2{%
    \rel@kern{0.8}%
    \overline{\rel@kern{-0.8}\macc@nucleus\rel@kern{0.2}}%
    \rel@kern{-0.2}%
  }%
  \macc@depth\@ne{}
  \let\math@bgroup\@empty{} \let\math@egroup\macc@set@skewchar{}
  \mathsurround\z@ \frozen@everymath{\mathgroup\macc@group\relax}%
  \macc@set@skewchar\relax
  \let\mathaccentV\macc@nested@a{}
  \macc@nested@a\relax111{#1}%
  \endgroup
}
\makeatother

\usepackage[acronym]{glossaries}
\makeglossaries{}

\newacronym{lhc}{LHC}{Large Hadron Collider}
\newacronym[description={A Toroidal LHC ApparatuS},
            first={ATLAS (A Toroidal LHC ApparatuS)}]
           {atlas}{ATLAS}{A Toroidal LHC ApparatuS}
\newacronym[description={Hadron-Electron Ring Accelerator},
            first={HERA (Hadron-Electron Ring Accelerator)}]
           {hera}{HERA}{Hadron-Electron Ring Accelerator}

\newacronym[description={Deep Inelastic Scattering}]
           {dis}{DIS}{deep inelastic scattering}
\newacronym[description={Neutral Current}]
           {nc}{NC}{neutral current}
\newacronym[description={Charged Current}]
           {cc}{CC}{charged current}
\newacronym[description={Quantum Chromodynamics},
            first={Quantum Chromodynamics (QCD)}]
           {qcd}{QCD}{quantum chromodynamics}
\newacronym[description={Perturbative Quantum Chromodynamics}]
           {pqcd}{pQCD}{perturbative quantum chromodynamics}
\newacronym[description={Parton Distribution Function}]
           {pdf}{PDF}{parton distribution function}
\newacronym[description={Quantum Electrodynamics}]
           {qed}{QED}{quantum electrodynamics}
\newacronym[description={Light-Front}]
           {lf}{LF}{Light-Front}
\newacronym[description={Leading Order}]
           {lo}{LO}{leading order}
\newacronym[description={Next-to-leading Order}]
           {nlo}{NLO}{next-to-leading order}
\newacronym[description={Next-to-next-to-leading Order}]
           {nnlo}{NNLO}{next-to-next-to-leading order}
\newacronym[description={Dokshitzer-Gribov-Lipatov-Altarelli-Parisi Evolution
                         Equation}]
           {dglap}{DGLAP}{Dokshitzer-Gribov-Lipatov-Altarelli-Parisi}
\newacronym[description={Transverse Momentum Dependent}]
           {tmd}{TMD}{transverse momentum dependent}
\newacronym[description={Operator Product Expansion}]
           {ope}{OPE}{operator product expansion}
\newacronym[description={Principle of Maximum Conformality}]
           {pmc}{pmc}{``principle of maximum conformality{}''}
\newacronym[description={Intrinsic charm}]
           {ic}{IC}{intrinsic charm}

\newglossaryentry{alphas}{%
  name = \ensuremath{\alpha_\mathrm{s}},
  description = Coupling constant of strong interaction
}
\newglossaryentry{lambdaqcd}{%
  name = \ensuremath{\Lambda_\mathrm{QCD}},
  description = The QCD scale
}
\newglossaryentry{pty}{%
  name = \ensuremath{p_\mathrm{T}^{\gamma}},
  description = The transverse momentum of the photon
}
\newglossaryentry{ety}{%
  name = \ensuremath{E_\mathrm{T}^{\gamma}},
  description = The transverse energy of the photon
}
\newglossaryentry{etiso}{%
  name = \ensuremath{E_\mathrm{T}^\mathrm{iso}},
  description = The calorimeter isolation energy of the photon
}
\newglossaryentry{etay}{%
  name = \ensuremath{\eta^{\gamma}},
  description = The pseudo-rapidity of the photon
}
\newglossaryentry{ptj}{%
  name = \ensuremath{p_\mathrm{jet}^{\gamma}},
  description = The transverse momentum of the jet
}
\newglossaryentry{etaj}{%
  name = \ensuremath{\eta^\mathrm{jet}},
  description = The pseudo-rapidity of the jet
}

\newglossaryentry{x}{%
  name = \ensuremath{x},
  description = The parton momentum fraction
}
\newglossaryentry{xf}{%
  name = \ensuremath{x_\mathrm{F}},
  description = The Feynman momentum fraction
}
\newglossaryentry{xbj}{%
  name = \ensuremath{x_\mathrm{Bj}},
  description = The Bjorken momentum fraction
}
\newglossaryentry{xQ}{%
  name = \ensuremath{x_\mathrm{Q}},
  description = The momentum fraction of heavy quark $Q$
}
\newglossaryentry{muz}{%
  name = \ensuremath{\mu_{0}},
  description = The threshold of the charm quark PDF evolution.
}
\newglossaryentry{pt}{%
  name = \ensuremath{p_\mathrm{T}},
  description = The transverse momentum
}
\newglossaryentry{kt}{%
  name = \ensuremath{k_\mathrm{T}},
  description = The transverse momentum
}
\newglossaryentry{ktvec}{%
  name = \ensuremath{\mathbf{k}_\mathrm{T}},
  description = The two-dimensional transverse momentum
}

\newglossaryentry{w}{%
  name = \ensuremath{w},
  description = The intrinsic charm probability in proton
}
\newglossaryentry{wc}{%
  name = \ensuremath{w_\mathrm{cent}},
  description = The fitted central value of intrinsic charm probability in proton
}
\newglossaryentry{wul}{%
  name = \ensuremath{w_\mathrm{ul}},
  description = The fitted upper limit of intrinsic charm probability in proton
}
\newglossaryentry{wins}{%
  name = \ensuremath{w_\mathrm{ins}},
  description = The inserted value of intrinsic charm probability in proton
}

\newcommand{\lhc}{\gls{lhc}}
\newcommand{\atlas}{\gls{atlas}}
\newcommand{\dis}{\gls{dis}}
\newcommand{\qcd}{\gls{qcd}}
\newcommand{\pqcd}{\gls{pqcd}}
\newcommand{\pdf}{\gls{pdf}}
\newcommand{\pdfs}{\glspl{pdf}}
\newcommand{\tmd}{\gls{tmd}}
\newcommand{\ope}{\gls{ope}}
\newcommand{\lf}{\gls{lf}}

\newcommand{\uudcc}{\ensuremath{|uudc\bar{c}\rangle}}
\newcommand{\uudqq}{\ensuremath{|uudq\bar{q}\rangle}}
\newcommand{\uudQQ}{\ensuremath{|uudQ\widebar{Q} \, \rangle}}
\newcommand{\alphas}{\gls{alphas}}
\newcommand{\lambdaqcd}{\gls{lambdaqcd}}

\newcommand{\gev}{GeV}
\newcommand{\tev}{TeV}
\newcommand{\sviii}{\mbox{${\sqrt{s} = 8}$~\tev}}
\newcommand{\sxiii}{\mbox{${\sqrt{s} = 13}$~\tev}}
\newcommand{\ycjet}{\mbox{$\gamma +c$-jet}}
\newcommand{\ppycjet}{\mbox{$pp \rightarrow \gamma + c$-jet}}
\newcommand{\ybjet}{\mbox{$\gamma +b$-jet}}

\newcommand{\sherpa}{\textsc{Sherpa}}
\newcommand{\lo}{\gls{lo}}
\newcommand{\nlo}{\gls{nlo}}
\newcommand{\nnlo}{\gls{nnlo}}

\newcommand{\pty}{\gls{pty}}
\newcommand{\ety}{\gls{ety}}
\newcommand{\etiso}{\gls{etiso}}
\newcommand{\etay}{\gls{etay}}
\newcommand{\aetay}{\ensuremath{|\gls{etay}|}}
\newcommand{\ptj}{\gls{ptj}}
\newcommand{\etaj}{\gls{etaj}}

\newcommand{\x}{\gls{x}}
\newcommand{\xf}{\gls{xf}}
\newcommand{\xbj}{\gls{xbj}}
\newcommand{\xQ}{\gls{xQ}}
\newcommand{\muz}{\gls{muz}}
\newcommand{\pt}{\gls{pt}}
\newcommand{\kt}{\gls{kt}}
\newcommand{\ktvec}{\gls{ktvec}}

\newcommand{\ic}{\gls{ic}}
\newcommand{\w}{\gls{w}}
\newcommand{\wul}{\gls{wul}}
\newcommand{\wc}{\gls{wc}}

\newcommand{\chis}{$\chi^{2}$}
\newcommand{\cl}{CL}

\newcommand{\HF}{\ensuremath{\mathrm{HF}}}
\newcommand{\VHF}{\ensuremath{V+\mathrm{HF}}}
\newcommand{\ZHF}{\ensuremath{Z+\mathrm{HF}}}

\newcommand{\jet}{\ensuremath{\mathrm{jet}}}

\newlength{\spw}
\newlength{\mpw}
\newlength{\lpw}
\begin{document}
\begin{frontmatter}

\title{Novel Heavy-Quark Physics Phenomena}

\author[slac]{S.J.~Brodsky}
\author[jinr]{G.I.~Lykasov\corref{correspondingauthor}}
\cortext[correspondingauthor]{Corresponding author}
\ead{lykasov@jinr.ru}
\author[jinr,mgu]{A.V.~Lipatov}
\author[sas]{J.~Smiesko}

\address[slac]{SLAC National Accelerator Laboratory,
               Stanford University, Menlo Park, CA 94025, United States}
\address[jinr]{Joint Institute for Nuclear Research,
               Dubna 141980, Moscow region, Russia}
\address[mgu]{Skobeltsyn Institute of Nuclear Physics, Moscow State University,
              119991 Moscow, Russia}
\address[sas]{Slovak Academy of Sciences, Institute of Experimental Physics,
              Watsonova 47, 040 01 Kosice, Slovakia}

\begin{abstract}
  We review the current understanding of heavy quark parton distributions in
  nucleons and their impact on deep inelastic scattering, collider physics, and
  other processes at high energies. The determination of the heavy-quark parton
  distribution functions is particularly significant for the analysis of hard
  processes at LHC energies, including the forward rapidity high $x_\mathrm{F}$
  domain. The contribution of ``intrinsic'' heavy quarks, which are multiply
  connected to the valence quarks of nucleons, is reviewed within non-perturbative
  physics which provides new information on the fundamental structure of hadrons
  in QCD\@. A new prediction for the non-perturbative intrinsic charm-anticharm
  asymmetry of the proton eigenstate has recently been obtained from a QCD
  lattice gauge theory calculation of the proton's $G_\mathrm{E}^p(Q^2)$ form
  factor~\cite{Sufian:2020coz}. This form factor only arises from non-valence
  quarks and anti-quarks if they have different contributions in the proton's
  eigenstate. This result, together with the exclusive and inclusive connection
  and analytic constraints on the form of hadronic structure functions from
  Light-Front Holographic QCD (LFHQCD) predicts a significant non-perturbative
  $c(x,Q) - \bar{c}(x,Q)$ asymmetry in the proton structure function at high
  $x$, consistent with the dynamics predicted by intrinsic charm models. Recent
  ATLAS data on the associated production of prompt photons and charm-quark jets
  in $pp$ collisions at $\sqrt{s} = 8$~TeV has provided new constraints on
  non-perturbative intrinsic charm and tests of the LGTH predictions. We also
  focus on other experimental observables which have high sensitivity to the
  intrinsic heavy contributions to PDFs.
\end{abstract}

\begin{keyword}
Heavy flavor quarks \sep\ gluons \sep\ charm \sep\ QCD \sep\ PDF\@.
\end{keyword}
\end{frontmatter}

\newpage
\tableofcontents{}
\newpage

\newpage


%
%
\section{Introduction}%
\label{sec:intro}

\subsection{Motivation of this review}

\qcd, the underlying theory of strong interactions, with
quarks and gluons as the fundamental degrees of freedom, predicts that the heavy
quarks in the nucleon-sea to have both perturbative ``extrinsic'' and
non-perturbative ``intrinsic'' origins. The extrinsic sea arises from gluon
splitting which is triggered by a probe in the reaction. It can be calculated
order-by-order in perturbation theory. In contrast, the intrinsic sea is
encoded in the non-perturbative wave functions of the nucleon eigenstate. The
existence of non-perturbative \ic{} was originally proposed in the
BHPS model~\cite{Brodsky:1980pb} and developed further in subsequent
papers~\cite{Brodsky:1984nx,Harris:1995jx,Franz:2000ee}. The intrinsic
contribution to the heavy quark distributions of hadrons at high \x{}
corresponds to Fock states such as \uudQQ{} where the heavy quark pair
is multiply connected to two or more valence quarks of the proton. It is
maximal at minimal off-shellness; i.e., when the constituents all have the same
rapidity  $y_I$, and thus
$x_i \propto \sqrt{m_i^2 + \mathbf{k}_{\mathrm{T}i}^2}$.
Here ${x = {k^{+} / P^{+}} = (k^0 + k^3) / (P^0 + P^3)}$ is the frame-independent
light-front momentum fraction carried by the heavy quark in a hadron with
momentum $P^\mu$. In the case of deep inelastic lepton-proton scattering, the
\lf{} momentum fraction variable \x{} in the proton structure functions can be
identified with the Bjorken variable $\xbj = {Q^2 / 2 p \cdot q}$. These heavy
quark contributions to the nucleon's \pdf{} thus peak at large \xbj{} and thus
have important implications for LHC and EIC collider phenomenology, including
Higgs and heavy hadron production at high \xf~\cite{Royon:2015eya}.

The existence of the non-perturbative intrinsic heavy quarks in the hadronic
eigenstates of hadrons and nuclei highlights the importance of experiments for
studying the high \xf{} and threshold domains of heavy particle production both
at colliders and fixed target facilities. Measurements of the strong asymmetry
of the intrinsic quark and antiquark distributions predicted by the lattice
gauge theory (LGTH) is particularly important. As we will review here, the
presence of intrinsic heavy quark degrees of freedom in hadrons also illuminates
many new and subtle aspects of \qcd{} phenomena. It also opens up new
opportunities to study heavy quark phenomena in fixed target experiments such as
the proposed AFTER~\cite{Brodsky:2015fna} fixed target facility at CERN\@. The
existence of intrinsic heavy quarks also illuminates fundamental aspects of
non-perturbative \qcd.

Thus \qcd{} predicts two separate and distinct contributions to the heavy quark
distributions $q(\x, Q^2)$ of the nucleons at low and high \x.
In the case of deep inelastic lepton-proton scattering at small \x, heavy-quark
pairs are dominantly produced via gluon-splitting subprocess
$g \to Q \widebar{Q}$. The presence of the heavy quarks in nucleon from this
standard contribution is a result of the \qcd{} evolution of the light quark and
gluon \pdfs. Unlike the conventional $\log m^2_\mathrm{Q}$ dependence of the low
\x{} extrinsic gluon-splitting contributions, the probabilities for the
intrinsic heavy quark Fock states at high \x{} scale as
$1 / m_\mathrm{Q}^2$ in non-Abelian \qcd\@. Thus the relative probability of
intrinsic bottom to charm is of order
${m_\mathrm{c}^2 / m_\mathrm{b}^2} \sim 1 / 10$. In contrast, the
probability for a higher Fock state containing heavy leptons in a \gls{qed} atom
scales as $1 / m_\ell^4$, corresponding to the twist-8 Euler-Heisenberg
light-by-light self-energy insertion. Detailed derivations based on the \ope{}
have been given in Ref.~\cite{Brodsky:1984nx,Franz:2000ee}.
light-by-light self-energy insertion. Detailed derivations based on the \ope{}
have been given in Ref.~\cite{Brodsky:1984nx,Franz:2000ee}.

In \lf{} Hamiltonian theory, the intrinsic heavy quarks of the
proton are associated with non-valence Fock states, such as \uudQQ{} in the
hadronic eigenstate of the \lf{} Hamiltonian; this implies that the heavy quarks
are multi-connected to the valence quarks. Since the LF wavefunction is maximal
at minimum off-shell invariant mass; i.e., at equal rapidity, the intrinsic
heavy quarks carry large momentum fraction \xQ. A key characteristic is
different momentum and spin distributions for the intrinsic $Q$ and
$\widebar{Q}$ in the nucleon; for example the charm-anticharm asymmetry, since
the comoving quarks are sensitive to the global quantum numbers of the
nucleon~\cite{Brodsky:2015fna}. Furthermore, since all of the intrinsic quarks
in the \uudQQ{} Fock state have similar rapidities they can re-interact, leading
to significant $Q$ vs $\widebar{Q}$ asymmetries. The concept of intrinsic heavy
quarks was also proposed in the context of meson-baryon fluctuation
models~\cite{Navarra:1995rq,Pumplin:2005yf} where intrinsic charm was
identified with two-body state $\bar{D}^0(u\bar{c})\Lambda^{+}_\mathrm{c}(udc)$
in the proton. This identification predicts large asymmetries in the charm
versus charm momentum and spin distributions, since these heavy quark
distributions depend on the correlations determined by the valence quark
distributions, they are referred to as \emph{intrinsic} contributions to the
hadron's fundamental structure. A specific analysis of the intrinsic charm
content of the deuteron is given in Ref.~\cite{Brodsky:2018zdh}. In contrast,
the contribution to the heavy quark PDFs arising from gluon splitting are
symmetric in $Q$ vs $\widebar{Q}$. The contributions generated by DGLAP
evolution at low \x{} can be considered as \emph{extrinsic} contributions since
they only depend on the gluon distribution. The gluon splitting contribution to
the heavy-quark degrees of freedom is perturbatively calculable using DGLAP
evolution. To first approximation, the perturbative extrinsic heavy quark
distribution falls as $(1 - x)$ times the gluon distribution and is limited to
low \xbj. However, \qcd{} also predicts additional Fock state contributions to
proton structure at high \x, such as \uudQQ{} where the heavy quark pair is
multiply connected to two or more valence quarks of the proton.
The heavy quark contributions to the nucleon's \pdf{} thus peak at large \x.
Since they depend on the correlations determined by the valence quark
distributions, these heavy quark contributions are \emph{intrinsic}
contributions to the hadron's fundamental structure.
Furthermore, since all of the  intrinsic quarks in the \uudQQ{} Fock
state have similar rapidities they
can re-interact, leading to significant $Q$ vs $\widebar{Q}$ asymmetries.  
In contrast, the contribution to the heavy quark PDFS arising from gluon
splitting are symmetric in $Q$ vs $\widebar{Q}$, because they only depend on the
gluon distribution.

We also emphasize that the \emph{intrinsic} $Q\widebar{Q}$ contributions to
\pdf{} can give a non-zero signal not only in the fragmentation processes of
colliding hadrons, but also in the hard inclusive or semi-inclusive processes.
As is shown in
Ref.~\cite{Lykasov:2012hf,Bednyakov:2017vck,Brodsky:2016fyh,Beauchemin:2014rya,
Lipatov:2018oxm} and references therein, the signal of the intrinsic charm
contribution can be observed in hard $pp$ inclusive $D$-meson production or
semi-inclusive $pp$ production of prompt photons or gauge baryons $Z$,$W$
accompanied by $c$- or $b$-jets at high transverse momenta and mid-rapidity in
the \lhc{} energy range.

The hard production of prompt photons and vector bosons accompanied by heavy
flavor\footnote{Here and below heavy flavor implies charm and bottom quarks.}
jets (\VHF) in $pp$ collisions at LHC energies can be considered as an
additional tool to study the quark and gluon \pdfs{} compared to the deep
inelastic scattering of electrons on protons. In these processes, in the
rapidity region $|y| < 2.5$, which corresponds to the kinematics of ATLAS and
CMS experiments, one can study these \pdfs{}, not only at low parton momentum
fractions $\x < 0.1$ but also at larger \x{} values~\cite{Brodsky:2016fyh}.
Therefore, such \VHF~processes can provide new information on the \pdfs{} at
large $x > 0.1$, where the non-trivial proton structure (for example, the
contribution of valence-like \emph{intrinsic} heavy quark components)
can be revealed~\cite{Brodsky:1980pb,Brodsky:1981se,Harris:1995jx,Franz:2000ee}.

A new prediction for the non-perturbative intrinsic charm-anticharm asymmetry of
the proton eigenstate has recently been obtained from a \qcd{} lattice gauge
theory calculation of the proton's $G_\mathrm{E}^p(Q^2)$ form
factor~\cite{Sufian:2020coz}. This form factor only arises from non-valence
quarks and anti-quarks if they have different contributions to the proton's
eigenstate. This result, together with the exclusive and inclusive connection
and analytic constraints on the form of hadronic structure functions from
Light-Front Holographic QCD (LFHQCD) predicts a significant non-perturbative
$c(\x, Q) - \bar{c}(\x,Q)$ asymmetry in the proton structure function at high
\x, consistent with the dynamics predicted by intrinsic charm models. A detailed
discussion of these results is presented in this review.

\subsection{Outline of this review}

The review consists of 9 sections. In Subsection~\ref{sec:collinear_pdf} we
present a brief overview about nucleon structure functions within the collinear
QCD approach. Then, in Subsection~\ref{sec:in_vs_ex} the concepts of intrinsic
and extrinsic quark components in a nucleon are discussed. In
Subsection~\ref{sec:lattice} we focus on the sensitivity of the charm
electro-magnetic form factors to the intrinsic $(c\bar{c})$ pairs in nucleon
calculated within the lattice \qcd. The Higgs production in $pp$ collisions at
LHC energies and heavy quark
distributions in proton is discussed in Subsection~\ref{sec:higgs}.
Sections~\ref{sec:fock} and~\ref{sec:overview} are devoted to general
theoretical aspects of the Fock state structure of hadrons
within the non-perturbative \qcd. In Section~\ref{sec:ktfact} we present a brief
overview on nucleon structure functions within the non-collinear \qcd{}
approach. In Section~\ref{sec:soft_hard} we discuss how the distribution of gluons
at starting point of $\muz^2$, which is the main input in the evolution equation of the
non-collinear QCD approach, could be calculated. The interplay between soft and
hard $pp$ processes is also discussed.
Section~\ref{sec:gamma_cjet} is devoted to the analysis of prompt photon production in $pp$
collision at $\sqrt{s} = 8$~TeV accompanied by $c$-jets. This investigation is
performed within two methods: the use of Monte Carlo generator SHERPA
including the NLO corrections of collinear QCD and the ``combined'' QCD approach,
which includes both collinear and non-collinear sets of \qcd. From comparison
of these theoretical calculations with the first LHC (ATLAS) data about
$pp \to \gamma + c + X$ process the constraints on the intrinsic charm
content in proton are found.
In Section~\ref{sec:z_c_jet} we present the theoretical analysis of $Z$-boson
production in $pp$ collision at $\sqrt{s} = 8$ TeV and 13 TeV accompanied by
$c$-jets. It is also performed within the SHERPA NLO Monte Carlo generator and
the combined \qcd. The intrinsic charm contributions to the proton \pdf{} are
taken into account. Therefore, the results presented in
Section~\ref{sec:z_c_jet} could be considered as theoretical predictions for
incoming ATLAS search for the \ic{} signal in the process $pp \to Z + c + X$ at
$\sqrt{s} = 13$ TeV. In Section~\ref{sec:future} we discuss future experiments,
which can give more precise information on intrinsic heavy quark distributions
in hadron.

%
%
\section{Structure of the Proton}%
\label{sec:proton_structure}

\subsection{Nucleon structure functions within the collinear QCD approach}%
\label{sec:collinear_pdf}

\begin{figure}[h]
  \centering
  \includegraphics[width=\spw]{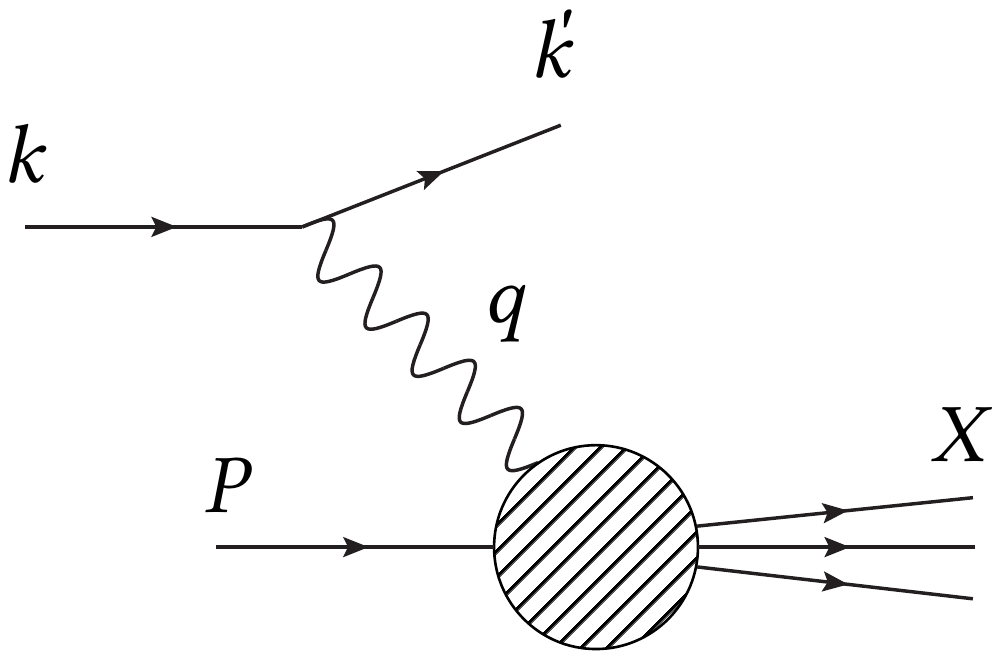}
  \caption{Illustration of process $lN \to l'X$.}%
  \label{fig:physics_dis}
\end{figure}

The structure of the proton is traditionally studied in process called
\gls{dis}, i.e.\ $lN \to l'X$. The illustration of this process is shown in
Fig.~\ref{fig:physics_dis}.

In order to understand the kinematics of this process one needs to describe
several kinematic variables. This variables are invariants.

\begin{equation}
  Q^{2} = -q^{2} = - {(k - k')}^{2} \qquad
  \nu = \frac{q \cdot P}{M} \qquad
  x = \frac{Q^{2}}{2 q \cdot P} \qquad
  y = \frac{q \cdot P}{k \cdot P}
\end{equation}

Starting with $Q^{2}$, it is the size of the momentum transferred to the proton.
Incident lepton's energy loss in nucleon rest frame is $\nu$. The \x{} is in
quark-parton model, described later, identified with the fraction of the
proton's momentum carried by the struck parton (quark or gluon). The $y$ is the
fraction of the lepton's energy lost in the nucleon rest frame, also known as
inelasticity, and $s$ is the center-of-mass energy squared of the lepton-nucleon
system. Furthermore, $M$ is the mass of the nucleon before
interaction~\cite{Tanabashi:2018oca}.

In the limit $M^{2}/Q^{2} \to 0$, the double differential cross-section for
\gls{nc} \gls{dis} on unpolarized nucleons can be expressed in terms of three
structure functions~\cite{Glazov:2007dis}

\begin{equation}
  \frac{\mathrm{d}^{2}\sigma^\mathrm{NC}_{e^\mp p}}{\mathrm{d}x \mathrm{d}Q^{2}} =
  \frac{2 \pi \alpha^{2} Y_{+}}{x Q^{4}} \left(
    F_{2} - \frac{y^{2}}{Y_{+}} F_{\mathrm{L}} \pm \frac{Y_{-}}{Y_{+}} x F_{3}
  \right).%
  \label{eq:physics_dis_variables}
\end{equation}

Here $\alpha = 1/137$ is the fine structure constant and $Y_{\pm} = 1 \pm {(1 -
y)}^{2}$. The structure functions are $F_{2}$, $F_{\mathrm{L}}$ and $xF_{3}$ out
of which the main source of information on the proton structure comes from the
$F_{2}$ structure function.

A leading contribution to the \gls{dis} cross-section has the $F_{2}$ structure
function. It can also be easily experimentally accessed across a broad kinematic
plane and its extraction from \gls{dis} cross-section is usually not complicated
by the other structure functions. The $xF_{3}$ structure function arises from
the $\gamma Z$ interference and becomes significant only at higher $Q^2$.
Experimentally it is much harder to access it. The $F_{\mathrm{L}}$ contribution
becomes negligible compared to the experimental uncertainties for $y < 0.35$. It
vanishes in \gls{lo} \gls{qcd} for spin 1/2 quarks and is known as Callan-Gross
relation~\cite{Callan:1969uq}. It is challenging to measure it and at high $y$
it is hard to decouple the $F_{2}$ and the $F_{\mathrm{L}}$ functions.

Two experimentally observed features lead to the establishment of the
quark-parton model~\cite{Bjorken:1969ja,Feynman:102074}. First it is the
observation, that the structure functions $F_{i}$ scale, i.e.\ they become scale
$Q^{2}$ independent in the Bjorken limit: $Q^{2} \to \infty$ and $\nu \to
\infty$ with fixed \x. This practically means, that the functions $F_{i}$ depend
only on a dimensionless variable $Q^{2}/M\nu$ and not $Q^{2}$ and $\nu$
independently and suggests that there are point-like objects within the
proton. Second, it is the already mentioned vanishing of the $F_{L}$ structure
function, which suggests that these objects have spin 1/2.

In \gls{qcd}, due to strong force, quarks and gluons of proton can radiate
additional gluons, which can convert into $q\bar{q}$ pairs. This circumstance
entails in a logarithmic violation of Bjorken scaling, which can be particularly
large at small $x$. The \gls{qcd} describes the structure functions $F_{i}$ in
terms of scale dependent parton distribution functions $f_{a} (x, \mu^{2})$,
which correspond to the probability to find a parton $a$ with particular
momentum fraction $x$ at scale $\mu$, where $a = g$ or $q$ ($q = u$, $\bar{u}$,
$d$, $\bar{d}$, \dots) and $\mu$ is typically scale of the probe $Q$ (size of
the transferred momentum between lepton and parton). For the $Q^{2} \gg M^{2}$,
the proton structure functions have following form~\cite{Tanabashi:2018oca}
\begin{equation}
  F_i = \sum_a C_{i}^{a} \otimes f_{a},%
  \label{eq:physics_pdf_factorization}
\end{equation}
where $\otimes$ denotes the convolution integral
\begin{equation}
  C \otimes f = \int_{x}^{1} \frac{\mathrm{d} y}{y}\, C(y)\,
                f \left( \frac{x}{y} \right).%
  \label{eq:physics_convolution_definition}
\end{equation}
The coefficient functions $C^{a}_{i}$ are calculated as a perturbation series in
$\alpha_\mathrm{s}$ --- the running coupling of the strong interaction. It is
worth mentioning, that the $C^{a}_{i}$ functions are, apart from being functions
of kinematic variables, also dependent on two scales --- the factorization
scale $\mu_\mathrm{F}$ and renormalization scale $\mu_\mathrm{R}$. At the same
time the scale of the $f_{a}$ function is, in fact, the factorization scale
$\mu_\mathrm{F}$. Typically, the simplifying assumption of a single scale $\mu =
\mu_\mathrm{F} = \mu_\mathrm{R}$ is made. The factorization scale
$\mu_\mathrm{F}$ is the scale determining parton structure and the
renormalization scale $\mu_\mathrm{R}$ is the scale determining the size of
$\alpha_\mathrm{s}$.

Since quasi-free quarks radiate gluons the parton distribution functions evolve
in $\mu$. With increasing $Q^{2}$ more and more gluons are radiated, those in
turn split into $q \bar{q}$ pairs. This process leads to the growth of the
gluon density and the $q \bar{q}$ sea as $x$ decreases. The evolution in $\mu$
of the parton distribution functions is in \gls{qcd} described by the
\gls{dglap} evolution equation~\cite{Gribov:1972ri,Altarelli:1977zs,
Dokshitzer:1977sg}, which has following schematic form~\cite{Tanabashi:2018oca}
\begin{equation}
  \frac{\partial f_{a}}{\partial \ln \mu^{2}} \sim
  \frac{\alpha_\mathrm{s}(\mu^{2})}{2 \pi} \sum_{b} (P_{ab} \otimes f_{b}).%
  \label{eq:physics_dglap}
\end{equation}
Here, the $P_{ab}$ describes the parton splitting $b \to a$ and is also given as
a power series in $\alpha_\mathrm{s}$.

Although the \gls{dglap} can be used to calculate (evolve) the \glspl{pdf} at
any scale above scale \muz{}, which allows application of perturbative theory,
and at any level of precision, it cannot predict them a priory. They have to be
determined by the \gls{qcd} fits to the cross-section data. In the fits any
observable involving a hard hadronic interaction can be used, however \gls{cc}
\gls{dis} cross-section data and data on lepton scattering off the deuteron are
most useful, because they are most sensitive to different quark flavors. One of
the most important experiments in this regard is the \gls{hera}, which covers
the range $0.0005 < \x < 0.05$. Additionally, the \gls{lhc} extends the reach in
both directions of \x, but mainly towards low \x at overall higher $Q^{2}$.

The gauge theory of the strong interaction, i.e.\ \qcd, plays a key role in all
aspects of modern high energy physics. In the theoretical picture the
constituents of the protons (the quarks and gluons, generally called partons)
collide and the interaction between them produces new states, which can be
observed experimentally. The necessary framework to separate hard and soft
partonic physics is provided by the QCD factorization theorem. According to the
factorization theorem, the physical cross-section of any process $\sigma$ can
be decomposed into the universal parton density functions (\pdfs)
$f_a(x,\mu^2)$, describing the distribution of partons inside the initial
state protons, and the perturbatively calculable hard scattering coefficients
$\hat \sigma$ describing the parton-parton collision:

\begin{equation}
  \sigma = \sum_{a,b} f_a(x_1,\mu^2) f_b(x_2, \mu^2) \otimes
                      \hat{\sigma}_{ab}(x_1, x_2, \mu^2),
\label{eq:collfact}
\end{equation}

\noindent where $a,b = q$ or $g$ and $x_1$ and $x_2$ are the longitudinal
momentum fractions of initial protons carried by the interacting partons $a$
and $b$. The QCD factorization theorem is essential to formulate and apply
methods of perturbative resummation at all orders in the QCD coupling constant
and provides theoretical ground for determining parton densities at some energy
scale $\mu_0^2$ from collider data. Then, their QCD evolution (i.e.\ scale
$\mu^2$ dependence) could be obtained by perturbative methods.

\subsection{Intrinsic quark states in nucleon}%
\label{sec:in_vs_ex}

The QCD also predicts additional Fock state contributions to the proton
structure at high \x, such as \uudQQ{} where the heavy quark pair is multiply
connected to two or more valence quarks of the proton.
As it is mentioned above, the heavy quark contributions to the nucleon's \pdf{}
have an enhancement at large \x. Since they depend on the correlations
determined by the valence quark distributions, these heavy quark contributions
are \emph{intrinsic} contributions to the hadron's fundamental structure.
Furthermore, since all of the intrinsic quarks in the \uudQQ{} Fock state have
similar rapidities they can re-interact, leading to significant $Q$ vs
$\widebar{Q}$ asymmetries. In contrast, the contribution to the heavy quark
\pdfs{} arising from gluon splitting are symmetric in $Q$ vs $\bar{Q}$. Since
they only depend on the gluon distribution, the contributions generated by DGLAP
evolution can be considered as \emph{extrinsic} contributions.

\begin{figure}[h]
  \centering
  \includegraphics[width=.65\textwidth]{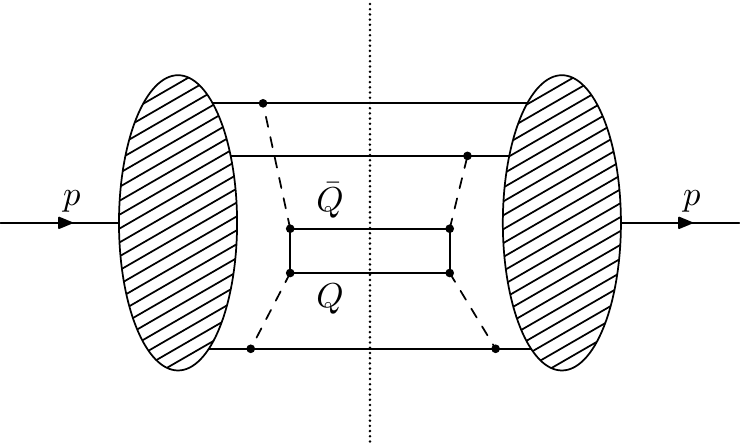}
  \caption{Schematic graph of the $Q\bar{Q}$ pair creation in a nucleon.}%
  \label{fig:QQbar}
\end{figure}

The \pdfs{} at a fundamental level are computed from the squares of the hadrons'
light-front wavefunctions, the frame-independent eigensolutions of the \qcd{}
Light-Front Hamiltonian. The intrinsic contributions are associated with
amplitudes such as $g g \to Q \bar Q \to g g$ in the self energy of the proton,
the analogs of light-by light scattering
$\gamma \gamma \to \ell \bar \ell \to \gamma \gamma$ in \gls{qed}, i.e., twist-6
contributions proportional to the gluon field strength cubed the operator
product expansion (OPE). This is illustrated in Fig.~\ref{fig:QQbar}. Thus the
OPE provides a first-principle derivation for the existence of intrinsic heavy
quarks. Unlike the conventional $\log m^2_\mathrm{Q}$ dependence of the low \x{}
extrinsic gluon-splitting contribution, the probabilities for the intrinsic
heavy quark Fock states at high \x{} scale as $1 / m_\mathrm{Q}^2$ in
non-Abelian \qcd.  In contrast the probability for a higher Fock state in an
atom such as $|e^+ e^- \ell \bar{\ell} \, \rangle$ in positronium scales as
$1 / m_\ell^4$ in Abelian \gls{qed}, corresponding to the twist-8
Euler-Heisenberg light-by-light insertion. Detailed derivations based on the
\ope{} have been given in Refs.~\cite{Brodsky:1984nx,Franz:2000ee}.

\subsection{Validation of Intrinsic Heavy Quarks from QCD Lattice Gauge Theory}%
\label{sec:lattice}

\begin{figure}[h]
\centering
  \includegraphics[width=.6\textwidth]{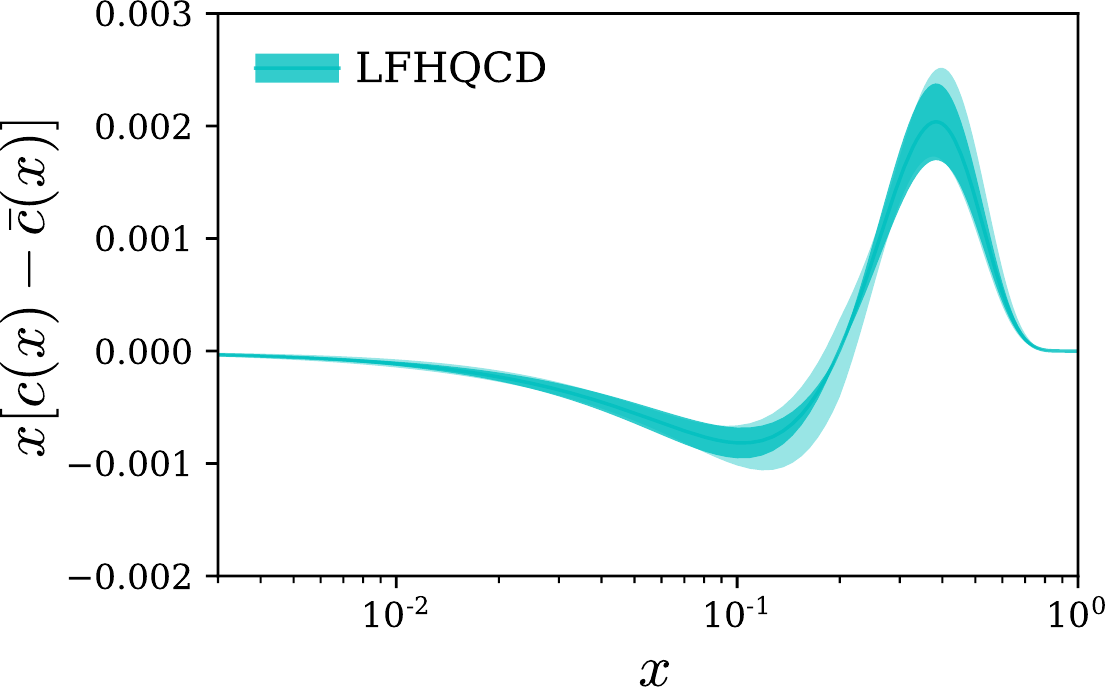}
  \caption{The distribution function $x[c(x)-\bar{c}(x)]$ obtained from the
           LFHQCD formalism using the lattice QCD input of charm electromagnetic
           form factors $G^c_{E,M}(Q^2)$. The outer cyan band indicates an
           estimate of systematic uncertainty in the $x[c(x)-\bar{c}(x)]$
           distribution obtained from a variation of the hadron scale $\kappa_c$
           by 5\%. It was taken from Ref.~\cite{Sufian:2020coz}.}%
           \label{fig:cbarc_asymm}
\end{figure}

In an important recent development~\cite{Sufian:2020coz}, the difference of the
charm and anticharm quark distributions in the proton $\Delta c(\x) = c(\x) -
\bar{c}(\x)$ has been computed from first principles in \qcd{} using lattice
gauge theory. The results are remarkable. The predicted $c(\x) - \bar{c}(\x)$
distribution is large and nonzero at large at $\x \sim 0.4$, consistent with the
expectations of intrinsic charm. The $c(\x)$ vs. $\bar{c}(\x)$ asymmetry can be
understood physically by identifying the \uudcc{} Fock state with the
$|\Lambda_{udc} D_\mathrm{u\bar{c}} \rangle$ off shell excitation of the proton.
See Fig.~\ref{fig:cbarc_asymm}. A related application of lattice gauge theory to
the non-perturbative strange-quark sea from lattice \qcd{} is given in
Ref.~\cite{Sufian:2018cpj}.

A key theoretical tool in the LGTH analysis of intrinsic charm is the
computation of the charm and anticharm quark contribution to the $G^p_E(Q^2)$
form factor of the proton which would vanish if $c(x) =\bar c(x)$. All of the
heavy quark contributions to $G^p_E(Q^2)$ --- both extrinsic and intrinsic ---
involve a heavy quark loop which couples the photon field tensor $F^{\mu \nu}$
to any odd number of gluon fields $G^{\mu \nu}$. The gauge invariant amplitude
is thus linear in both the photon and gluon momenta --- like light-by-light
scattering which is proportional to $F^4$. The form factor $G_E(t)$ thus
vanishes at $t = 0$. The extrinsic (DGLAP `gluon-splitting') contributions to
$G^p_E(Q^2)$ come from a heavy quark loop in which gluons attach to the same
valence quark in the hadron. The intrinsic contributions come from a heavy quark
loop in which the gluons attach to more than one valence quark in the proton.
One is thus sensitive to the intrinsic structure of the hadron. The multiple
valence quark couplings allows the transfer of the entirety of the hadron's
valence quark momenta to the heavy quark.

The intrinsic heavy quark contribution is maximum in the LFWF at minimum
off-shellness of the invariant mass; i.e.\ at equal rapidity $y_i$, when $x_i$
is proportional to the quark's transverse mass
$\sqrt{k_{\mathrm{T}i}^2 +m_i^2}$. The $c(\x)$ and $\bar{c}(\x)$ are thus large
at $\x \simeq 0.4$ in the proton structure function.

There have been many phenomenological calculations involving the existence of
a non-zero \ic{} component to explain anomalies in the experimental data and to
predict its novel signatures of \ic{} in upcoming
experiments~\cite{Brodsky:2015fna}. The new LGTH results will make these
predictions precise.

\subsection{Higgs production at High $x_\mathrm{F}$ and the Intrinsic
            Heavy-Quark Distributions of the Proton}%
\label{sec:higgs}

\begin{figure}
  \centering
  \includegraphics[width=.6\textwidth]{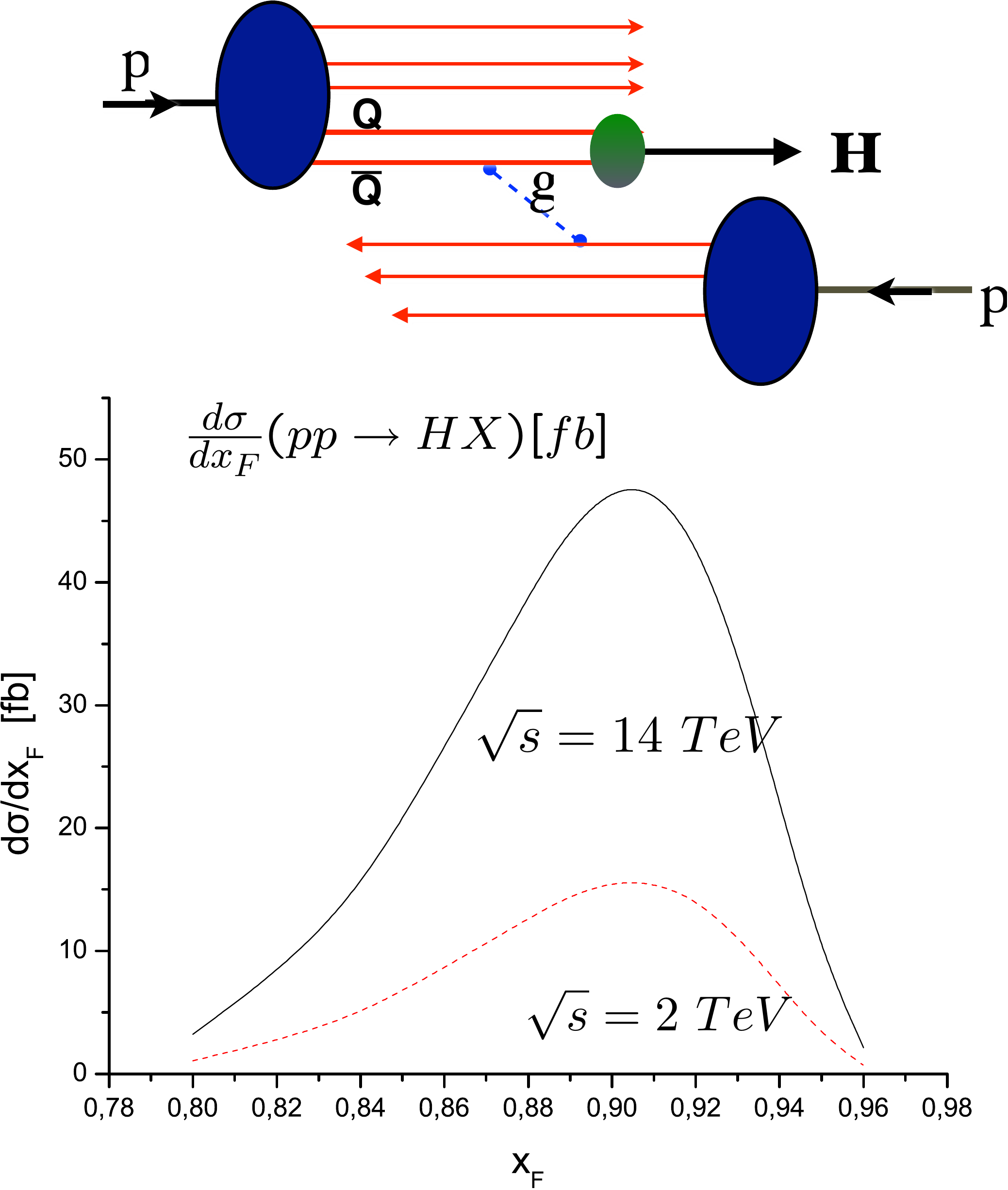}
  \caption{Intrinsic Heavy Quark Mechanism and cross-section for Higgs
           production at LHC and Tevatron energies. It was taken from
           Ref.~\cite{Brodsky:2007yz}.}%
  \label{fig:higgs}
\end{figure}

The conventional \pqcd{} mechanisms for Higgs production at the \lhc, such as
gluon fusion $gg \to H$, lead to Higgs boson production in the central rapidity
region. However, the Higgs can also be produced at very high \xf{} by the
process $[Q\widebar{Q}] + g \to H$~\cite{Brodsky:2007yz}, where both heavy
quarks from the proton's five quark Fock state \uudQQ{} couple directly to the
Higgs, see Fig.~\ref{fig:higgs}. Since the Higgs couples to each quark proportional
to its mass, one has roughly equal contributions from intrinsic
$s\bar{s}$, $c\bar{c}$, $b\bar{b}$ and even $t\bar{t}$ Fock states. The
intrinsic heavy-quark distribution of the proton at high \x{} leads to Higgs
production with as much as 80\% of the beam momentum. One can also use the \xf{}
distribution of the produced Higgs boson to discriminate Higgs production from
strange, charm, and bottom quarks. The same intrinsic mechanism produces the
$J/\psi$ at high \xf{} as observed in fixed-target experiments such as
NA3.

The decay of the high-\xf{} Higgs to muons could be observed using very forward
detectors at the \lhc. The predicted cross-section ${d\sigma/d \xf}(pp \to HX)$
for Higgs production at high $\xf \sim 0.8$ computed in
Ref.~\cite{Brodsky:2007yz} is of order of 50~fb. The corresponding
double-diffractive rate for $pp \to HppX$ was computed in
Ref.~\cite{Brodsky:2006wb}. Testing these diffractive Higgs production
predictions would open up a new domain of Higgs physics at the
LHC~\cite{Brodsky:2006wb}.

%
%
\section{The Fock State Structure of Hadrons from Non-Perturbative QCD}%
\label{sec:fock}

The masses of hadrons and their quark and gluon composition predicted by quantum
chromodynamics are given by the eigenvalues and eigenfunctions of the \qcd{}
\lf{} Hamiltonian:

\begin{equation}
  H^{\mathrm{QCD}}_{\mathrm{LF}} |\Psi_\mathrm{H}\rangle =
  M^2_\mathrm{H} |\Psi_\mathrm{H}\rangle.
\end{equation}

\noindent The light-front formalism, which is based on Dirac's Front Form
(quantization at fixed LF time $\tau = t+z/c$) is causal and frame-independent.

Each  hadronic (and nuclear) eigenstate of the QCD light-front (LF) Hamiltonian
is built on LF Fock states $|n\rangle$, the color-singlet eigenstates of the
free LF Hamiltonian $H^0_\mathrm{LF}$.

\begin{equation}
  |\Psi_\mathrm{H}\rangle =
  \sum_n \Psi^\mathrm{H}_n(\x[_i], k_{\mathrm{T} i}, \lambda_i) |n\rangle
\end{equation}

\noindent where $\x = {k^{+} / P^{+}}$ is the LF momentum fraction. The
coefficients $\langle n|H \rangle$ in the Fock state expansion, starting with
the valence Fock state of the hadron ($n = 3$ for baryons) are the LF
wavefunctions $\Psi^\mathrm{H}_n(\x[_i], k_{\mathrm{T} i}, \lambda_i)$ which
underlay hadronic observables such as form factors, structure functions,
distribution amplitudes, etc.

In principle the hadronic eigenstates can be computed by diagonalizing
$H^\mathrm{QCD}_\mathrm{LF}$ on the Fock basis, as in the DLCQ and BLFQ methods
they have been developed based on LF holography, the duality of $AdS_5$ space
with physical 3+1 LF spacetime at fixed $\tau$. Lattice gauge theory methods
have been developed using the correspondence of LF Hamiltonian theory with
ordinary instant-time quantization in a Lorentz frame where the observer moves
at infinite momentum $P^z\to \-\infty$.

The \lf{} Hamiltonian theory for holographic AdS/QCD has led to novel
perspectives for the non-perturbative \qcd{} structure of hadrons such as: the
quark-antiquark structure of mesons, the quark-diquark structure of baryons, and
the diquark-antidiquark structure of tetraquarks. For example, the \lf{}
holographic AdS/QCD approach, combined with superconformal
algebra~\cite{Brodsky:2015oia} predicts that the three-quark valence state of
the proton has the configuration $|u [ud] \rangle$ where $[ud]$ is a scalar
diquark with color $\bar{3}_\mathrm{c}$. This approach gives systematic
accounting of observed hadron spectroscopy including the massless pion in the
chiral limit. It also predicts supersymmetric 4-plet relations between the
meson, baryon and tetraquark eigenstates and their Regge trajectories with
universal slopes. In the case of the deuteron, the \lf{} Fock expansion of the
$I = 0$, $J = 1$ deuteron valence Fock state is expanded on five color-singlet
combinations $|uuuddd\rangle$ of six $u$ and $d$ quarks. Only one of these Fock
state corresponds to the standard two nucleon state $|np\rangle$. The other
``hidden color'' Fock states are relevant for deuteron phenomena at short
distances such as the deuteron form factor at large momentum transfer or the
deuteron structure function at large \xbj.

The higher Fock states of hadrons with additional quarks, antiquarks and gluons
are particularly interesting and are the focus of this review. For example, the
five-quark Fock components of the proton eigenstate, such as \uudqq{} where
$q=u$, $d$, $s$, $c$ or $b$, are non-valence quark contributions to the proton's
five-quark \lf{} wavefunction. The square of this LFWF gives the sea-quark
contributions to the proton's structure function
$F^{2p}_{q} (x_\mathrm{Q}, Q^2) $.

Part of the sea quark structure function arise from perturbative \qcd{}
gluon-splitting processes $g \to q \bar{q}$ corresponding to DGLAP evolution.
These contributions are called \emph{extrinsic}, since they arise from \pqcd{}
processes which are independent of the parent hadron's structure.

The extrinsic and intrinsic sea contributions have distinctive properties. Since
the extrinsic contributions arise dominantly from perturbative subprocesses such
as ${u \to u + g \to u + c \bar c}$, the resulting extrinsic $c(x)$ and
$\bar{c}(x)$ distributions are identical (at least at leading order) and are
dominantly produced at low \x. In contrast, the intrinsic contributions to
$c(x)$ and $\bar{c}(x)$ are coupled to all of the valence quarks of the proton
thus are not symmetric. Moreover, since the intrinsic contribution to \uudcc{}
is maximal when the mass $M^2$ of the five-quark Fock state is minimally
off-shell where

\begin{equation}
  M^2 = \sum_{q = uudc\bar{c}}
        \frac{k^2_{\mathrm{T}\,q} + m_q^2}{x_q}.
\end{equation}

\noindent This occurs when $x_q$ is proportional to its transverse mass
$x_q \propto \sqrt {k^2_{\mathrm{T}\,q} + m_q^2}$; i.e., when the five
quarks have equal rapidity. The intrinsic charm quarks in the proton are thus
predicted to have most of the proton's LF momentum, typically
$x_\mathrm{c} \simeq x \bar{c} \simeq 0.4$. This is consistent with the EMC
measurement of the proton's structure function, which is approximately 30 times
larger than the extrinsic DGLAP contribution at $\x = 0.42$ and
$Q^2 = 75$~GeV$^2$.

There have been many phenomenological calculations involving the existence of a
non-zero \ic{} component to explain anomalies in the experimental data and to
predict its novel signatures in upcoming experiments~\cite{Brodsky:2015fna}. The
new LGTH results will make these predictions more precise.

%
%
\section{The Hadronic Phenomenology of Intrinsic Heavy Quarks: An Overview}%
\label{sec:overview}

The existence of intrinsic heavy quarks in the proton leads to a broad array of
heavy hadron production processes in the high \xf{} forward domain at the EIC
and \lhc{} colliders. When a proton collides with other protons at the \lhc{} or
in a fixed target experiment, the heavy quark Fock states in the proton, such as
\uudcc{} are materialized and can produce open or hidden charm states at high
momentum fraction \xf. For example, the comoving $udc$ quarks in a Fock state
such as \uudcc{} can coalesce to produce a $\Lambda_\mathrm{c}(udc)$ baryon with
a high Feynman momentum fraction
$\xf = x_\mathrm{c} + x_\mathrm{u} + x_\mathrm{d}$ or produce a $J/\psi$ with
$\xf = c \bar{c}$. Such high \xf{} heavy hadron
events have been observed and measured with substantial cross-sections at the
ISR proton-proton collider and at fixed target experiments such as NA3 at CERN
and SELEX at Fermilab. Intrinsic charm components in the proton can
explain~\cite{Brodsky:1980pb} the large cross-section for the forward open charm
production in $pp$ collision at ISR energies~\cite{Drijard:1978gv,Giboni:1979rm,
Lockman:1979aj,Drijard:1979vd}. The $\Lambda_\mathrm{b}(udb)$ baryon was first
observed at the ISR in forward $pp \to \Lambda_\mathrm{b} X$ reactions at high
\xf{} as expected from intrinsic bottom.

The first direct experimental indication for the intrinsic heavy quarks in a
nucleon was observed in the EMC deep inelastic muon experiment at CERN\@.
The measurement of the charm structure function at high \xbj{} by the EMC
experiment at CERN using deep inelastic muon-nucleus scattering showed a
significant contribution to the proton structure function at large
\xbj~\cite{Aubert:1982tt}. In fact, the charm structure function $c(x,Q)$
measured by the EMC collaboration was approximately 30 times higher than
expected from gluon splitting and at $\xbj = 0.42$ and $Q^2 = 75~GeV^2$.

The effect of whether the \ic{} parton distribution is either included or
excluded in the determinations of charm parton distribution functions (PDFs)
can induce changes in other parton distributions through the momentum sum
rule, which can indirectly affect the analyses of various physical processes
that depend on the input of various PDFs. On the experimental side, an
estimate of intrinsic charm ($c$) and anticharm ($\bar{c}$) distributions can
provide important  information to the understanding of charm quark production
in deep inelastic $l p\to l'cX$ scattering in the EMC
experiment~\cite{Aubert:1982tt}. The enhancement of charm distribution in the
measurement  the charm quark structure function $F_2^c$ compared to the
expectation from the gluon splitting mechanism in the EMC experimental data
has been interpreted as evidence for nonzero \ic{} in several
calculations~\cite{Brodsky:1984nx,Harris:1995jx,Hoffmann:1983ah,Brodsky:1991dj}.
A precise determination of charm and anticharm PDFs by considering both the
perturbative and non-perturbative contributions is important in understanding
charmonia and open charm productions, such as the $J/\psi$ production at large
momentum from $pA$ collisions at CERN~\cite{Badier:1983dg}, from $\pi A$
collisions at FNAL~\cite{Leitch:1999ea}, from $pp$ collisions at
LHC~\cite{Aaij:2018ogq}, and charmed hadron or jet production from $pp$
collisions at ISR, FNAL, and
LHC~\cite{Aaij:2018ogq,Chauvat:1987kb,Aitala:2000rd,Aitala:2002uz}. 

The cross-sections for forward heavy quark or quarkonium production include
contributions from diffractive reactions such as $\gamma^* p\to Q + X + p$,
where the proton target remains intact. The final-state interactions of the
outgoing state can lead to additional strong nuclear effects not associated with
shadowing of the nuclear structure functions~\cite{Vogt:1992ki}. The
interference of different amplitudes leads to shadowing and flavor-specific
antishadowing of the DIS cross-section on nuclei. An important consequence is
the inapplicability of the OPE and the violation of the momentum sum rule for
nuclear structure functions, see Ref.~\cite{Brodsky:2019jla}.

An investigation of prompt photon and $c(b)$-jet production in $p\bar{p}$
collisions at ${\sqrt{s} = 1.96}$~TeV was carried out at the
Tevatron~\cite{Abazov:2009de,D0:2012gw,Aaltonen:2009wc,Abazov:2012ea}. The
observed cross-section for $p\bar{p} \to \gamma cX$ is significantly larger than
predictions without the \ic{} contribution at photon transverse momenta above
110~GeV --- by a factor of 3. The ratio of the cross-section using the NLO
calculations of the \pt-spectrum is consistent with the BHPS model and CTEQ66c
with intrinsic charm probability in the proton about
3.5~\%~\cite{Brodsky:2016fyh}. In the case of the prompt photon production
accompanied by the $b$-jet in $p\bar{p}$ annihilation, the Tevatron data do not
show any signal of the intrinsic $b$ contribution, as expected from the small
intrinsic beauty probability in a proton.

\lhc{} measurements associated with cross-section of inclusive production of
Higgs, $Z$, $W$ bosons via gluon-gluon fusion, and productions of charm jet
and $Z^0$~\cite{Aad:2016naf,Aad:2014xaa,Aad:2015auj,Khachatryan:2015oaa},
$J/\psi$ and $D^0$ mesons at  LHCb experiment~\cite{Aaij:2018ogq} can also be
sensitive to the  intrinsic charm distribution. The $J/\psi$ photo- or
electro-productions near the charm threshold is sensitive to intrinsic charm;
experiments have been proposed at JLab
as well as  for the future EIC to measure the production cross-section near
the threshold. The existence of \ic{} in the proton will provide additional
production channels and thus enhance the cross-section for both open and
hidden charm, especially near threshold~\cite{Brodsky:2000zc}.
If the $c$ and $\bar{c}$ quarks have different distributions in the proton,
the enhancements on $D$ and $\bar{D}$ productions will appear at slightly
different kinematics. \ic{} has also been proposed to have an impact on
estimating the astrophysical neutrino flux observed at the IceCube
experiment~\cite{Laha:2016dri}.

A recent calculation of the intrinsic charm contribution to the production of
double charm baryons at both colliders and fixed target experiments is given in
Ref.~\cite{Vogt:2019kbd}. The resolution of the SELEX-LHCb double-charm baryon
conflict between SELEX and LHCb due to intrinsic heavy-quark hadroproduction is
given in Ref.~\cite{Brodsky:2017ntu}.

An earlier review of collider tests of heavy quark distributions is given in
Ref.~\cite{Brodsky:2016fyh}. The constraints on the intrinsic charm content of
the proton that can be obtained from ATLAS data is given in
Ref.~\cite{Bednyakov:2017vck}. A global analysis of intrinsic charm signals in
the nucleon is given in Ref.~\cite{Brodsky:2015uwa}.

The elimination of renormalization scale and scheme ambiguities in \pqcd{}
predictions for hard \qcd{} processes will greatly improve predictions for
intrinsic heavy quark cross-sections, especially for EIC tests. Recent
applications of the BLM/PMC method to jet production and Heavy Quark Pair
Production in $e^{+}e^{-}$ annihilation are given in
Refs.~\cite{Wang:2019isi,Wang:2020ckr}. The presence of intrinsic heavy quarks
in the Fock states of light hadrons can also lead to new signals such as novel
effects in $B$ decay~\cite{Brodsky:2001yt} and the resolution of issues, such as
the $\rho-\pi$ puzzle~\cite{Brodsky:1997fj}.

%
%
\section{Nucleon structure functions within the non-collinear QCD approach}%
\label{sec:ktfact}

\subsection{High energy factorization in QCD}%

The most familiar evolution strategy is based on
Dokshitzer-Gribov-Lipatov-Altarelli-Parisi (DGLAP)~\cite{Gribov:1972ri,
Lipatov:1974qm,Altarelli:1977zs,Dokshitzer:1977sg} evolution equations mentioned
above. In this way large logarithmic terms proportional to
$\alphas[^n] \ln^n \mu^2/\lambdaqcd[^2]$ are resummed to all orders, thus
rearranging the perturbative expansion into a more rapidly converging series.
The dominant contributions come from diagrams where parton emissions in initial 
state are strongly ordered in virtuality. This is called collinear QCD
factorization, as the strong ordering means that the virtuality of the
parton entering the hard scattering amplitude can be neglected compared to
the large scale $\mu^2$. Thus, the parton interaction proceeds in the plane
spanned by the initial protons and only their longitudinal momentum fractions
$x_1$ and $x_2$ and the scale $\mu^2$ are relevant. Such a one dimensional
collinear treatment is typically valid for single scale observables and has
formed the basis for \qcd{} applications at colliders. In particular, by using
input parton densities which are sufficiently singular at $\x \to 0$, this
formalism reproduces the strong rise of deep inelastic 

For many processes studied at the modern colliders, the hard scattering
coefficients $\hat{\sigma}_{ab}$ are calculated not only at \lo{} in the
perturbative expansion, but also at higher orders --- \nlo{} and, in some cases,
even at \nnlo. However, the collinear DGLAP-based scenario meets some
difficulties in the description of multi-scale and/or non-inclusive collider
observables. A classical example could be given by Drell-Yan hadroproduction of
electroweak gauge bosons. So, in the transverse momentum distributions of $Z$
bosons one can distinguish the three kinematical regions, namely, high-\pt{}
region, the peak region and the low-\pt{} region. In the high-\pt{} region the
measured cross-sections are well reproduced by the \nlo{} \pqcd{} calculations
performed using \textsc{madgraph5}\_a\textsc{mc@nlo}~\cite{Alwall:2014hca},
\textsc{powheg}~\cite{Nason:2004rx,Frixione:2002ik,Alioli:2010xd,Alioli:2008gx}
and \textsc{powheg-minnlo}~\cite{Hamilton:2012rf} tools. On the other hand, if
fixed-order \pqcd{} calculations are performed to the region of decreasing \pt,
they will not be able to describe the data at the peak region, where
$\pt \sim 5$~GeV, nor the turn-over region, where $\pt \sim 1$~GeV, since they
diverge as \pt{} decreases. The reason for this is that the physical behavior of
the $Z$ boson transverse momentum distribution near the peak and below is
governed by multi-parton \qcd{} radiation~\cite{Dokshitzer:1978yd,Parisi:1979se}
(with terms proportional to $\alphas[^n] \ln^n m_\mathrm{Z}^2/\pt[^2]$), which is
not well approximated by truncating the \qcd{} perturbation series to any fixed
order. To describe the data in these kinematical regions the special methods to
resum arbitrarily many parton emissions are needed (so called soft gluon
resummation technique~\cite{Gawron:2003np}, that regularizes the infrared
divergences). Usually, the fixed-order \pqcd{} calculations are combined with
higher order parton radiation via parton showers~\cite{Alwall:2014hca,
Nason:2004rx,Frixione:2002ik,Alioli:2010xd,Alioli:2008gx,Hamilton:2012rf}, that
significantly improve an overall description of the data.

Taking into account higher order parton emissions can be performed in a
systematic manner via a generalized form of \qcd{} factorization, which involves
quark and gluon distribution functions that include information on the
transverse momenta~\cite{Collins:1984kg,Collins:1982wa,Collins:2011zzd}. These
\tmd{} parton densities obey Collins-Soper-Sterman evolution
equations\cite{Collins:2011zzd,Collins:1981uw,Collins:1981va}. The latter allow
one to resum logarithmically enhanced terms
$\alphas[^n] \ln^n m_\mathrm{Z}^2/\pt[^2]$ in perturbative expansion to all
higher orders in the \qcd{} coupling and generalize the ordinary renormalization
group evolution equations. This generalized factorization analysis (CSS
approach), going beyond the conventional (collinear) approximation, can
reproduce the physical behavior of the measured transverse momentum
distribution.

Next example concerns the hadroproduction processes in $pp$ collisions in
another kinematical limit, $s \to \infty$ for fixed momentum transfer. In this
limit, as we push forward the high-energy frontier, more and more events having
small momentum fraction $\x \sim \mu / \sqrt{s}$ contribute to processes probing
short distance physics, so that fraction of momentum carried by transverse
degrees of freedom becomes increasingly important. The perturbative higher-order
corrections to the proton structure functions at small \x{} are known to be
large. These corrections come from multiple radiation of gluons over long
intervals in rapidity~\cite{Gribov:1984tu,Levin:1991ry,Mueller:1993rr}, not
ordered in the transverse momenta \kt{} and are present beyond \nnlo{} to all
orders of perturbation theory~\cite{Catani:1993rn,Catani:1994sq}. Similar to the
CSS approach, the theoretical framework to resum the unordered multi-gluon
emissions is a generalized form of \qcd{} factorization~\cite{Gribov:1984tu,
Levin:1991ry,Catani:1990xk,Catani:1990eg,Collins:1991ty} in terms of \tmd{}
parton distribution functions obeying the appropriate evolution equation. In
the small \x{} region, the theoretically correct description is given by the
Balitsky-Fadin-Kuraev-Lipatov (BFKL)~\cite{Kuraev:1976ge,Kuraev:1977fs,
Balitsky:1978ic} equation, which allows one to resum logarithmically enhanced
corrections proportional to
$\alphas[^n] \ln^n s \sim \alphas[^n] \ln^n 1/\x$ to all higher orders in
the QCD coupling. The Catani-Ciafaloni-Fiorani-Marchesini (CCFM) evolution
equation~\cite{Ciafaloni:1987ur,Catani:1989yc,Catani:1989sg,Marchesini:1994wr}
resums large logarithmic terms proportional to $\alphas[^n] \ln^n 1/x$ and
$\alphas[^n] \ln^n 1/(1 - x)$ and, therefore, is valid at both small and large
\x. So that, these equations provide another generalization of the ordinary
renormalization-group evolution and add a new physical dimension, the
transverse momentum, to the factorization ansatz:

\begin{equation}
  \sigma = \sum_{a,b} f_a(x_1, \mathbf{k}_\mathrm{T\,2}^2, \mu^2)
                      f_b(x_2, \mathbf{k}_\mathrm{T\,2}^2, \mu^2) \otimes
                      \hat{\sigma}_{ab}^{*}(x_1, x_2,
                                            \mathbf{k}_\mathrm{T\,1}^2,
                                            \mathbf{k}_\mathrm{T\,2}^2, \mu^2).
  \label{eq:ktfact}
\end{equation}

\noindent The hard scattering is no longer collinear with the colliding protons
and both the parton density functions $f_a$, $f_b$ and hard scattering
coefficients $\hat{\sigma}_{ab}^{*}$ depends on the non-zero transverse momenta
of interacting quarks and gluons\footnote{Calculation of transverse momentum
dependent hard scattering coefficients (off-shell partonic amplitudes) is
explained in the Section~\ref{sec:soft_hard}.}. The gauge-invariant operator
definitions of the \tmd{} parton distributions can be given in terms of nonlocal
operator combinations~\cite{Collins:2011zzd,Bomhof:2007xt,Dominguez:2011wm,
Buffing:2012sz,Buffing:2013kca,Boer:2015kxa}. Unlike the CSS approach, valid at
low transverse momenta at fixed invariant masses, the high-energy factorization,
or \kt-factorization, based on the BFKL or CCFM equations, is valid for
arbitrarily large momentum transfer. In particular, it allows one to obtain the
structure of logarithmic scaling violations in \dis{} at high energies (see
Refs.~\cite{Vogt:2004mw,Vermaseren:2005qc,Moch:2004xu}) and resum logarithmic
corrections of higher order in \qcd{} coupling to Higgs and top quark production
cross-sections~\cite{Luisoni:2015xha,Czakon:2013goa}. Besides that,
\kt-factorization can be applied to a variety of processes studied at the \lhc.
Special area for its applications concerns single spin asymmetries and azimuthal
asymmetries in polarized collisions (see, for example,
Refs.~\cite{Sivers:1989cc,Brodsky:2002cx,Collins:2002kn}). Thus, nowadays it has
become a widely exploited tool and it is of interest and importance to test it
in as many cases as possible.

\subsection{Off-shell partonic amplitudes} \indent

The calculation of partonic amplitudes follows the standard Feynman rules, with
the exception that the initial quarks and gluons are off-shell. Off-shell gluons
may have nonzero transverse momentum and an admixture of longitudinal component
in the polarization vector. In accordance with the \kt-factorization
prescriptions, the initial gluon spin density matrix is taken in the
form~\cite{Gribov:1984tu,Levin:1991ry}:

\begin{equation} \label{gauge}
  \sum \epsilon_g^\mu \epsilon_g^{*\nu} =
    \kt[^\mu] \kt[^\nu]/|\kt|^2.
\end{equation}

\noindent In the collinear limit, when $\kt \to 0$, this expression converges
to the ordinary
${\sum \epsilon_g^\mu \epsilon_g^{*\nu} = -g^{\mu\nu}/2}$.
This property provides continuous on-shell limit for the partonic amplitudes.

The off-shell amplitudes for quark induced subprocesses can be derived in the
framework of the reggeized parton approach~\cite{Lipatov:2000se,Bogdan:2006af,
Hentschinski:2011tz,Hentschinski:2011xg}. The latter is based on the effective
action formalism~\cite{Lipatov:1995pn,Lipatov:1996ts}, that ensures the gauge
invariance of obtained amplitudes despite the off-shell initial interacting
partons. One can also use Britto-Cachazo-Feng-Witten (BCFW) recursion for
off-shell gluons~\cite{vanHameren:2014iua} and method of auxiliary quarks for
off-shell quarks~\cite{vanHameren:2013csa}, implemented in the Monte-Carlo
generator \textsc{katie}~\cite{vanHameren:2016kkz}.

\subsection{CCFM evolution equation}%
\label{sec:ccfm}

As it was noted above, the CCFM gluon evolution equation resums large logarithms
$\alphas[^n] \ln^n 1/(1 - \x)$ in addition to BFKL ones $\alphas[^n] \ln^n 1/\x$
and introduces angular ordering of initial emissions to correctly treat gluon
coherence effects. In the limit of asymptotic energies, it is almost equivalent
to BFKL, but also similar to the DGLAP evolution for large
\x~\cite{Ciafaloni:1987ur,Catani:1989yc,Catani:1989sg,Marchesini:1994wr}. In the
leading logarithmic approximation (LLA), the CCFM equation for \tmd{} gluon
density with respect to the evolution (factorization) scale $\mu^2$ can be
written as

\begin{align}
  \nonumber f_g(\x, \ktvec[^2], \mu^2) &=
      f_g^{(0)}(\x, \ktvec[^2], \muz[^2])
      \Delta_\mathrm{s} (\mu, \mu_0) \\
    &\quad + \int\limits_x^{z_\mathrm{M}} \frac{\mathrm{d}z}{z}
             \int \frac{\mathrm{d} q^2}{q^2}
             \Theta(\mu - zq)\Theta(q - \muz) \Delta_\mathrm{s}(\mu, zq)
             \tilde{P}_{gg}(z, \ktvec[^2], q^2)
             f_g \left(\frac{x}{z}, \ktvec[^{\prime \, 2}],
                                q^2\right),
  \label{eq:ccfm}
\end{align}

\noindent where
$\ktvec[^\prime] = \mathbf{q}(1 - z) + \ktvec$, \muz{} is the soft starting
scale of the evolution, $z_\mathrm{M} = 1 - \muz/\mu$ is a resolution
parameter\footnote{It was shown in Ref.~\cite{Hautmann:2017fcj} that with
$z_\mathrm{M}$ parameter virtual and resolvable branchings are treated
consistently at $z_\mathrm{M} \to 1$.} and
$\tilde{P}_{gg}(z, \ktvec[^2], q^2)$ is the CCFM splitting
function:

\begin{align}
  \nonumber \tilde{P}_{gg}(z, \ktvec[^2], q^2) &=
      \bar{\alpha}_\mathrm{s}(q^2{(1 - z)}^2)
      \left[\frac{1}{1 - z} + \frac{z(1 - z)}{2}\right] \\
    &\quad + \bar{\alpha}_\mathrm{s}(\ktvec[^2])
      \left[\frac{1}{z} - 1 + \frac{z(1 - z)}{2}\right]
      \Delta_\mathrm{ns}(z, \kt[^2], q^2).
  \label{eq:ccfmPgg}
\end{align}

\noindent
The Sudakov and non-Sudakov form factors read:
\begin{equation}
  \ln \Delta_\mathrm{s}(\mu, \muz) =
    - \int\limits_{\muz[^2]}^{\mu^2}
      \frac{\mathrm{d}\mu^{\prime \, 2}}{\mu^{\prime \, 2}}
      \int\limits_0^{z_\mathrm{M}} \mathrm{d}z\,
      \frac{\bar{\alpha}_\mathrm{s}(\mu^{\prime \, 2}{(1 - z)}^2)}{1 - z},
  \label{eq:ccfmSudakov}
\end{equation}

\begin{equation}
  \ln \Delta_\mathrm{ns}(z, \ktvec[^2], \mathbf{q}_\mathrm{T}^2) =
    - \bar{\alpha}_\mathrm{s}(\ktvec[^2])
      \int\limits_0^1 \frac{\mathrm{d}z^\prime}{z^\prime}
      \int\frac{\mathrm{d}q^2}{q^2}
      \Theta(\ktvec[^2] - q^2)
      \Theta(q^2 - z^{\prime\,2} \mathbf{q}^2_\mathrm{T}).
  \label{eq:ccfmNonSudakov}
\end{equation}

\noindent
where $\bar{\alpha}_\mathrm{s} = 3 \alpha_\mathrm{s}/\pi$ and the
$z^\prime$-integral in Eq.~\ref{eq:ccfmNonSudakov} is finite due to the theta
functions~\cite{Kwiecinski:1995pu}. The first term in the CCFM equation, which
is the initial \tmd{} gluon density multiplied by the Sudakov form factor,
corresponds to the contribution of non-resolvable branchings between the
starting scale $\muz[^2]$ and scale $\mu^2$. The second term describes the
details of the \qcd{} evolution expressed by the convolution of the CCFM gluon
splitting function with the gluon density and the Sudakov form factor. The theta
function introduces the angular ordering condition. The evolution scale $\mu^2$
is defined by the maximum allowed angle for any gluon
emission~\cite{Ciafaloni:1987ur,Catani:1989yc,Catani:1989sg,Marchesini:1994wr}.
A similar equation also can be written~\cite{Deak:2010gk} for valence quark
densities\footnote{The sea quarks are not defined in CCFM\@. However, they can
be obtained from the gluon densities in the last gluon splitting approximation,
see Ref.~\cite{Hautmann:2012sh}.} (with replacement of the gluon splitting
function by the quark one). Usually, the initial \tmd{} gluon and valence quark
distributions are taken as

\begin{equation}
  \x f_g^{(0)}(\x, \ktvec[^2], \muz[^2]) =
      N \x^{-\mathrm{B}}{(1 - \x)}^\mathrm{C}
      \exp(-\ktvec[^2]/\sigma^2),
  \label{eq:ccfmfg0}
\end{equation}

\begin{equation}
  \x f_{q_v}^{(0)}(\x, \ktvec[^2], \muz^2) =
      \x q_v(\x, \muz[^2]) \exp(-\ktvec[^2]/\sigma^2)/\sigma^2,
  \label{eq:ccfmfq0}
\end{equation}

\noindent
where $\sigma = \muz/\sqrt{2}$ and $q_{v}(\x, \mu^2)$ is the standard
(collinear) density function. The parameters \muz, $N$, $B$ and $C$ can be
fitted from the collider data (see, for example, Ref.~\cite{Hautmann:2013tba}
and references therein).

The CCFM equation can be solved numerically using the \textsc{updfevolv}
program~\cite{Hautmann:2014uua}, and the TMD gluon and valence quark densities
can be obtained for any \x, $\ktvec[^2]$ and $\mu^2$ values. The main advantage
of this approach is the ease of including into the predictions higher-order
radiative corrections (namely, a part of \nlo{} + \nnlo{} + \dots{} terms
corresponding to the initial-state real gluon emissions) even within \lo.

\subsection{Kimber-Martin-Ryskin approach}

The Kimber-Martin-Ryskin (KMR) approach~\cite{Kimber:2001sc,Watt:2003mx}
provides a technique to construct \tmd{} gluon and quark densities from
conventional \pdfs{} by loosing the DGLAP strong ordering condition at the last
evolution step, that results in \kt{} dependence of the parton distributions.
This procedure is believed to take into account effectively the major part of
next-to-leading logarithmic (NLL) terms
$\alpha_\mathrm{s} {(\alpha_\mathrm{s} \ln\mu^2)}^{n-1}$ compared to the LLA,
where terms proportional to $\alpha_\mathrm{s}^n \ln^n \mu^2$ are taken into
account.

At the \lo, the KMR method, defined for
$\ktvec[^2] \geq \muz[^2] \sim 1$~GeV$^2$, results in expressions for \tmd{}
quark and gluon distributions~\cite{Kimber:2001sc,Watt:2003mx}:

\begin{align}
  \nonumber f_q(\x, \ktvec[^2], \mu^2) &=
      T_q(\ktvec[^2], \mu^2)
      \frac{\alpha_\mathrm{s}(\ktvec[^2])}{2\pi} \\
    &\quad \times \int\limits_x^1 \mathrm{d}z \left[
      P_{qq}^\mathrm{LO}(z)\frac{\x}{z}q
      \left(\frac{\x}{z}, \ktvec[^2]\right)
      \Theta\left(\Delta - z\right) +
      P_{qg}^\mathrm{LO}(z)\frac{\x}{z}g
      \left(\frac{\x}{z}, \ktvec[^2]\right)
      \right],
\end{align}

\begin{align}
  \nonumber f_g(\x, \ktvec[^2],\mu^2) &=
      T_g(\ktvec[^2], \mu^2)
      \frac{\alpha_\mathrm{s}(\ktvec[^2])}{2\pi} \\
    &\quad \times \int\limits_x^1 \mathrm{d}z \left[
      \sum_q P_{gq}^\mathrm{LO}(z)\frac{\x}{z}q
      \left(\frac{\x}{z}, \ktvec[^2]\right) +
      P_{gg}^\mathrm{LO}(z)\frac{\x}{z}g
      \left(\frac{\x}{z}, \ktvec[^2]\right)
      \Theta\left(\Delta - z\right)
      \right],
\end{align}

\noindent where $P_\mathrm{ab}^\mathrm{LO}(z)$ are the usual DGLAP splitting
functions at \lo{} and $\muz^2$ is the minimum scale for which DGLAP evolution
is valid. The theta functions introduce the specific ordering conditions in the
last evolution step, thus regulating the soft gluon singularities. The cut-off
parameter $\Delta$ usually has one of two forms, $\Delta = \mu/(\mu + |\ktvec|)$
or $\Delta = |\ktvec|/\mu$, that reflects the angular or strong ordering
conditions. In the case of angular ordering, the parton densities are extended
into the $\ktvec[^2] > \mu^2$ region, whereas the strong ordering condition
leads to a steep drop of the parton distributions beyond the scale $\mu^2$. At
low $\ktvec[^2] < \muz[^2]$ the behavior of the \tmd{} parton densities has to
be modeled. Usually it is assumed to be flat under strong normalization
condition:

\begin{equation}
  \int\limits_0^{\mu^2}
  f_\mathrm{a}(\x, \ktvec[^2], \mu^2) \mathrm{d}\ktvec[^2] =
  \x a(\x, \mu^2).
\end{equation}

\noindent where $a(\x, \mu^2)$ are the conventional (collinear) \pdfs. The
Sudakov form factors allow one to include logarithmic virtual (loop)
corrections, they take the form:

\begin{equation}
  T_q(\ktvec[^2], \mu^2) = \exp\left[
    -\int\limits_{\ktvec[^2]}^{\mu^2}
    \frac{\mathrm{d}\mathbf{q}_\mathrm{T}^2}{\mathbf{q}_\mathrm{T}^2}
    \frac{\alpha_\mathrm{s}(\mathbf{q}_\mathrm{T}^2)}{2\pi}
    \int\limits_0^{z_{\max}}\mathrm{d}\zeta\,P_{qq}^\mathrm{LO}(\zeta)
  \right],
\end{equation}

\begin{equation}
  T_g(\ktvec[^2], \mu^2) = \exp\left[
    -\int\limits_{\ktvec[^2]}^{\mu^2}
    \frac{\mathrm{d}\mathbf{q}_\mathrm{T}^2}{\mathbf{q}_\mathrm{T}^2}
    \frac{\alpha_\mathrm{s}(\mathbf{q}_\mathrm{T}^2)}{2\pi}
    \left(
      \int\limits_{z_{\min}}^{z_{\max}}\mathrm{d}\zeta\,\zeta
      P_{gg}^\mathrm{LO}(\zeta) +
      n_\mathrm{f} \int\limits_0^1\mathrm{d}\zeta\,
      P_{gq}^\mathrm{LO}(\zeta)
    \right)
  \right],
\end{equation}

\noindent with
$z_{\max} = 1 - z_{\min} = \mu / (\mu + |\mathbf{q}_\mathrm{T}|)$. These form
factors give the probability of evolving from a scale $\ktvec[^2]$ to a scale
$\mu^2$ without parton emission. At the \nlo, the \tmd{} parton densities can be
written as~\cite{Martin:2009ii}:

\begin{equation}
  f_a(\x, \ktvec[^2], \mu^2) =
    \int\limits_0^1 \mathrm{d}z\,T_a(\mathbf{p}_\mathrm{T}^2, \mu^2)
    \frac{\alpha_\mathrm{s}(\mathbf{p}_\mathrm{T}^2)}{2\pi}
    \sum_{b=q,g}P_{ab}^\mathrm{NLO}(z)\frac{\x}{z}b
    \left(\frac{\x}{z}, \mathbf{p}_\mathrm{T}^2\right)
    \Theta(\Delta - z),
\end{equation}

\noindent where $\mathbf{p}_\mathrm{T}^2 = \ktvec[^2]/(1-z)$. Note that both
DGLAP splitting functions and conventional parton distributions should be taken
with \nlo{} accuracy. The Sudakov form factors at \nlo{} read:

\begin{equation}
  T_q(\ktvec[^2], \mu^2) =
    \exp\left[
      -\int\limits_{\ktvec[^2]}^{\mu^2}
      \frac{\mathrm{d}\mathbf{q}_\mathrm{T}^2}{\mathbf{q}_\mathrm{T}^2}
      \frac{\alpha_\mathrm{s}(\mathbf{q}_\mathrm{T}^2)}{2\pi}
      \int\limits_0^1d\zeta\, \zeta \left(
        P_{qq}^\mathrm{NLO}(\zeta) + P_{gq}^\mathrm{NLO}(\zeta)
      \right)
    \right],
\end{equation}

\begin{equation}
  T_g(\ktvec[^2], \mu^2) =
    \exp\left[
      -\int\limits_{\ktvec[^2]}^{\mu^2}
      \frac{\mathrm{d}\mathbf{q}_\mathrm{T}^2}{\mathbf{q}_\mathrm{T}^2}
      \frac{\alpha_\mathrm{s}(\mathbf{q}_\mathrm{T}^2)}{2\pi}
      \int\limits_0^1 \mathrm{d}\zeta\,
      \zeta \left(
        P_{gg}^\mathrm{NLO}(\zeta) +
        2n_\mathrm{f} P_{qg}^\mathrm{NLO}(\zeta)
      \right)
    \right].
\end{equation}

\noindent
It was demonstrated in Ref.~\cite{Martin:2009ii} that the \nlo{} prescription,
with a good accuracy, can be significantly simplified to keep only the \lo{}
splitting functions while the main effect is related to the Sudakov form
factors.

%
%
\section{Interplay between soft and hard $pp$ processes}%
\label{sec:soft_hard}

\subsection{Gluon TMD at low $\mu_0^2$}

The question arises on the form of the \tmd{} gluon distribution
$f_g^{(0)}(\x, \kt[^2], \muz[^2])$ at initial $\muz[^2]$, which enters
into the CCFM evolution equation Eq.~\ref{eq:ccfm}. Usually, the initial \tmd{}
gluon distribution is taken as a product of two functions, each of them depends
on \x{} or \kt, for example in the form given by
Eq.~\ref{eq:ccfmfg0}~\cite{Hautmann:2013tba}. On the other hand, the \tmd{}
gluon distribution is directly related to the dipole-nucleon cross-section
within the model proposed in Refs.~\cite{GolecBiernat:1998js,GolecBiernat:1999qd}
(see also Refs.~\cite{Hautmann:2013tba,Nikolaev:1990ja,Ivanov:2000cm,
Barone:1993sy,Kopeliovich:2003cn,Gotsman:2002yy,Albacete:2010bs}), which is
saturated at low $\mu$ or large transverse distances $r \sim 1/\mu$ between
quark $q$ and antiquark $\bar{q}$ in the $q\bar{q}$ dipole created from the
splitting of the virtual photon $\gamma^{*}$ in the $ep$ \dis. Here we find a
new parametrization for this dipole-nucleon cross-section, as a function of
$r$, using the saturation behavior of the gluon density.

The \tmd{} obtained by Golec-Biernat and W\"usthof
(GBW)~\cite{GolecBiernat:1998js,GolecBiernat:1999qd} within the dipole-nucleon
approach had the unfactorized form as a function of \x{} and \kt:
\begin{equation}
  \x g(\x, \kt, \muz) =
    \frac{3\sigma_0}{4\pi^2\alpha_\mathrm{s}(\muz)} R_0^2(\x) \kt[^2]
    \exp \left(-R_0^2(\x) \kt[^2] \right),
  \quad
  R_0 (x) = \frac{1}{\muz}{\left(\frac{\x}{\x[_0]}\right)}^{\lambda/2}.
  \label{def:GBWgl}
\end{equation}

\begin{figure}[h]
  \centering
  \includegraphics[width=0.8\textwidth]{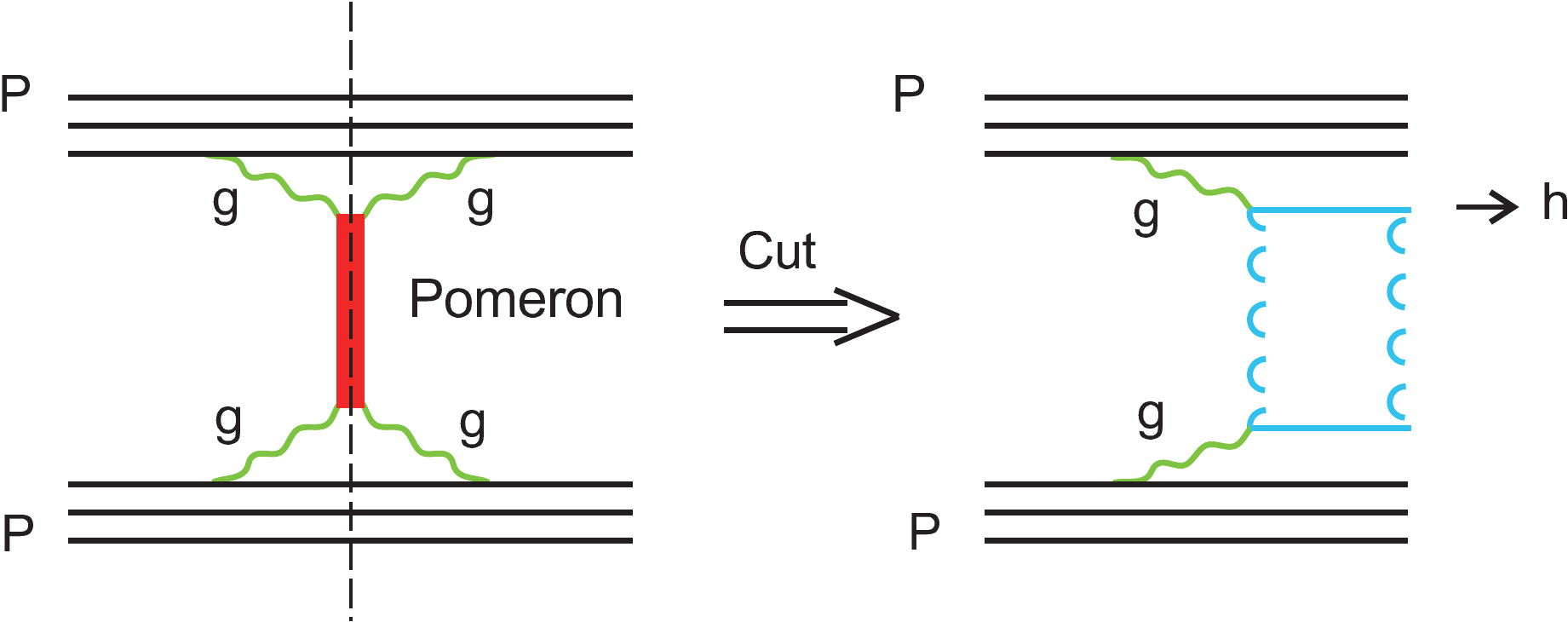}
  \caption{The one-pomeron exchange graph between two gluons in the elastic $pp$
           scattering (left) and the cut one-pomeron due to the creation of two
           colorless strings between quarks/antiquarks that decay into
           $q{\bar q}$ pairs, which are drawn as the semi-circles
           (right)~\cite{Kaidalov:1980bq}.}%
  \label{fig:pomeron}
\end{figure}

\noindent However, this GBW \tmd{} given by Eq.~\ref{def:GBWgl} does not allow
to describe satisfactorily the LHC data on inclusive hadron spectra in $pp$
collisions at low transverse hadron momenta $p_{\mathrm{T}\,h}$ in the central
rapidity range $y \simeq 0$. Therefore, another parametrization of the u.g.d.\
$g(\x, \kt, \muz)$ was suggested in Ref.~\cite{Grinyuk:2013tt}. A possible
existence of the non-perturbative gluons in the proton was suggested. The
contribution of gluon-gluon interaction to the production of
hadrons in $pp$ collision was calculated as the cut graph
(Fig.~\ref{fig:pomeron}, right) of the one-pomeron exchange in the gluon-gluon
interaction (Fig.~\ref{fig:pomeron}, left) using the splitting of the gluons
into the $q\bar{q}$ pair. The right diagram of Fig.~\ref{fig:pomeron}
corresponds to the creation of two colorless strings between the quark/antiquark
($q / \bar{q}$) and antiquark/quark ($\bar{q} / q$). Then,
after their brake, $q\bar{q}$ are produced and fragmented into the hadron $h$.
Actually, the calculation can be made in a way similar to the calculation of the
sea quark contribution to the inclusive spectrum within the
quark-gluon string model (QGSM)~\cite{Kaidalov:1980bq}.

From the best description of inclusive spectra of hadrons produced in $pp$
collisions at different LHC energies $\sqrt{s} = 540$~GeV, 900~GeV, 2.36~TeV,
7~TeV at $p_{ht}\leq$ 2.5 GeV$/$c and $y\simeq$ 0 the following form for the
$\x g(\x, \kt, Q_0)$ was found in Ref.~\cite{Grinyuk:2013tt}:

\begin{align}
  \nonumber \x g(\x, \kt, Q_0) &=
    \frac{3 \sigma_0}{4 \pi^2 \alpha_\mathrm{s}(Q_0)}
    C_1 i{(1-x)}^{b_g} \\
    &\quad \times \left(R_0^2(x) \kt[^2] + C_2{(R_0(x)\kt)}^a \right)
    \exp\left[-R_0(x) \kt - d{(R_0(x)\kt)}^3\right],
  \label{def:gldistrnew}
\end{align}

\noindent where $R_0 (\x)$ is defined in Eq.~\ref{def:GBWgl}.

\begin{figure}[h]
  \centering
  \includegraphics[width=\mpw]{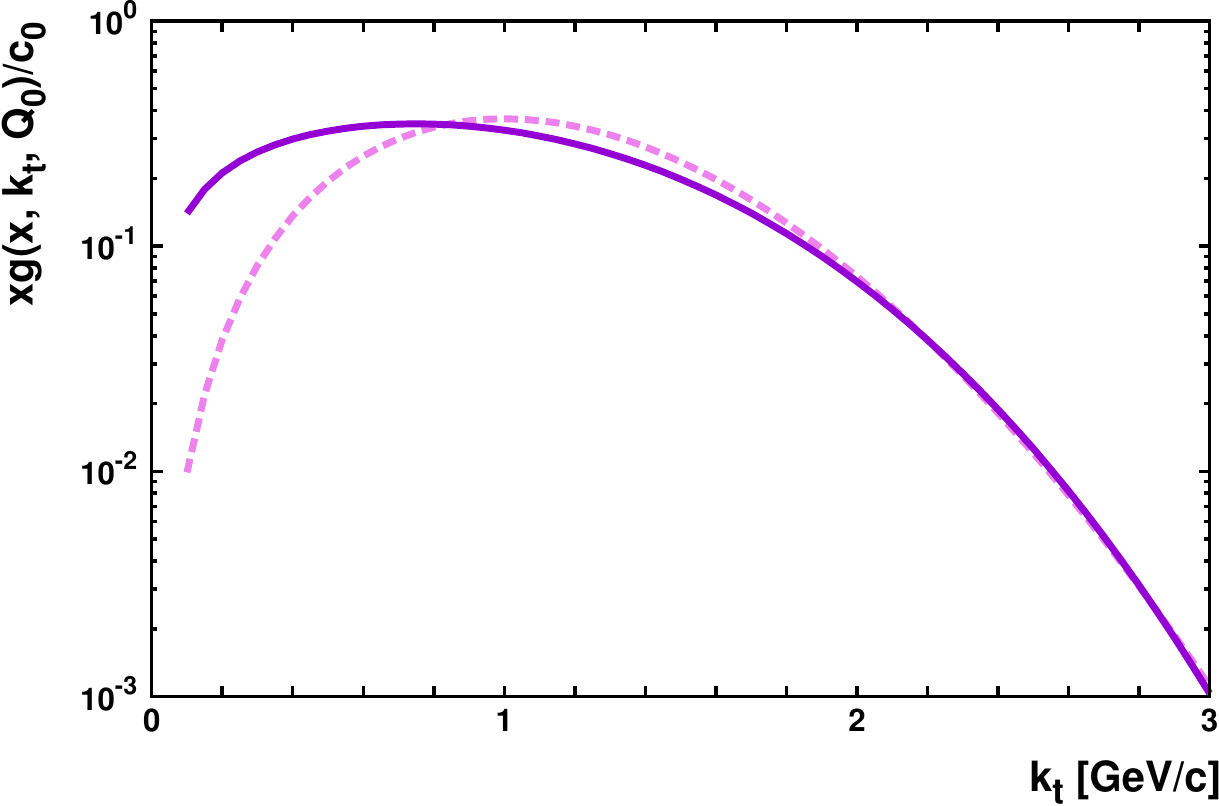}
  \includegraphics[width=\mpw]{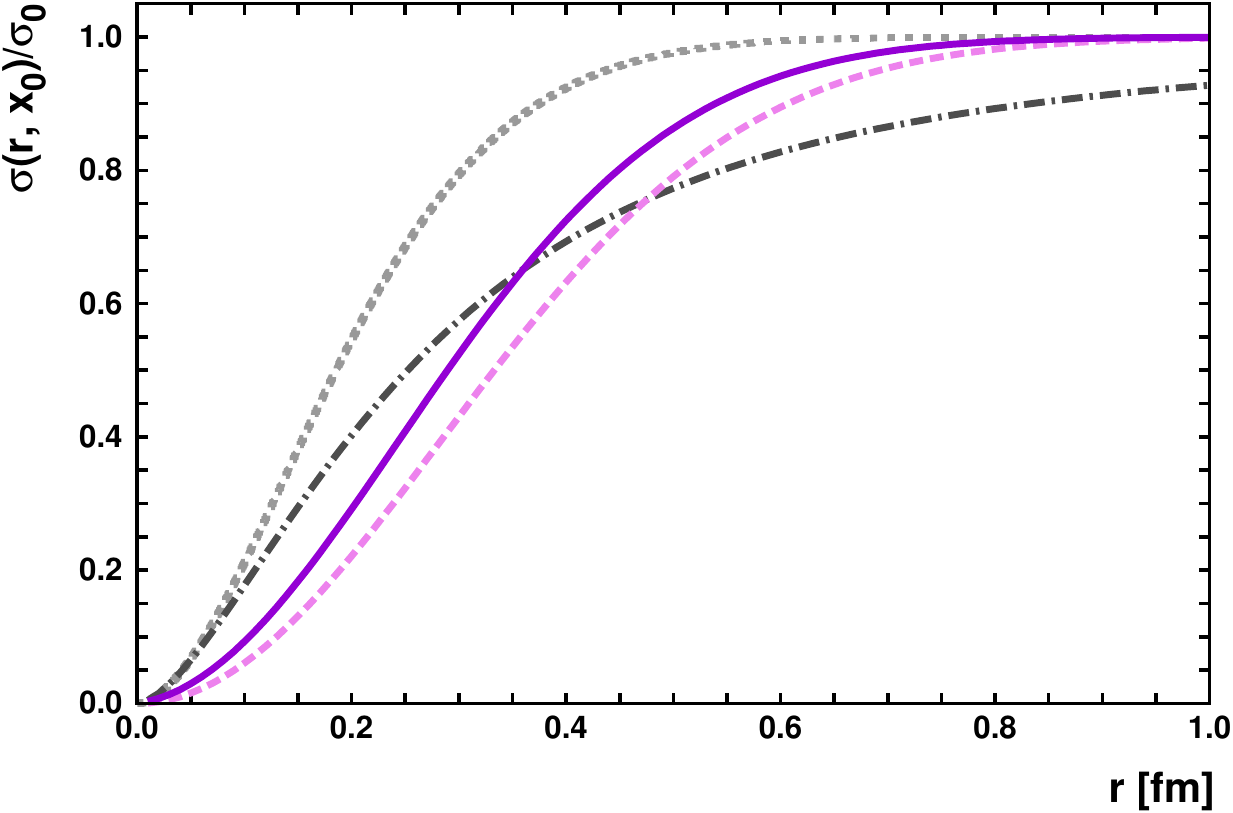}
  \caption{Left: The \tmd{} gluon density $\x g(\x, \kt, Q_0) / c_0$ (where
           $c_0 = 3 \sigma_0 / (4 \pi^2 \alpha_\mathrm{s}(Q_0))$) as a function
           of \kt{} at $\x = x_0$ and $Q_0 = 1$~GeV. The solid and dashed curves
           correspond to the modified \tmd{} (Eq.~\ref{def:gldistrnew}) and the
           original GBW gluon density
           (Eq.~\ref{def:GBWgl})~\cite{GolecBiernat:1998js}, respectively.
           Right: The dipole cross-section $\hat{\sigma} / \sigma_0$ at
           $\x = x_0$ as a function of $r$. The solid, dashed, dash-dotted and
           dotted curves correspond to our calculations of
           Eq.~\ref{def:sigGBW}~\cite{GolecBiernat:1998js}, calculations of
           Ref.~\cite{Nikolaev:1990ja} and Ref.~\cite{Albacete:2010bs},
           respectively.}%
  \label{Fig_5_6}
\end{figure}

In Fig.~\ref{Fig_5_6} (left) we present the modified TMD obtained by calculating
the cut one-pomeron graph of Fig.~\ref{fig:pomeron} and the original GBW
\tmd~\cite{GolecBiernat:1998js} as a function of the transverse gluon momentum
\kt. One can see that the modified \tmd, the solid line in Fig.~\ref{Fig_5_6}
(left), is different from the original GBW gluon
density~\cite{GolecBiernat:1998js} at small $\kt < 1.5$ GeV and coincides with
it at larger \kt.

\begin{figure}[h]
  \centering
  \includegraphics[width=7.8cm]{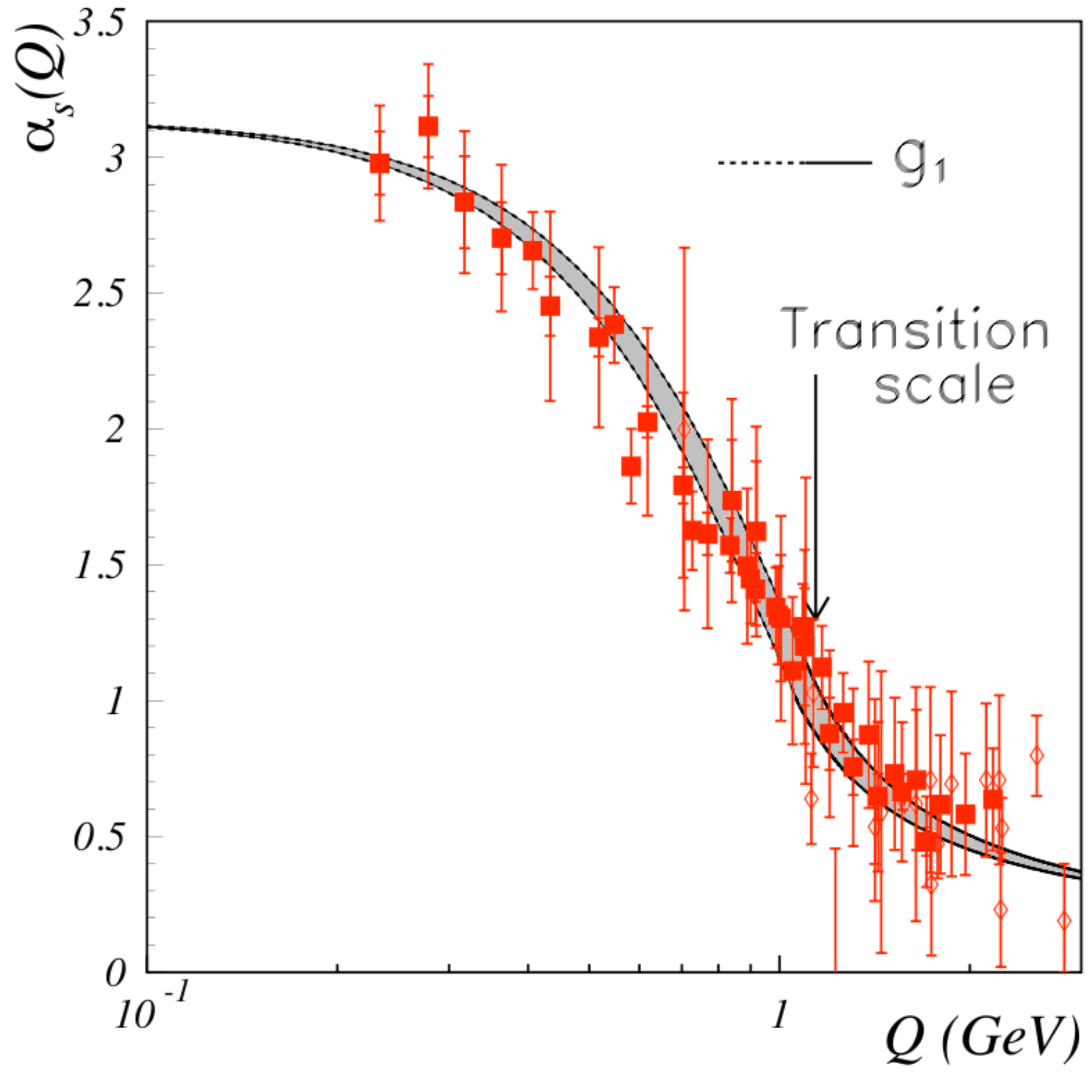}
  \caption{Matching the non-perturbative and perturbative couplings regimes at
           5-loop $\beta$-function in the $\bar{MS}$ renormalization scheme and
           comparison with $\alpha_\mathrm{s}$
           measurements~\cite{Deur:2005cf,Deur:2008rf} from the Bjorken sum
           rule. With $\sqrt{\lambda} = 0.523 \pm 0.024$ GeV result
           $\Lambda_{\bar{MS}} = 0.339 \pm 0.019$~GeV is obtained, compatible
           with with the world average
           ($\Lambda_{\bar{MS}} = 0.332 \pm 0.017$~GeV~\cite{Deur:2016opc}).}%
  \label{Fig_matchalphas}
\end{figure}

It was shown in Ref.~\cite{Grinyuk:2013tt}, that the modified GBW TMD given by
Eq.~\ref{def:gldistrnew} describes reasonably well the HERA data on the proton
longitudinal structure function. Let us note that the serious question on the
value of $\alpha_\mathrm{s} (Q_0)$ at $Q_0 \simeq 1$~GeV. Usually, it was taken
$\alpha_\mathrm{s} (Q_0) \simeq 0.2$~\cite{Grinyuk:2013tt,Abdulov:2018ccp},
which, in principle, corresponds to the perturbative regime of
$\alpha_\mathrm{s}$ as a function of $Q$. However, recently the matching of the
non-perturbative and perturbative couplings was analyzed in
Ref.~\cite{Brodsky:2020ajy}, see Fig.~\ref{Fig_matchalphas}.

Fig.~\ref{Fig_matchalphas} shows big experimental uncertainty of
$\alpha_\mathrm{s}$ at $Q\simeq$ 1 GeV. It would be very interesting to
investigate the sensitivity of the modified input gluon density
$\x g(\x, \kt, Q_0)$ to $\alpha_\mathrm{s}(Q_0)$.

%
%
\subsection{Saturation dynamics}

According to Ref.~\cite{GolecBiernat:1998js,GolecBiernat:1999qd} (see also
Ref.~\cite{Nikolaev:1990ja,Ivanov:2000cm,Barone:1993sy,Kopeliovich:2003cn,
Gotsman:2002yy,Albacete:2010bs}), the gluon \tmd{} can be related to the
cross-section $\hat{\sigma}(\x, r)$ of the $q\bar{q}$ dipole with the nucleon.
This dipole is created from the split of the virtual exchanged photon
$\gamma^{*}$ to $q\bar{q}$ pair in $ep$ deep inelastic scattering (DIS). The
relation at the fixed $Q_0^2$ is following\cite{GolecBiernat:1999qd}:

\begin{equation}
  \hat{\sigma}(\x, r) =
    \frac{4 \pi \alpha_\mathrm{s}(Q_0^2)}{3}
    \int \frac{\mathrm{d}^2 \kt}{\kt[^2]}
    \left[ 1 - J_0(r \kt) \right] \x g(\x, \kt).
  \label{def:sigxr}
\end{equation}

\noindent Using the simple form for $\x g(\x, \kt)$ given by Eq.~\ref{def:GBWgl}
as input to Eq.~\ref{def:sigxr} one can get the following form for the dipole
cross-section:

\begin{equation}
  \hat{\sigma}_\mathrm{GBW} (\x, r) =
    \sigma_0 \left[1 - \exp \left( -\frac{r^2}{4R_0^2(\x)}\right) \right].
  \label{def:sigGBW}
\end{equation}

\noindent However, the modified u.g.d.\ given by Eq.~\ref{def:gldistrnew}
inputted to Eq.~\ref{def:sigxr} results in a more complicated form for
$\hat{\sigma}(\x, r)$~\cite{Grinyuk:2013tt}:

\begin{equation}
  \hat{\sigma}_\mathrm{modif.}(\x, r) =
    \sigma_0 \left[1 - \exp\left(-\frac{b_1 r}{R_0(\x)} -
      \frac{b_2 r^2}{R_0^2(\x)}\right) \right],
  \label{def:sigxrmod}
\end{equation}

\noindent where $b_1 = 0.045$ and $b_2 = 0.3$. Cross-sections
$\hat{\sigma}_\mathrm{GBW}(\x = \x[_0], r)$~\cite{GolecBiernat:1999qd} and
${\hat{\sigma}_\mathrm{modif.}(\x = \x[_0], r)}$ obtained from the modified
u.g.d.\ given by Eq.~\ref{def:gldistrnew}.

There are different forms of the dipole cross-sections suggested
in Refs.~\cite{Nikolaev:1990ja,Ivanov:2000cm,Barone:1993sy,Kopeliovich:2003cn,
Gotsman:2002yy,Albacete:2010bs}. The dipole cross-section can be presented in
the general form~\cite{GolecBiernat:1998js}:

\begin{equation}
  \hat{\sigma}(\x, r) = \sigma_0 g(\hat{r}^2),
  \label{def:sigdipgen}
\end{equation}

\noindent where $\hat{r} = r / (2R_0(\x))$. The function $g(\hat{r}^2)$ can be
written in the form~\cite{Nikolaev:1990ja}

\begin{equation}
  g(\hat{r}^2) = \hat{r}^2 \log \left(1 + \frac{1}{\hat{r}^2}\right),
  \label{def:sigdipNZ90}
\end{equation}

\noindent or in the form~\cite{Albacete:2010bs}
\begin{equation}
  g(\hat{r}^2) = 1 - \exp \left[- \hat{r}^2 \log \left(
    \frac{1}{\Lambda r} + e\right) \right],
  \label{def:sigdipMV98}
\end{equation}

\noindent where saturation occurs for larger $r$.

In Fig.~\ref{Fig_5_6} (left) we illustrate the dipole cross-sections
$\hat{\sigma}/\sigma_0$ at $\x = \x[_0]$ which are saturated at $ r > 0.6$~fm,
obtained in Refs~\cite{GolecBiernat:1998js,Nikolaev:1990ja,Albacete:2010bs,
McLerran:1997fk}. They are compared with the results of our calculations (solid
line) given by Eq.~\ref{def:sigxrmod}. The solid curve in Fig.~\ref{Fig_5_6}
(right) corresponds to the modified u.g.d.\ given by Eq.~\ref{def:gldistrnew},
which allowed us to describe the LHC data on inclusive spectra of hadrons
produced in the mid-rapidity region of $pp$ collision at low \pt. Therefore,
the form of the dipole-nucleon cross-sections presented in Fig.~\ref{Fig_5_6}
(right) can be verified by the description of the last LHC data on hadron
spectra in soft kinematical region.

Comparing the solid curve (``Modified $\sigma$'') and dashed curve
(``GBW $\sigma$'') in Fig.~\ref{Fig_5_6} (right) one can see that
$\hat{\sigma}_\mathrm{modif.}(\x, r)$ given by Eq.~\ref{def:sigxrmod} is
saturated earlier than $\hat{\sigma}_\mathrm{GBW}(\x, r)$ given by
Eq.~\ref{def:sigGBW} with increasing the transverse dimension $r$ of the
$q\bar{q}$ dipole. If ${R_0 = (1/\mathrm{GeV})\,{(\x /\x[_0])}^{\lambda/2}}$,
according to Ref.~\cite{GolecBiernat:1998js}, then the saturation scale has the
form $Q_\mathrm{s} \sim 1 / R_0 = Q_{\mathrm{s}\,0}{(\x[_0]/\x)}^{\lambda/2}$,
where $Q_{\mathrm{s}\,0} = 1~\mathrm{GeV} = 0.2$~fm$^{-1}$.
The saturation of the dipole cross-section, Eq.~\ref{def:sigGBW}, sets in when
$r \sim 2 R_0$ or
$Q_\mathrm{s} \sim (Q_{\mathrm{s}\,0} / 2) {(\x[_0]/\x)}^{\lambda/2}$.
Comparing the saturation properties of the modified cross-section
$\sigma_\mathrm{modif.}$ and GBW cross-section $\sigma_\mathrm{GBW}$ presented
in Fig.~\ref{Fig_5_6} (right) one can get slightly larger value for
$Q_{\mathrm{s}\,0}$ in comparison with $Q_{\mathrm{s}\,0} = 1$~GeV.

\subsection{Non-perturbative TMD gluon input and its evolution}

The determination of the parameters of the initial \tmd{} gluon density in
proton can be split into the two almost independent parts, which refer to
the regions of small and large \x, respectively. Let's consider the small-\x{}
region first and start from the simple analytical expression for the starting
\tmd{} gluon distribution function $f_g^{(0)}$ at some fixed scale
$\muz \sim 1$~GeV. It can be presented in following form~\cite{Abdulov:2018ccp}:

\begin{equation}
  f_g^{(0)} (\x, \ktvec[^2], \muz[^2]) =
    \tilde{f}_g^{(0)}(\x, \ktvec[^2], \muz[^2]) +
    \lambda_1 (\x, \ktvec[^2], \muz^2)
            f_g(\x, \ktvec[^2]),
  \label{def:fgold13}
\end{equation}

\noindent where \x{} and \ktvec{} are the proton longitudinal momentum
fraction and two-dimensional gluon transverse momentum, respectively.
The first term,
$\tilde{f}_g^{(0)}(\x, \ktvec[^2], \muz^2)$, was
calculated~\cite{Grinyuk:2013tt} within the soft \qcd{} model and reads:

\begin{align}
  \nonumber \tilde{f}_g^{(0)}(\x, \ktvec[^2], \muz^2) &=
    c_0 c_1 {(1 - x)}^b \\
  &\quad \times \left[R_0^2(\x) \ktvec[^2] +
      c_2 {\left(R_0^2(\x) \ktvec[^2]\right)}^{a/2}\right]
  \exp\left[-R_0(\x) |\ktvec| -
      d{\left(R_0^2(\x)\ktvec[^2]\right)}^{3/2}\right],
  \label{def:fgold18}
\end{align}

\noindent
where $R_0^2(\x) = {(\x / \x[_0])}^\lambda/\muz[^2]$ and
$c_0 = 3 \sigma_0 / 4\pi^2 \alpha_\mathrm{s}$. The parameters
$\sigma_0 = 29.12$~mb, $\lambda = 0.22$, $x_0 = 4.21 \cdot 10^{-5}$ and
$\alpha_\mathrm{s} = 0.2$ come from the Golec-Biernat-W\"usthoff (GBW)
saturation model~\cite{GolecBiernat:1998js}, while other parameters $a$, $b$,
$c_1$, $c_2$ and $d$ were fitted from LHC data on inclusive spectra of charged
hadrons. The numerical values of these parameters, details of the calculations
and the relation between the \tmd{} gluon density and the inclusive hadron
spectra are given in our previous papers, see Refs.~\cite{Grinyuk:2015lna,
Lipatov:2013yra,Grinyuk:2013tt}. The gluon density
$\tilde{f}_g^{(0)}(\x, \ktvec[^2], \muz[^2])$ differs from
the one obtained in the GBW model at $|\ktvec| < 1$~GeV and
coincides with the GBW gluon at larger
$|\ktvec| > 1.5$~GeV~\cite{Grinyuk:2015lna}. The second term,
$f_g(\x, \ktvec[^2])$, represents the analytical
solution~\cite{Kovchegov:1999ua} of the linear BFKL equation at low \x{}
weighted with a matching function
$\lambda_1(\x, \ktvec[^2], \muz[^2])$~\cite{Abdulov:2018ccp}:

\begin{equation}
  f_g(\x, \ktvec[^2]) =
    \alpha_\mathrm{s}^2 \, \x[^{-\Delta}] \, t^{-1/2} \frac{1}{v}
    \exp \left[- \frac{\pi \ln^2 v}{t}\right],
  \label{def:fgnew18}
\end{equation}

\begin{equation}
  \lambda_1(\x, \ktvec[^2], \muz^2) =
    c_0 {\left(\frac{\x}{\x[_0]}\right)}^{0.81}
    \exp \left[- k_0^2 \frac{R_0(\x)}{|\ktvec|} \right],
\label{def:match18}
\end{equation}

\noindent where
$t = 14\,\alpha_\mathrm{s} N_\mathrm{c} \, \zeta(3) \ln(1/\x)$,
$\Delta = 4 \, \alpha_\mathrm{s} N_\mathrm{c} \ln 2 / \pi$,
$v = |\ktvec|/\Lambda_\mathrm{QCD}$ and $k_0 = 1$~GeV. This term
allows one to describe LHC measurements of inclusive charged hadrons up to
$\pt \leq 4.5$~GeV~\cite{Grinyuk:2013tt}. It is important that the contribution
from $f_g(\x, \ktvec[^2])$ is only non-zero at
$|\ktvec| \ll \Lambda_\mathrm{QCD} \, {(1/x)}^{\delta}$
with $\delta = \alpha_\mathrm{s} N_\mathrm{c}$, resulting in an average
generated gluon transverse momentum of
$\langle |\ktvec| \rangle \sim 1.9$~GeV. The latter value is
close to the non-perturbative \qcd{} regime, that allows one to treat the \tmd{}
gluon density above as a starting one for the CCFM evolution.

Previously, the phenomenological parameters $a$, $b$, $c_1$, $c_2$ and $d$ in
Eqs.~\ref{def:fgold13}--\ref{def:match18} were determined in the small-\x region
only, where $\x \sim 10^{-4}$--$10^{-5}$ (see Refs.~\cite{Grinyuk:2015lna,
Grinyuk:2013tt,Lipatov:2013yra}). The fit was based on NA61 data on inclusive
cross-sections of $\pi^-$ meson production in $pp$ collisions at initial momenta
$31$ and $158$ GeV~\cite{Abgrall:2013qoa} and on CMS~\cite{Khachatryan:2010us}
and ATLAS~\cite{Aad:2010ac} data on inclusive hadron production in $pp$
collisions at the \lhc. In the present note we tested all these parameters using
the experimental data on the pion transverse mass distribution in Au + Au and
Pb + Pb collisions taken by the STAR Collaboration at the
RHIC~\cite{Abelev:2006cs,Abelev:2006jr} and ALICE Collaboration at the
LHC~\cite{Aamodt:2011zj,Aamodt:2010my,Aamodt:2010jj}. The details of the
calculations of hadron production cross-sections in AA collisions are given in
Ref.~\cite{Lykasov:2018ttr}. Let us stress that the possible higher-order
corrections (see Refs.~\cite{Fadin:1998py,Ciafaloni:1998gs,
Triantafyllopoulos:2002nz}) to the leading-order BFKL motivated \kt-dependence
of the proposed gluon input at low-\x{} (as well as saturation dynamics) are
effectively included.

\begin{figure}[h]
  \centering
  \includegraphics[width=\mpw]{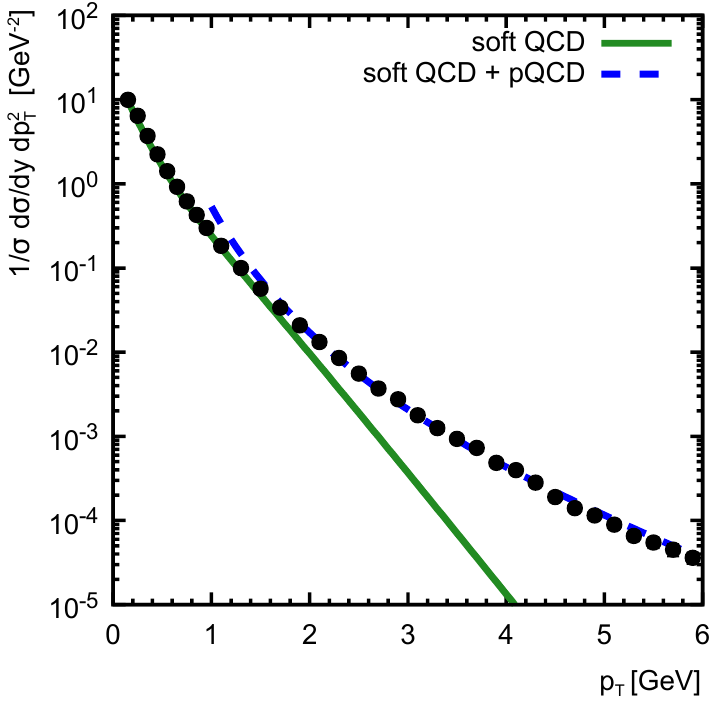}
  \includegraphics[width=\mpw]{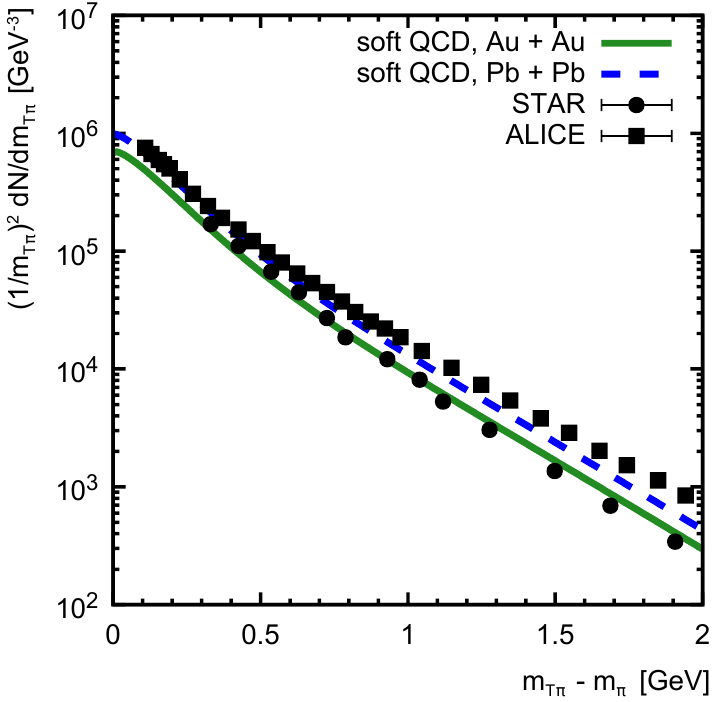}
  \caption{Left panel: the inclusive cross-section of charge hadrons produced in
           $pp$ collisions as a function of their transverse momentum at
           $\sqrt{s} = 7$~TeV. The experimental data are from CMS and
           ATLAS~\cite{Khachatryan:2010us,Aad:2010ac}. Right panel: pion
           transverse mass spectra in Au + Au and Pb + Pb collisions. The
           experimental data are from STAR~\cite{Abelev:2006cs,Abelev:2006jr}
           and ALICE~\cite{Aamodt:2011zj,Aamodt:2010my,Aamodt:2010jj}.}%
  \label{fig1}
\end{figure}

\begin{figure}[h]
\centering
  \includegraphics[width=\mpw]{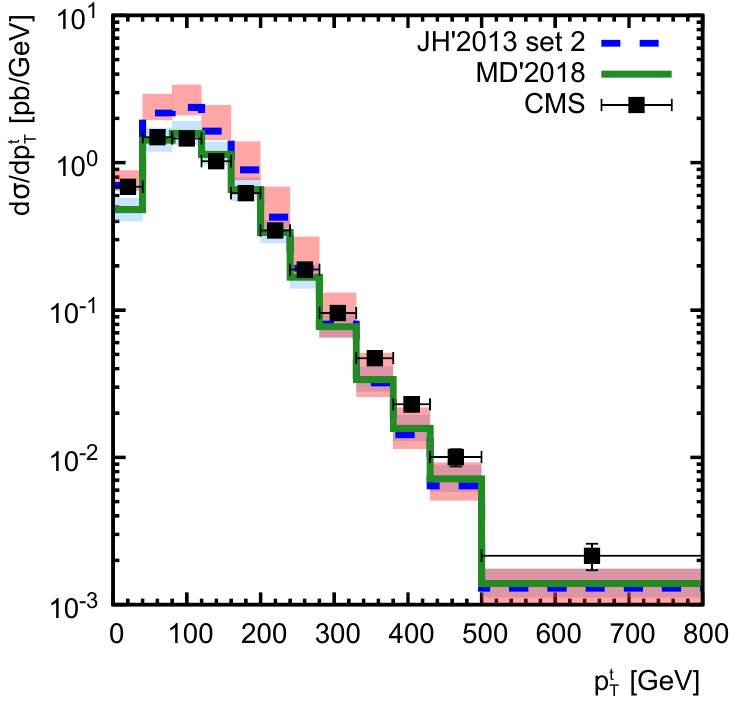}
  \includegraphics[width=\mpw]{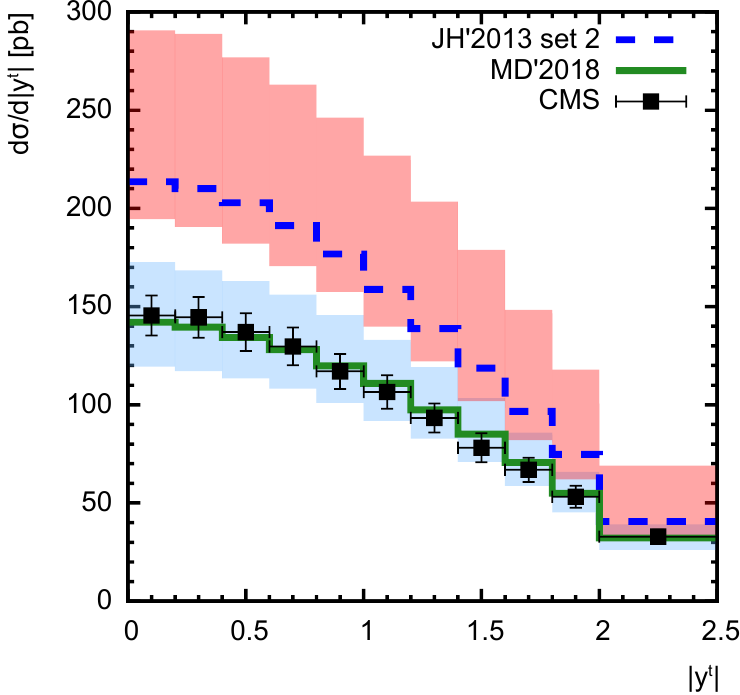}
  \caption{The transverse momentum and rapidity distributions of inclusive
           $t\bar t$ production in $pp$ collisions at $\sqrt{s} = 13$~TeV. The
           green (solid) and blue (dashed) curves correspond to the predictions
           obtained using the \emph{MD'2018} and \emph{JH'2013 set 2} gluons,
           respectively. The shaded bands represent their scale uncertainties.
           The experimental data are from CMS~\cite{Sirunyan:2018wem}.}%
  \label{fig2}
\end{figure}

\begin{figure}[h]
  \centering
  \includegraphics[width=\mpw]{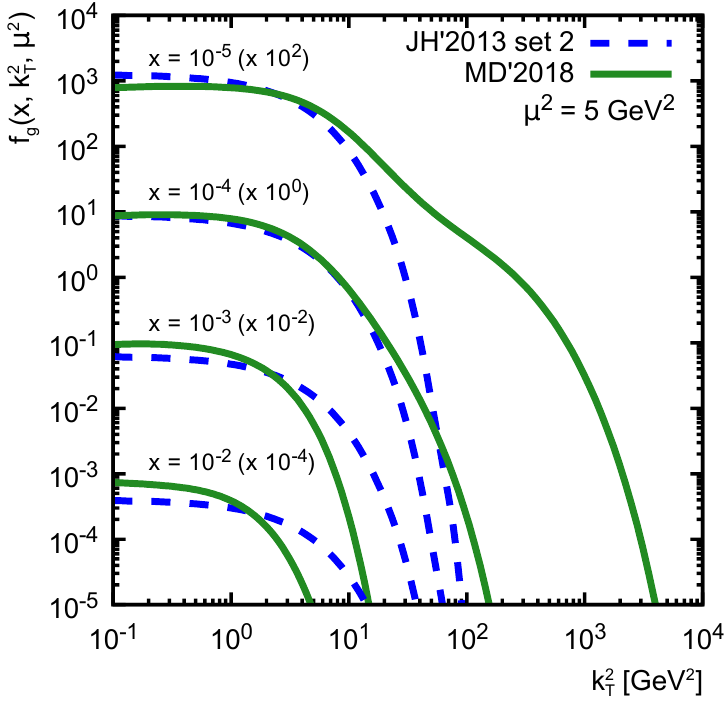}
  \includegraphics[width=\mpw]{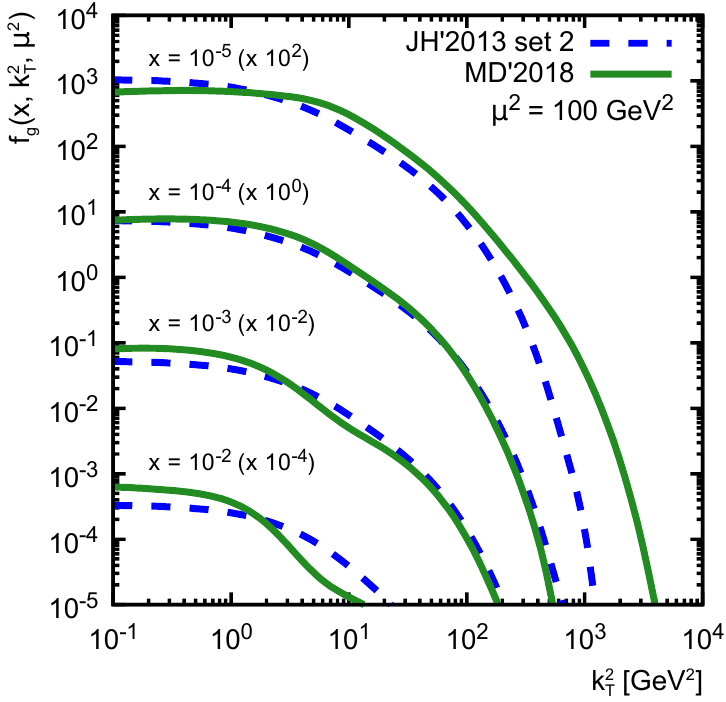}
  \caption{The \tmd{} gluon densities in the proton calculated as a function of
           the gluon transverse momentum $\ktvec[^2]$ at different longitudinal
           momentum fractions \x and $\mu^2$ values. The green (solid) and blue
           (dashed) curves correspond to the \emph{MD'2018} and
           \emph{JH'2013 set 2} gluon density functions, respectively.}%
  \label{fig3}
\end{figure}

\begin{figure}[h]
  \centering
  \includegraphics[width=\mpw]{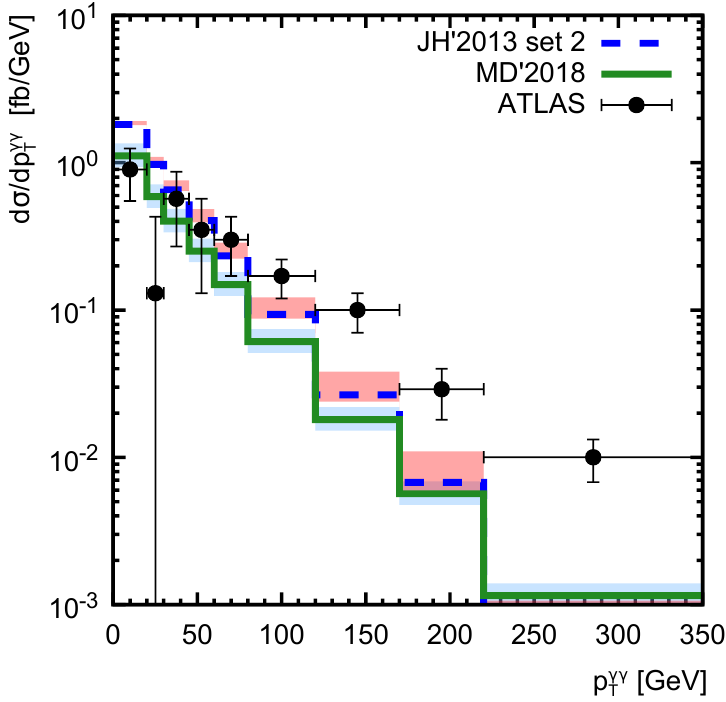}
  \includegraphics[width=\mpw]{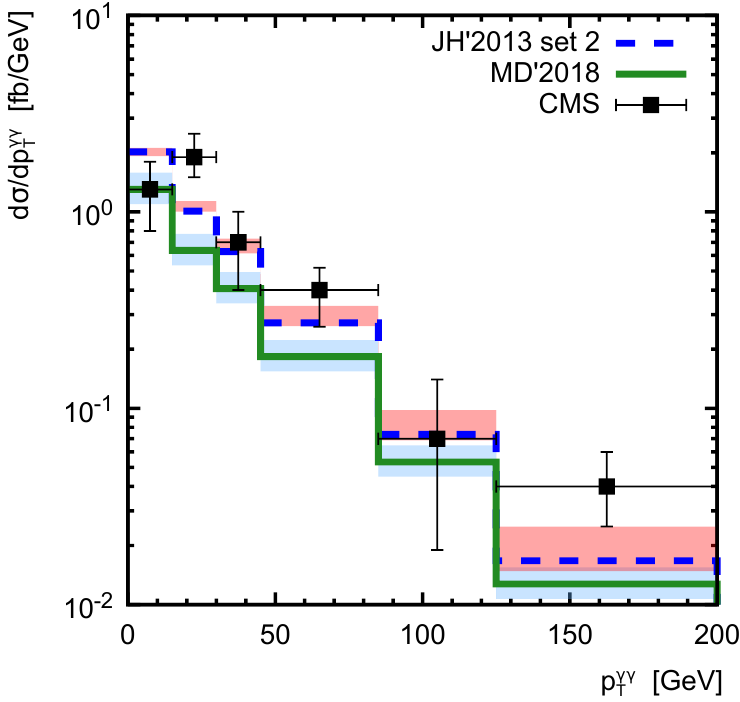}
  \includegraphics[width=\mpw]{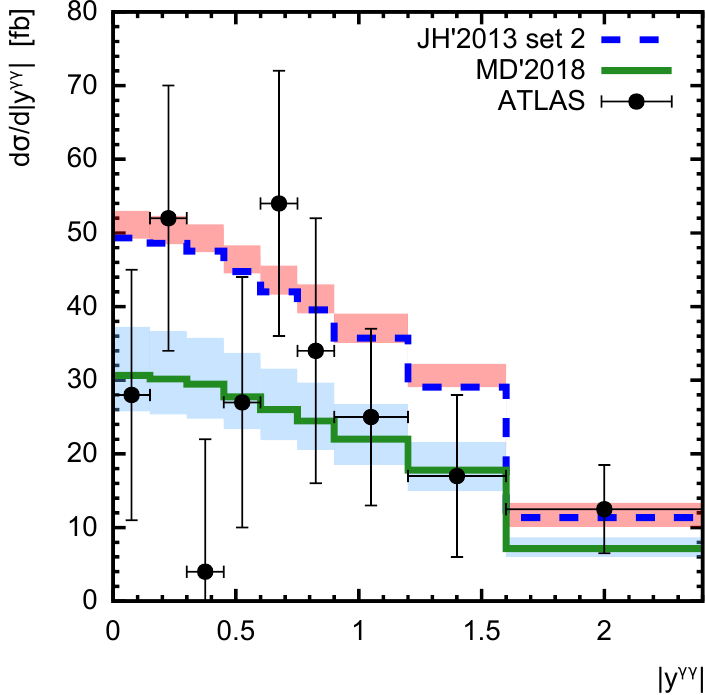}
  \includegraphics[width=\mpw]{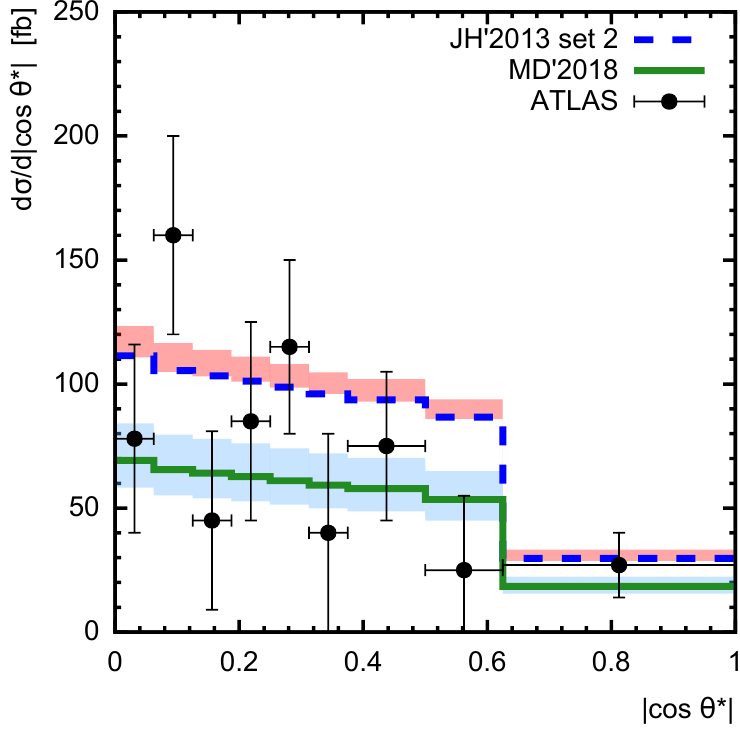}
  \caption{The differential cross-sections of inclusive Higgs boson production
           (in the diphoton decay mode) at $\sqrt{s} = 13$~TeV as functions of
           diphoton pair transverse momentum $p_\mathrm{T}^{\gamma \gamma}$,
           rapidity $|y^{\gamma \gamma}|$ and photon helicity angle
           $\cos \theta^*$ in the Collins-Soper frame. Notation of histograms
           is the same as in Fig.~\ref{fig2}. The experimental data are from
           CMS~\cite{CMS:2017nyv} and ATLAS~\cite{Aaboud:2018xdt}.}%
  \label{fig10}
\end{figure}

\begin{figure}[h]
  \centering
  \includegraphics[width=\mpw]{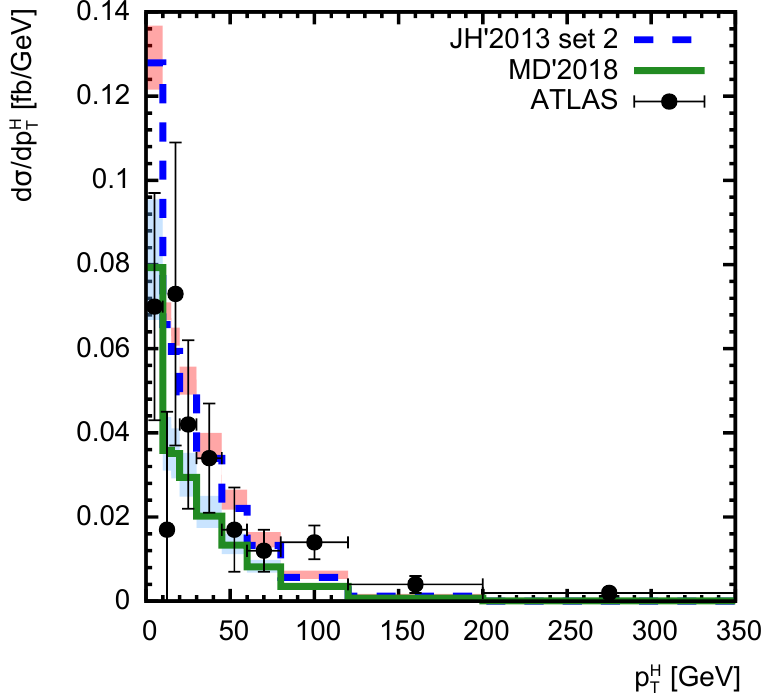}
  \includegraphics[width=\mpw]{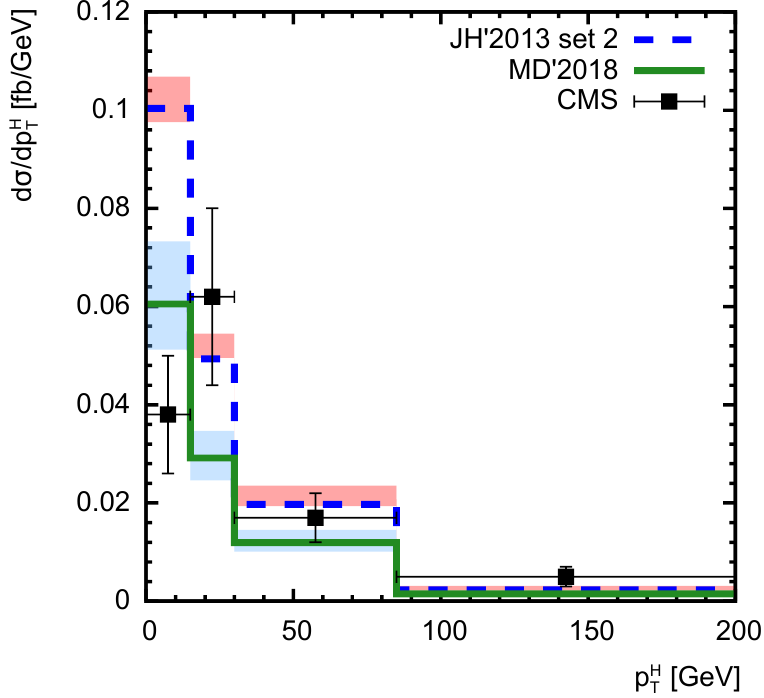}
  \includegraphics[width=\mpw]{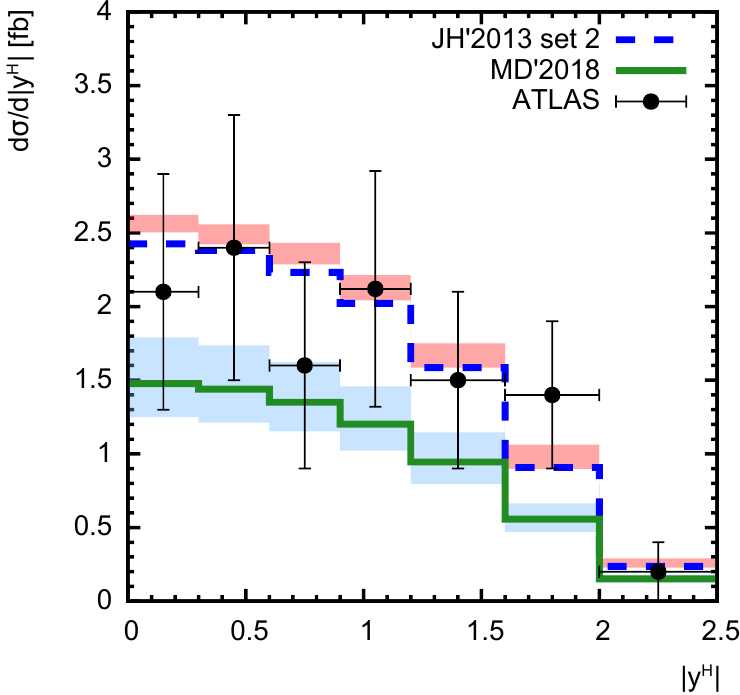}
  \includegraphics[width=\mpw]{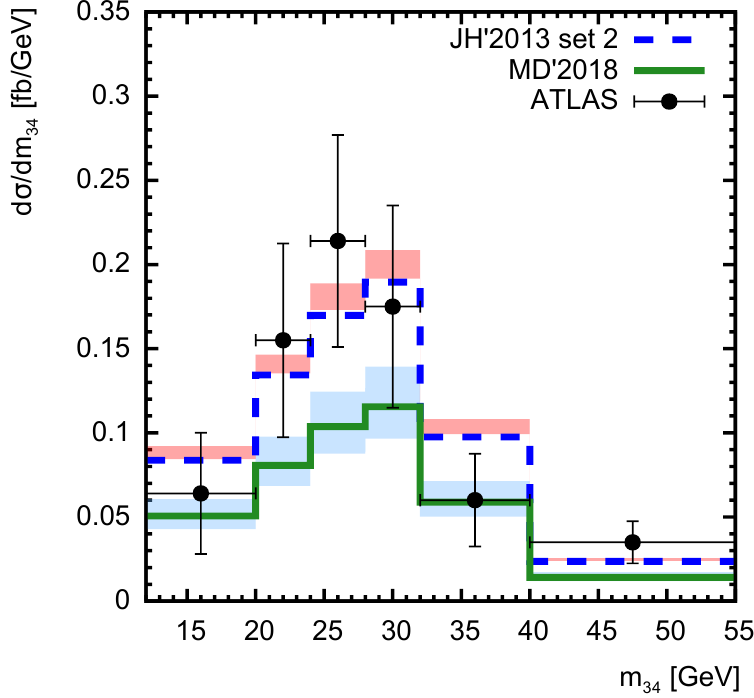}\\
  \includegraphics[width=\spw]{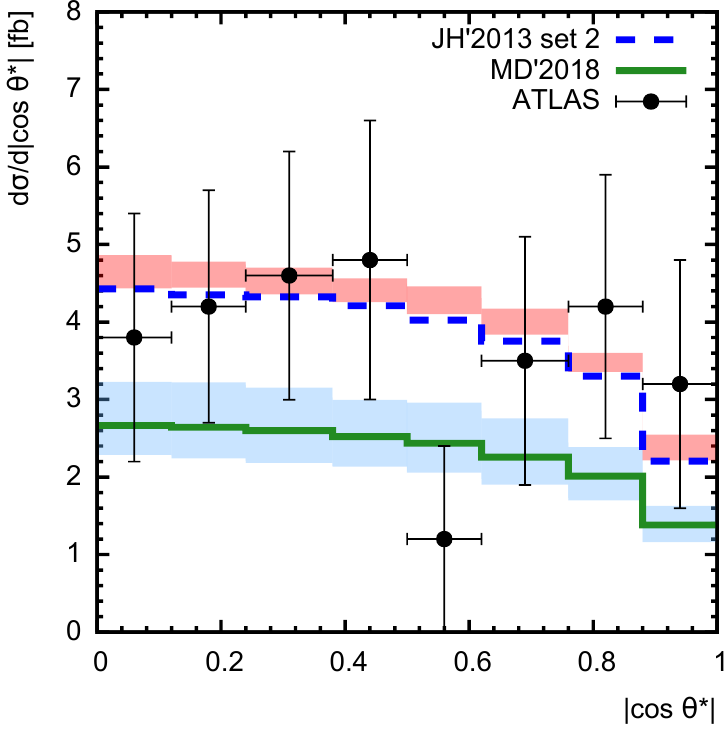}
  \caption{The differential cross-sections of inclusive Higgs production (in the
           $H \to ZZ^* \to 4l$ decay mode) at $\sqrt{s} = 13$~TeV as a function
           of Higgs boson transverse momentum $\pt[^\mathrm{H}]$, rapidity
           $|y^\mathrm{H}|$, leading lepton pair decay angle $|\cos \theta^{*}|$
           (in the Collins-Soper frame) and invariant mass $m_{34}$ of the
           subleading lepton pair. Notation of histograms is the same as in
           Fig.~\ref{fig2}. The experimental data are from
           CMS~\cite{CMS:2017jkd} and ATLAS~\cite{Aaboud:2018xdt}.}%
  \label{fig11}
\end{figure}

 In Fig.~\ref{fig1} (left) the inclusive cross-section of charge hadrons produced in
 $pp$ collisions as a function of their transverse momentum at $\sqrt{s} =
 7$~TeV obtained in \cite{Lykasov:2018ttr,Abdulov:2018ccp} is presented. In Fig.~\ref{fig1}
 (right) pion transverse mass spectra in Au + Au and Pb + Pb collisions
 obtained in \cite{Lykasov:2018ttr,Abdulov:2018ccp} are presented. Theoretical results
 presented in Fig.~\ref{fig1} were obtained using the gluon TMD given by 
 Eq.~\ref{def:fgnew18}. In Fg.~\ref{fig2} the transverse momentum and rapidity
 distributions of inclusive $t\bar t$ production in $pp$ collisions at
 $\sqrt{s} = 13$~TeV obtained in \cite{Abdulov:2018ccp} are presented. 
In Fig.~\ref{fig3} the \tmd{} gluon densities in the proton calculated as a function of
the gluon transverse momentum $\ktvec[^2]$ at different longitudinal
momentum fractions \x and $\mu^2$ values obtained in \cite{Abdulov:2018ccp} are
presnted. Fig.~\ref{fig10} illustrates the differential cross-sections of
inclusive Higgs boson production (in the diphoton decay mode) at $\sqrt{s} =
13$~TeV as functions of diphoton pair transverse momentum $p_\mathrm{T}^{\gamma \gamma}$,
 rapidity $|y^{\gamma \gamma}|$ and photon helicity angle
 $\cos \theta^*$ in the Collins-Soper frame. Notation of histograms
 is the same as in Fig.~\ref{fig2}. Theoretical results wde obtained in 
\cite{Abdulov:2018ccp}. In Fig.~\ref{fig11} the differential cross-sections of
inclusive Higgs production (in the $H \to ZZ^* \to 4l$ decay mode) at
$\sqrt{s} = 13$~TeV obtained in \cite{Abdulov:2018ccp} were presented. 

\clearpage{}

%
%
\section{Prompt photon production accompanied by $c$-jet in $pp$ collision at
         LHC}%
\label{sec:gamma_cjet}

The investigation of possible \ic{} in proton employing latest \gls{atlas}
semi-inclusive $pp \rightarrow \gamma +c$-jet measurement was carried out in
Ref~.\cite{Bednyakov:2017vck}, were two approaches to create simulated samples
with variable \ic{} contribution \w{} were used. First approach utilizes ability
of \sherpa~\cite{Gleisberg:2008ta} generator to reweigh generated spectra even
at the \nlo{} level and resulting spectra are show in
Fig.~\ref{fig:mc_y_c_sherpa}. Second approach, called the Combined QCD approach,
is a combination of \kt-factorization, presented in Sec.~\ref{sec:ktfact}, in
domain of small \x and conventional (collinear) \gls{qcd} factorization in
domain of large \x.

\subsection{Sherpa NLO Sample}

\subsubsection{Simulated Samples}

\begin{figure}[h]
  \centering
  \includegraphics[width=\spw]{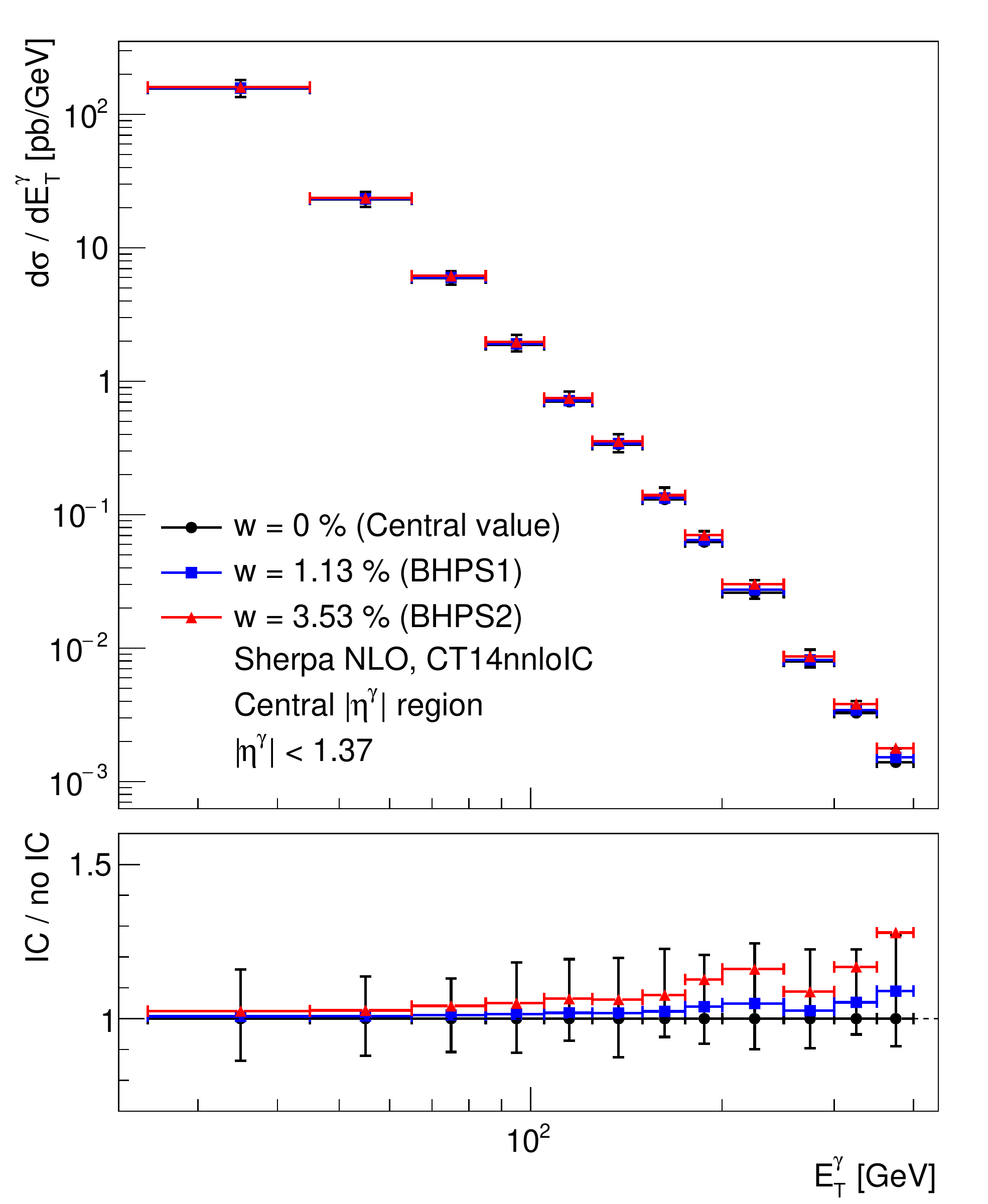}
  \includegraphics[width=\spw]{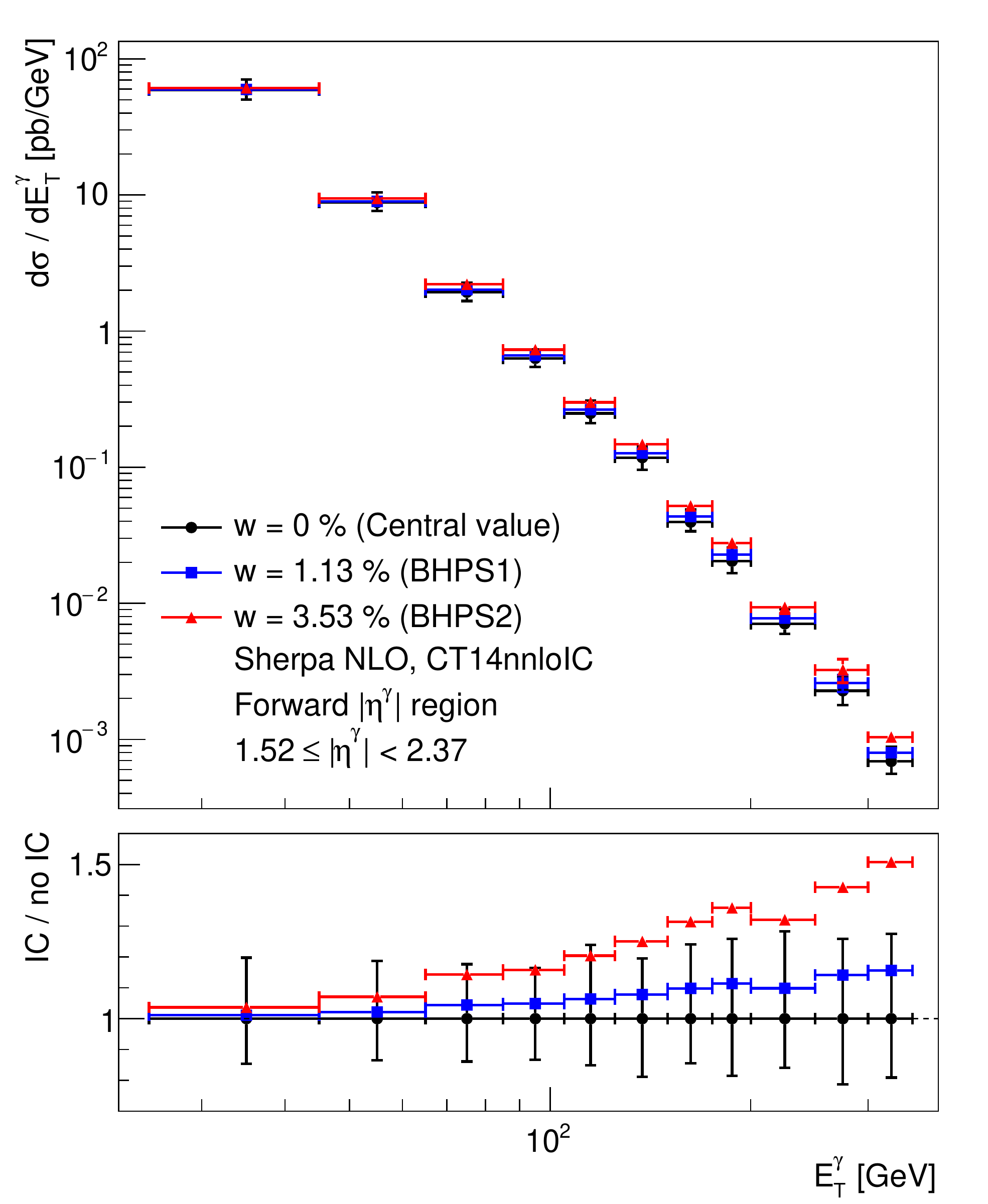}
  \includegraphics[width=\spw]{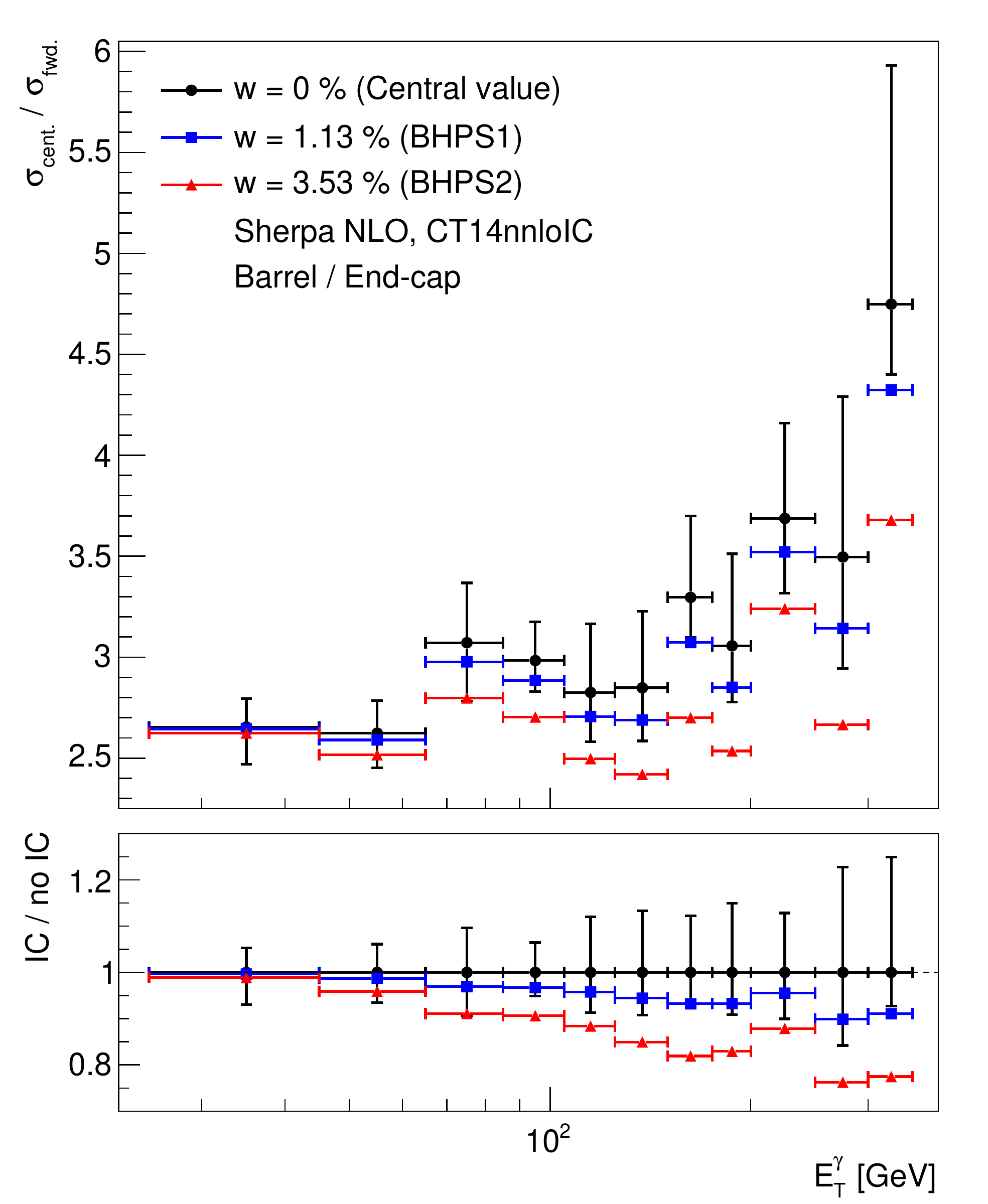}
  \caption{The \sherpa{} NLO simulated \ety{} spectra of \ppycjet{} process and
           the relative uncertainties of these spectra in two pseudo-rapidity
           \aetay{} regions central, $|\eta^\gamma| < 1.37$ (left),
           forward (middle) $1.52 < |\eta^\gamma| < 2.37$ and their ratio
           (Barrel/End-cap $\equiv$ Central/Forward, right) at various values
           of \ic{} probability \w.}%
  \label{fig:mc_y_c_sherpa}
\end{figure}

In Fig.~\ref{fig:mc_y_c_sherpa} the spectra of prompt photons produced in $pp$
collision at \sviii{} simulated using the Sherpa NLO generator as a function of
its transverse energy \ety{} at different \ic{} contributions to \gls{pdf} are
presented. The BHPS1 in Fig.~\ref{fig:mc_y_c_sherpa} corresponds to the mean
value of the $c\bar c$ fraction is
$\langle x_{{\mathrm{c}}\bar{\mathrm{c}}} \rangle \simeq 0.6$~\%, which is
equivalent to the \ic{} probability $\w \approx 1.14$\%. The BHPS2 in
Fig.~\ref{fig:mc_y_c_sherpa} corresponds to
$\langle x_{{\mathrm{c}}\bar{\mathrm{c}}} \rangle \simeq 2.1$~\%,
$\w \approx 3.54$\%~\cite{Brodsky:1980pb,Brodsky:1981se,Nadolsky:2008zw}.

Fig.~\ref{fig:mc_y_c_sherpa} illustrates the prediction for \ety{} spectra of
\ppycjet{} process at \sviii{} obtained within the Sherpa NLO generator. One can
see from this figure that the \ic{} signal in these spectra could be more
visible in the forward pseudo-rapidity region than in the central one.

In this section the detailed description of those two approaches is given, see
also Ref.~\cite{Bednyakov:2017vck,Brodsky:2016fyh}.

\subsubsection{Event Generation}

The Monte Carlo generator \sherpa~\cite{Gleisberg:2008ta,Schumann:2007mg,
Krauss:2001iv,Gleisberg:2008fv} (version 2.2.4) with \gls{nlo}, i.e.~$O(\alpha
\alpha_{s}^2)$, matrix elements generated by
\textsc{OpenLoops}~\cite{Cascioli:2011va,Ossola:2007ax,vanHameren:2010cp}
(version 1.3.1) within ME+PS@NLO model is employed to generate sample of
$\gamma$ + jet + up to 3 additional jets at \sviii. The heavy flavor quarks in
the calculation were considered massless. The calculation employs several
\gls{pdf} sets with the help of LHAPDF6~\cite{Buckley:2014ana}, main \gls{pdf}
set is CT14nnlo~\cite{Dulat:2015mca}, which is extended by
CT14nnloIC~\cite{Hou:2017khm} set. The CT14nnloIC set is an \ic{} addition to
the CT14nnlo set and contains only central value and several \ic{} values in two
models BHPS and SEA~\cite{Hou:2017khm}. Two BHPS sets were used designated
BHPS1 and BHPS2, which contain \ic{} with $w \approx 1.14$\% and $w \approx 3.53$\%
respectively. Additionally NNPDF 3.0~\cite{Ball:2014uwa},
CT10nlo~\cite{Lai:2010vv} and CTEQ66~\cite{Nadolsky:2008zw} \glspl{pdf} were
used to assess effects of different \glspl{pdf} on simulated \ety{} spectra of
\ycjet{} production.

The sample is generated in five \ety{} slices with boundaries of 25, 45, 85,
150, 300 and 450~\gev, which were chosen to match the bin boundaries of the
\gls{atlas} $pp \rightarrow \gamma + c$-jet measurement. The size of the whole
generated sample is about 1 million events. Additional sample for \sxiii{}
predictions is about 0.8 million events.

\subsubsection{Event selection}

For the event selection the custom Rivet~\cite{Buckley:2010ar} analysis is used,
which was later validated against the Rivet analysis of the \gls{atlas} $pp
\rightarrow \gamma + c$ measurement~\cite{Rivet:2017phf}.

The analysis starts with the search for leading photon with $\ety > 25$~\gev{}
and $\aetay < 2.37$, events with leading photon falling into the \gls{atlas}
Calorimeter gap $1.37 < \aetay < 1.56$ being discarded. The selected leading
photons are required to satisfy sliding calorimeter isolation criterion $\etiso
< 4.8~\mathrm{GeV} + 0.0042 \times \ety$. The \etiso{} variable is calculated as
a sum of transverse energy of all particles with a lifetime greater than 10~ps
with a separation in angle of $\Delta R < 0.4$ around the photon. Muons and
neutrinos are excluded because they deposit little or no energy in the
calorimeter. Also, the \etiso{} variable is corrected for the energy density of
the underlying event~\cite{Cacciari:2007fd}.

The jets are built using the anti-$k_t$ algorithm, which takes as input all
particles in the event with a lifetime greater than 10~ps and radius parameter
of $R = 0.4$. Only jets with \ptj{} greater than 20~\gev{} and separation from
the leading photon $\Delta R > 0.4$ are considered. The selected leading jet
must satisfy $|\etaj| < 2.5$ requirement, otherwise the event is discarded. In
addition, the leading photon and leading jet separation criteria are applied,
discarding all events where $\Delta R < 1.0$ between the two. If a $b$ hadron
with $\pt > 5$~\gev{} is found to be in a cone of $R = 0.3$ around the leading
jet, it will be considered to be a $b$-jet and the event is discarded. In
contrary, if a $c$ hadron with $\pt > 5$~\gev{} is found to be in a cone of
$\Delta R = 0.3$ around the leading jet, it will be considered to be a $c$-jet
and the event is kept.

\subsubsection{Uncertainties}%
\label{sec:mc_uncert_sherpa}

Four types of systematic uncertainties are considered in the \sherpa{} \gls{nlo}
sample. The scale uncertainty, which is assessed by multiplying or dividing the
renormalization $\mu_\mathrm{R}$ and factorization $\mu_\mathrm{F}$ scales by a
factor of two, both separately and simultaneously. The final uncertainty is
taken as an envelope of the deviations from the nominal prediction. The
uncertainty of the CT14nnlo \gls{pdf} set is assessed by varying spectra using
all 56 eigenvectors provided and the largest deviation, out of the all possible,
with respect to the nominal value is taken as the final \gls{pdf} uncertainty.
The uncertainty in the strong coupling $\alpha_\mathrm{s}$ value is assessed by
changing its value to 0.117 and 0.119. The uncertainty coming from the
uncertainty of the total fiducial cross-section calculation is taken into
account by varying the normalization of the \ety{} spectra by one standard
deviation of the \sherpa{} NLO total cross-section calculation.

There are several other methods to check the sensitivity of the observables to
the scale uncertainty, see, for example,~\cite{Wu:2013ei} and references there
in.


\subsection{Combined QCD Sample}

\begin{figure}[h]
  \centering
  \includegraphics[width=\spw]{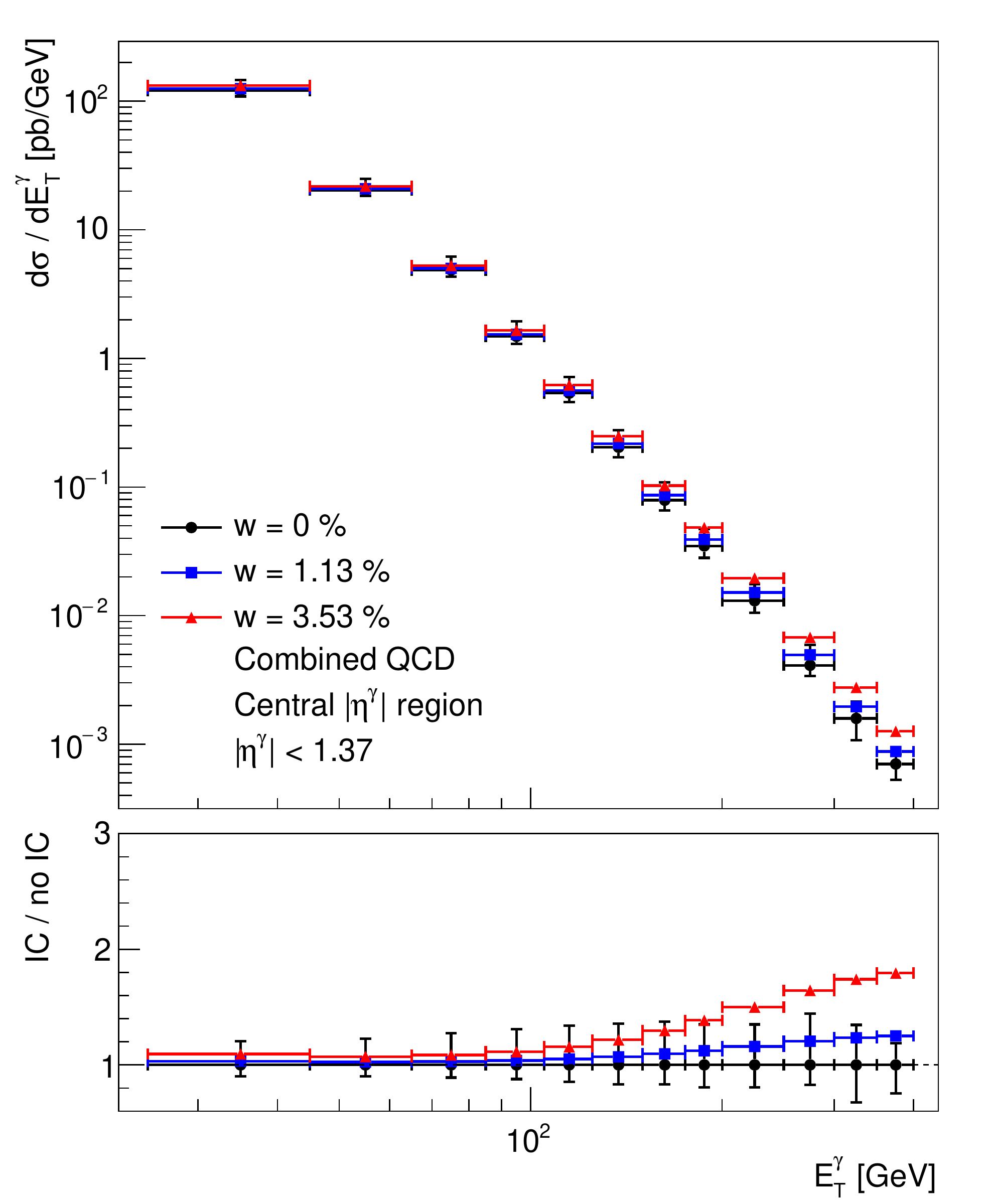}
  \includegraphics[width=\spw]{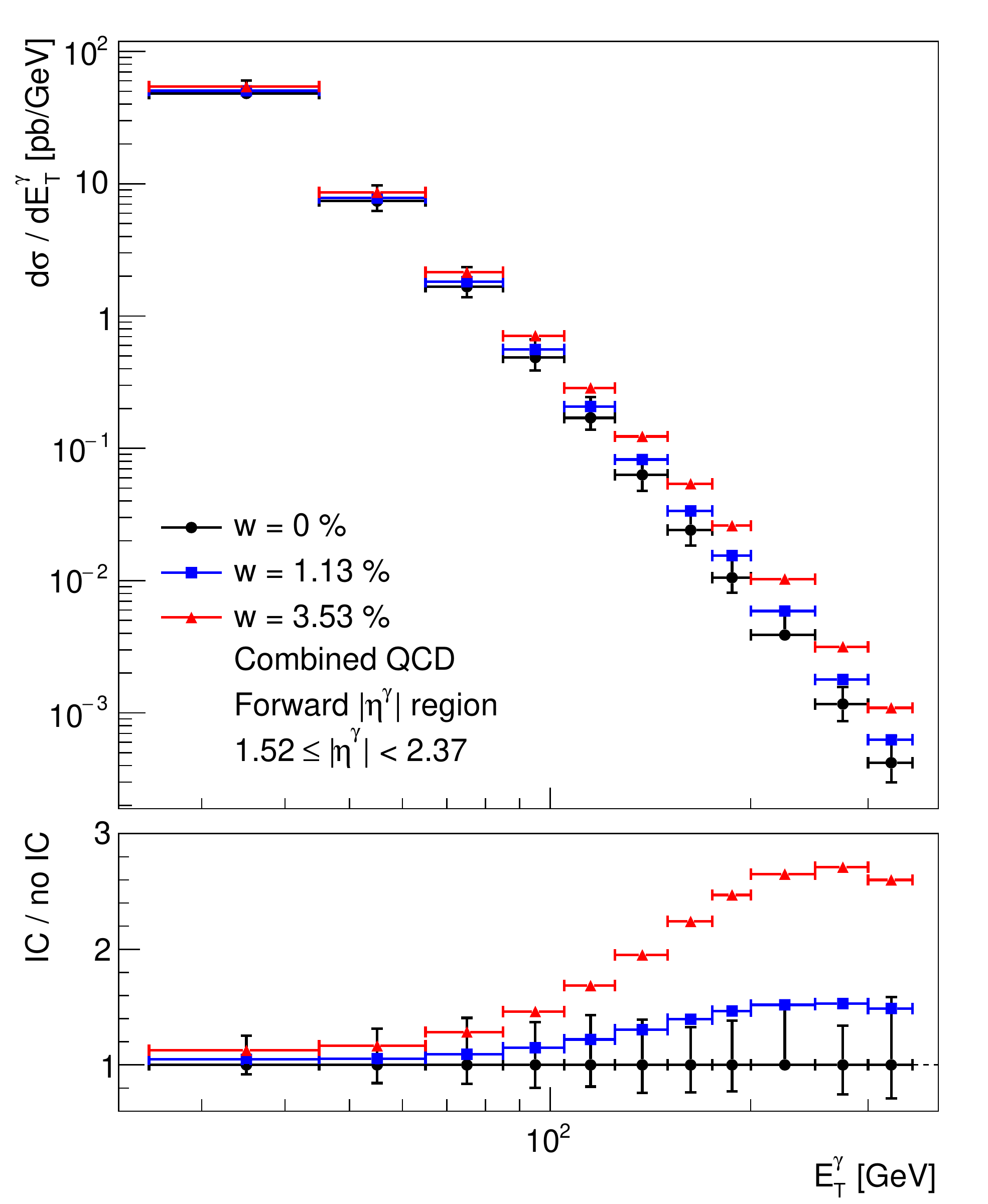}
  \includegraphics[width=\spw]{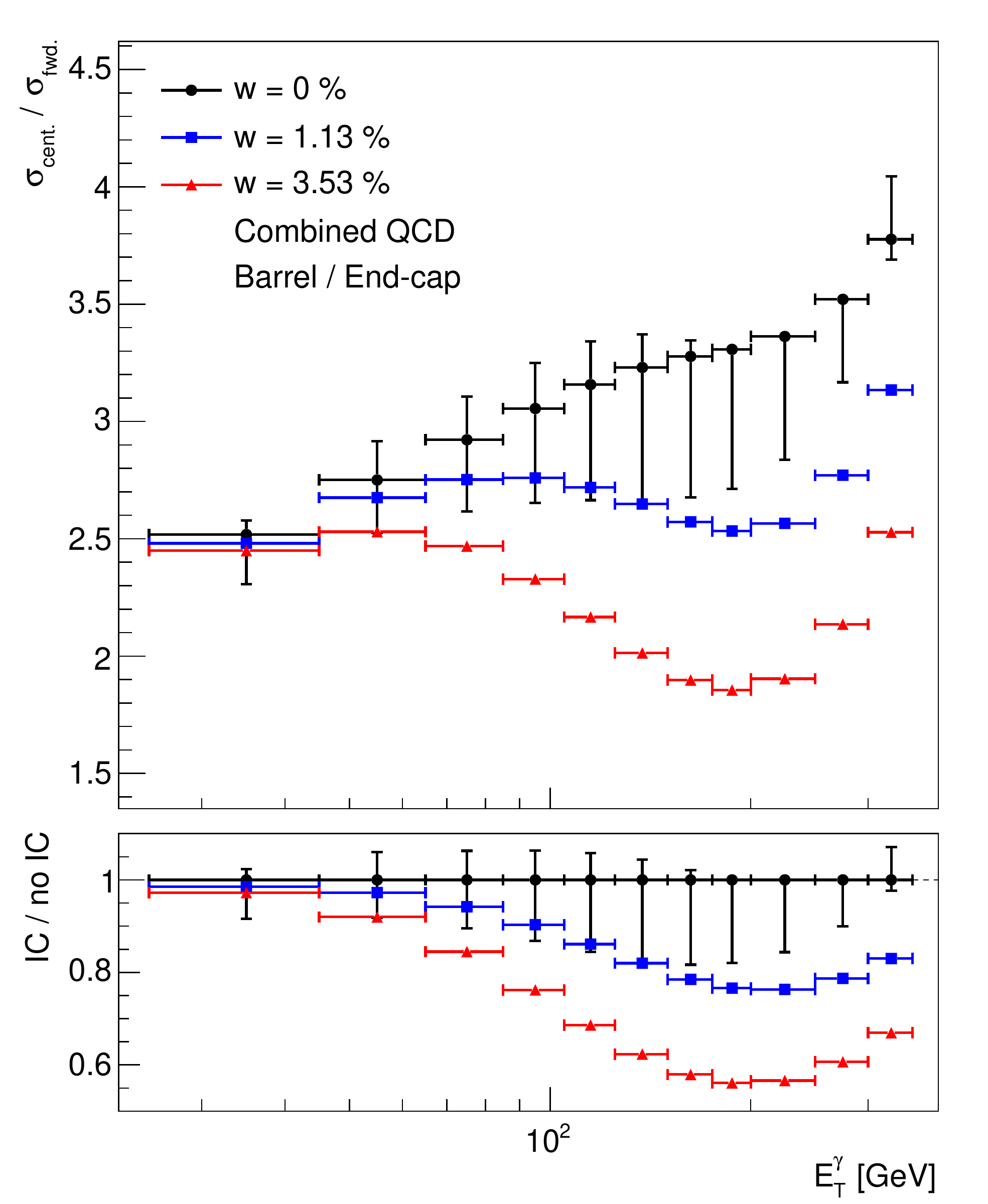}
  \caption{The Combined QCD simulated \ety{} spectra of \ppycjet{} process and
           the relative uncertainties of these spectra in two pseudo-rapidity
           \aetay{} regions central (left), forward (middle) and their ratio
           (Barrel/End-cap, right) at various values of \ic{} probability \w.}%
  \label{fig:mc_y_c_comb_qcd}
\end{figure}

The Combined QCD approach combines two techniques to analytically calculate
\ety{} spectra of \ppycjet{} process in the kinematic regime, where they are
most suitable~\cite{Baranov:2017tig}. First the \kt-factorization formalism is
employed to calculate the leading contributions from the ${\cal O}(\alpha
\alpha_s^2)$ off-shell gluon-gluon fusion $g^{*} g^{*} \to \gamma c \bar c$. In
this way one takes into account the conventional perturbative charm contribution
to associated $\gamma c$ production. In addition there are backgrounds from jet
fragmentation.\footnote{Here $\alpha$ is the electromagnetic coupling constant
and $\alpha_s$ is the \gls{qcd} coupling constant.}

The \ic{} contribution is computed using the
${\cal O}(\alpha \alpha_\mathrm{s})$ \gls{qcd} Compton scattering
$c g^* \to \gamma c$ amplitude, where the gluons
are kept off-shell and incoming quarks are treated as on-shell partons. This is
justified by the fact that the \ic{} contribution begins to be visible at the
domain of large $\x \geq 0.1$, where its transverse momentum can be safely
neglected.

The \kt-factorization approach has technical advantages, since one can include
higher-order radiative corrections using the TMD parton distribution of
the proton~\cite{Andersson:2002cf}. Technically, the numerical solution of the
Ciafaloni-Catani-Fiorani-Marchesini (CCFM) gluon evolution
equation~\cite{Ciafaloni:1987ur,Catani:1989sg} is employed~\cite{Catani:1990eg},
which resumes the leading logarithmic terms, proportional to $\log 1/x$, up to
all orders of perturbation theory.

In addition, several standard \gls{pqcd} subprocesses involving quarks in the
initial state are taken into account. These are the flavor excitation $c q \to
\gamma c q$, quark antiquark annihilation $q \bar q \to \gamma c \bar c$ and
quark gluon scattering subprocess $q g\to \gamma q c \bar c$. These processes
become important at large transverse momenta \ety{} or at large parton
longitudinal momentum fraction \x, which is the kinematics needed to produce
high \ety{} events; it is the domain where the quarks are less suppressed or can
even dominate over the gluon density. The calculation relays on the conventional
(DGLAP) factorization scheme, which should be reliable in the large \x{} region.

In the calculation, the relatively old \gls{pdf} is used,
CTEQ66c~\cite{Nadolsky:2008zw}. It is due to technical requirement on \gls{pdf}
to be \kt-factorization compatible and at the same time to provide few
distributions with non zero \ic{} contribution. The relative difference among
various \glspl{pdf} can be seen in Fig.~\ref{fig:results_pdf_var}.

The only systematic uncertainty of the calculation considered in this case is
the uncertainty coming from the variation of factorization $\mu_\mathrm{F}$ and
renormalization $\mu_\mathrm{R}$ scale. The scales are multiplied or divided by
a factor of two and the extreme cases of deviation from the nominal prediction
are taken as the final uncertainty.

The plots showing the results of the Combined QCD approach are presented in
Fig.~\ref{fig:mc_y_c_comb_qcd}.

\subsection{Optimal Forward-Central \texorpdfstring{$|\eta^\gamma|$}{etagamma}
            Split}

The relation between the photon transverse momentum \pty{} (or energy \ety), the
photon pseudo-rapidity \etay{} and photon momentum fraction \xf{}, see
Ref.~\cite{Brodsky:2016fyh}, implies that in order to search for \ic{}
contribution to $\gamma + c$ cross-section at large \xf{} one needs to probe
large \ety{} and high \aetay. With the help of the \sherpa{} \nlo{} sample the
investigation of optimal lower cut on photon pseudo-rapidity \aetay{} was
performed.

\begin{figure}[h]
  \centering
  \includegraphics[width=\mpw]{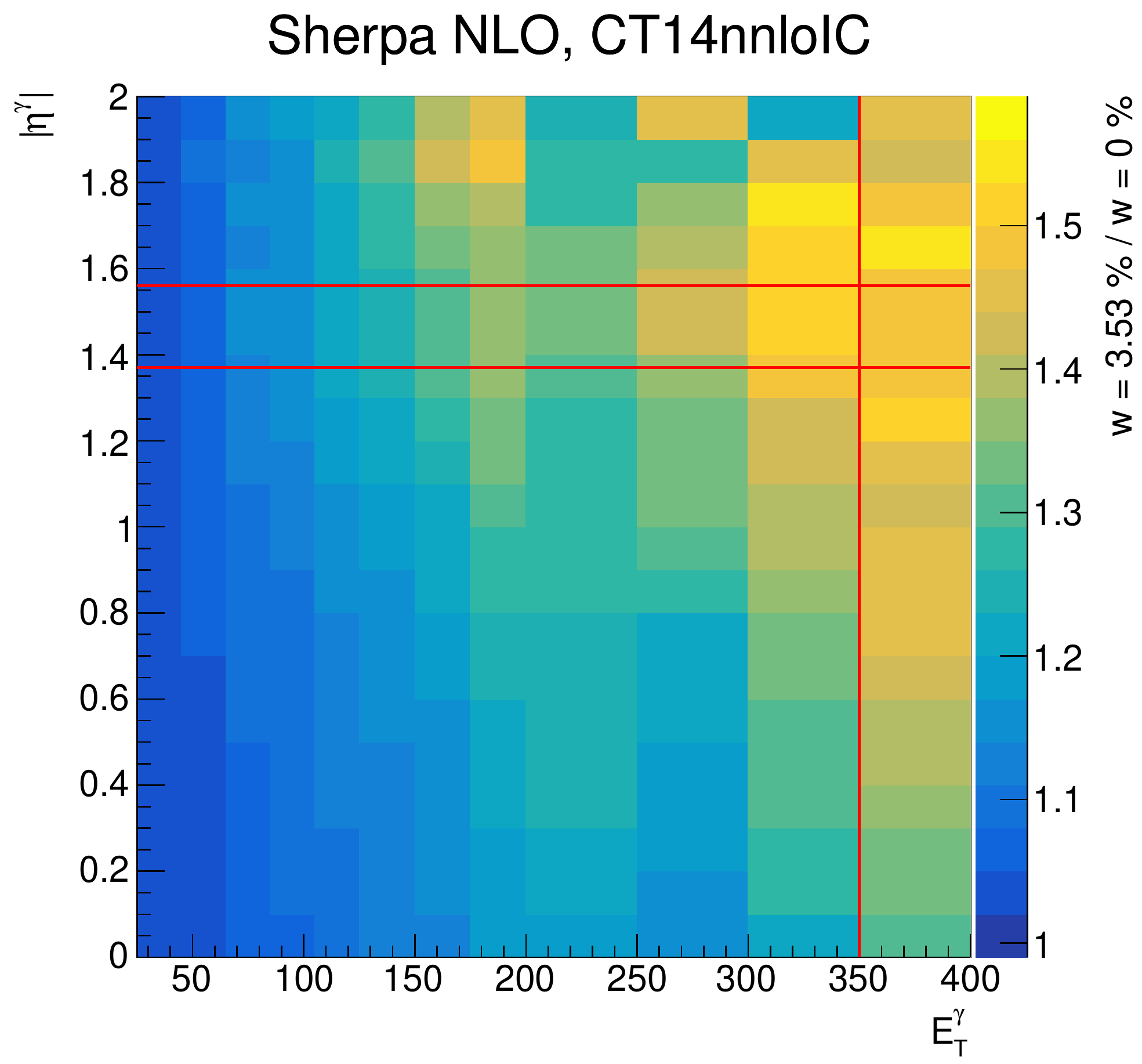}
  \caption{The ratio of the \ety{} spectra of \ycjet{} production with \ic{}
           contribution $\w = 3.53\%$ and no \ic{} contribution at different
           \aetay{} cuts using the \sherpa{} \nlo{} sample. The \atlas{}
           calorimeter gap is depicted by two horizontal red lines and \ety{}
           reach of the measurement in the forward \aetay{} region by vertical
           red line.}%
  \label{fig:results_eta_min}
\end{figure}

The decision of dividing the \atlas{} measurement of \ety{} spectra of
\ycjet{} production at the calorimeter gap is from the \ic{} point of view
optimal. The central \aetay{} region is defined as $\aetay < 1.37$ and forward
\aetay{} region as $1.56 \leq \aetay < 2.37$. Moving the \aetay{} cut towards
higher values would cause loss of statistics (in the \sherpa{} \nlo{} sample
only 30\% of all selected events end up in the forward \aetay{} region) and
moving it towards lower values might bring increase in systematic uncertainties,
since the forward \aetay{} region will need to combine events from two different
parts of the \atlas{} calorimeter. From the ratio of \ety{} spectra of
\ycjet{} with and without \ic{} contribution in
Fig.~\ref{fig:results_eta_min} is also visible, that the region of largest \ic{}
contribution only starts at around 250~\gev{} and it would be advantageous to
extend the \ety{} range of the measurement further above 350~\gev{}.

\subsection{Simulated Samples versus the Measurement}%
\label{sec:results_meas_vs_mc}

\begin{figure}[h]
  \centering
  \includegraphics[width=\spw]{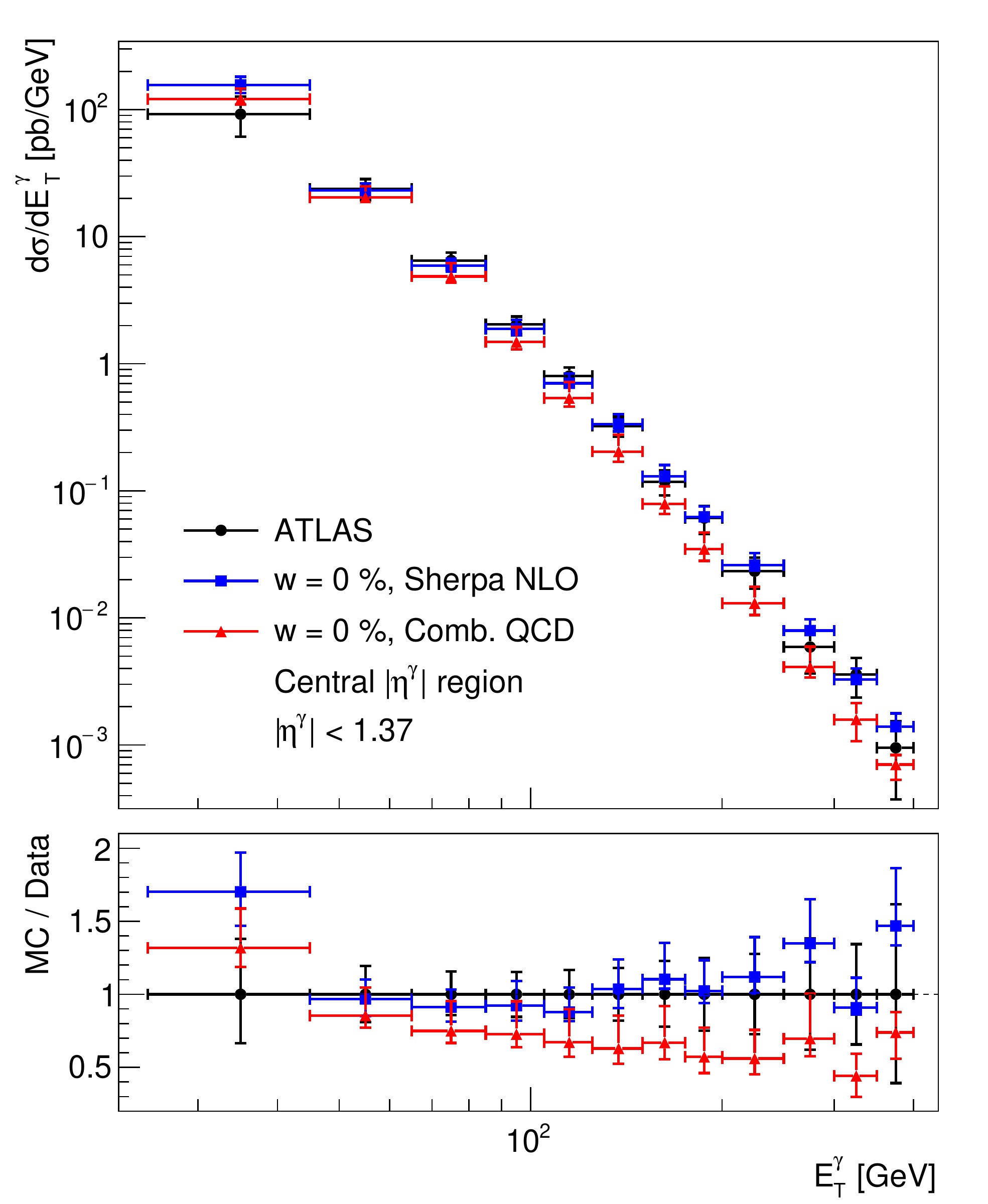}
  \includegraphics[width=\spw]{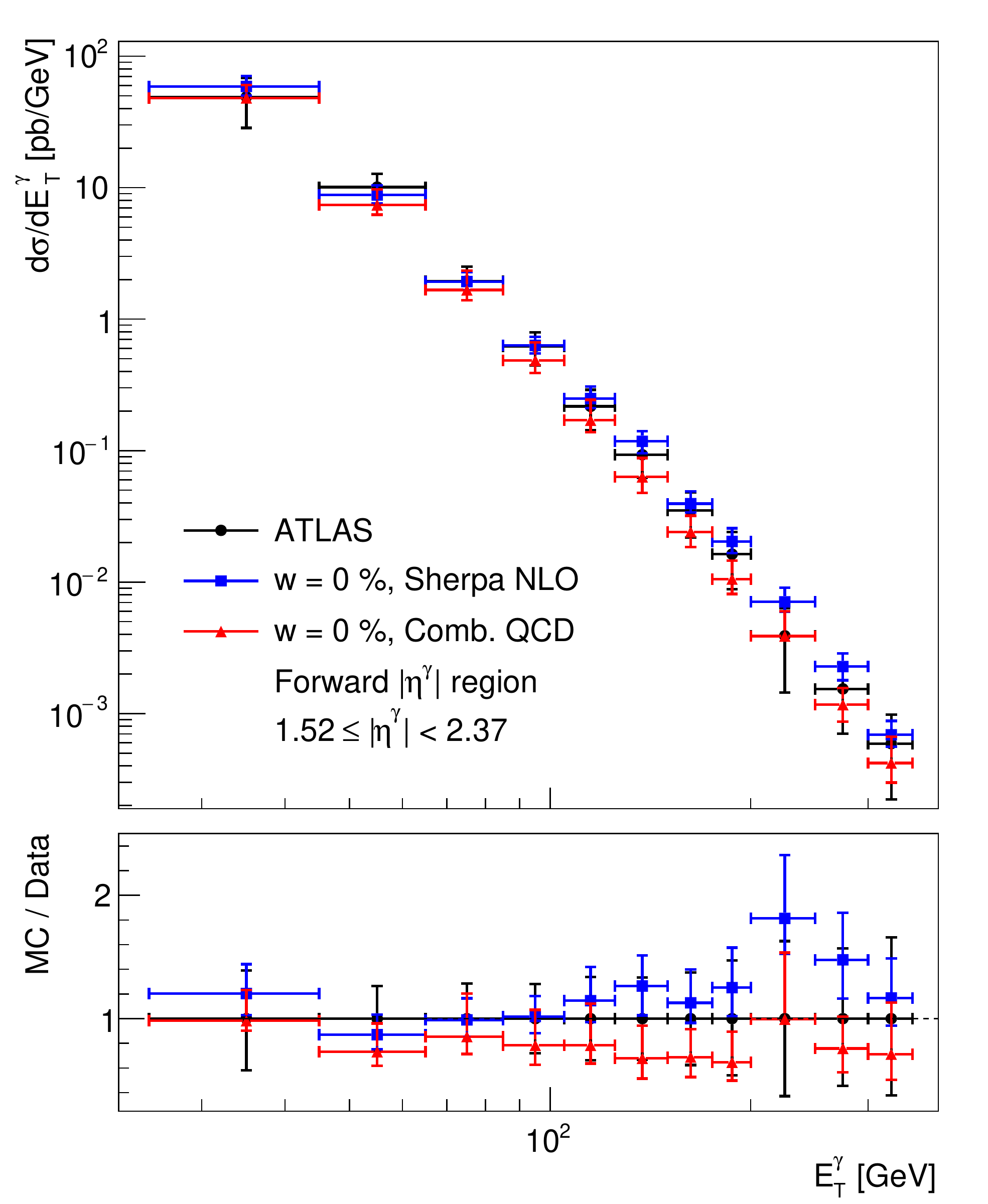}
  \includegraphics[width=\spw]{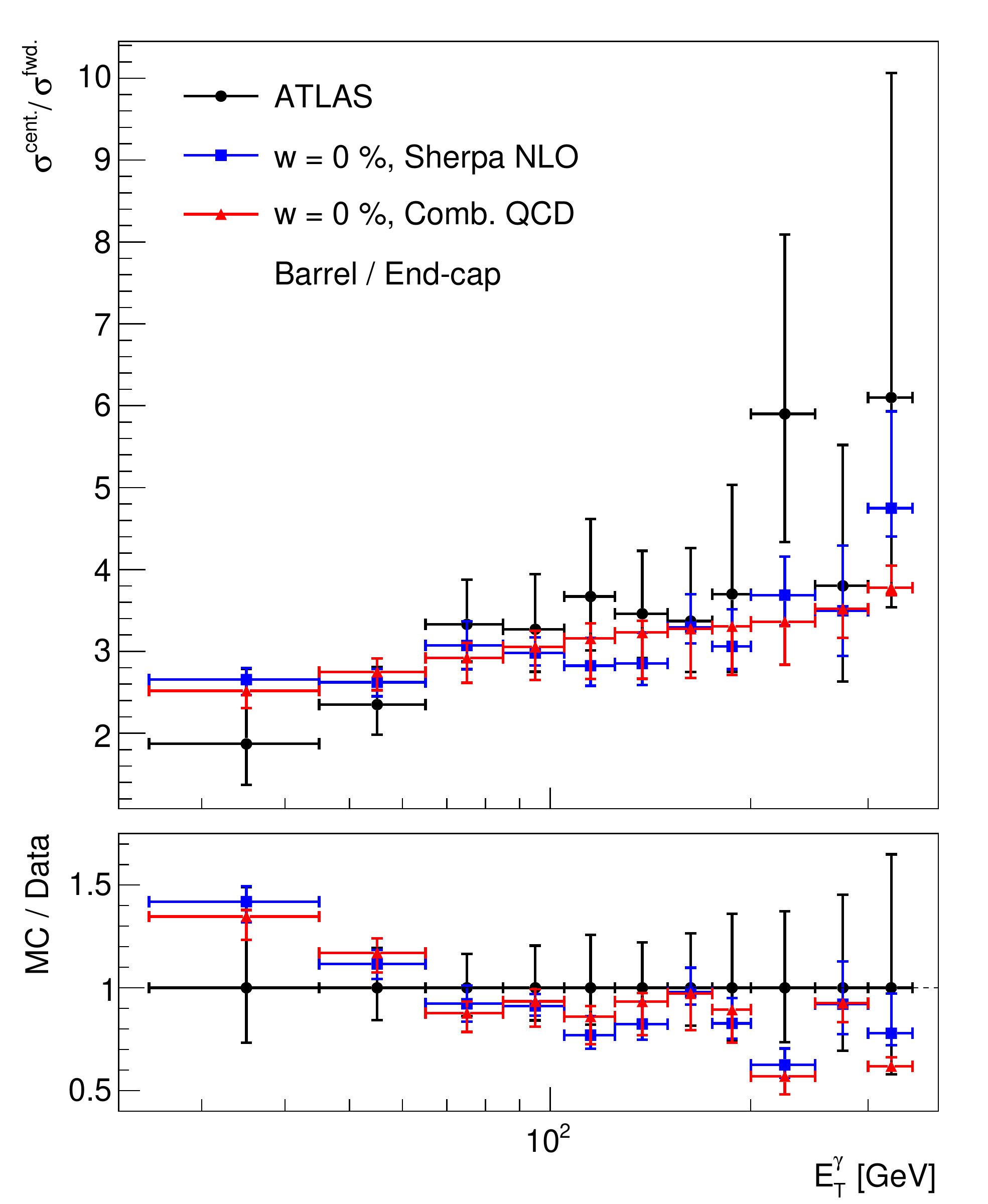}
  \caption{The \atlas{} measured \ety{} spectrum of \ppycjet{}
           process~\cite{Aaboud:2017skj} compared with the \sherpa{} \nlo{}
           and the Combined QCD simulated ones in two \aetay{} regions:
           central (left), forward (middle) and their ratio (Barrel/End-cap,
           right).}%
  \label{fig:results_yc_w0}
\end{figure}

The comparisons between \atlas{} \ycjet{} measurement and simulated
samples in different \aetay{} ranges are presented in
Fig~\ref{fig:results_yc_w0}. One can see that the \sherpa{} \nlo{} sample is
in agreement with the measurement in the central \aetay{} region within the
total uncertainties of the measurement. In case of the Combined QCD sample
one can observe underestimation of the measurement at high \ety{}. There is also
slight overestimation of the measurement in the first bin of the \ety{} spectra
in the both samples, which can be attributed to large scale uncertainties coming
from off-shell gluon-gluon fusion. This subprocess dominates the \ety{} region
below 100~\gev{}. The underestimation of the \ety{} spectra in the case of the
Combined QCD sample at large \ety{} can be explained by absence of the
effects of parton showers, hadronization and loop \nlo{} diagrams in this
calculation.

In case of the forward \aetay{} region the situation is similar, but there is a
smaller underestimation in the case of the Combined QCD sample at large \ety{}.
There is also a better agreement between the simulated samples and the
measurement at small \ety{}. The ratio of the two \aetay{} regions (central
\aetay{} region to forward \aetay{} region) exhibits good agreement between the
simulated samples and the measurement.

Determination of the optimal multiplication factor for the simulated \ety{}
spectra was performed in the central \aetay{} region. In the case of the
\sherpa{} \nlo{} simulated sample the found value of the factor is 0.97 and
in the case of the Combined QCD sample it is 1.28.

\subsection{Effect of Different PDFs}

\begin{figure}[h]
  \centering
  \includegraphics[width=\spw]{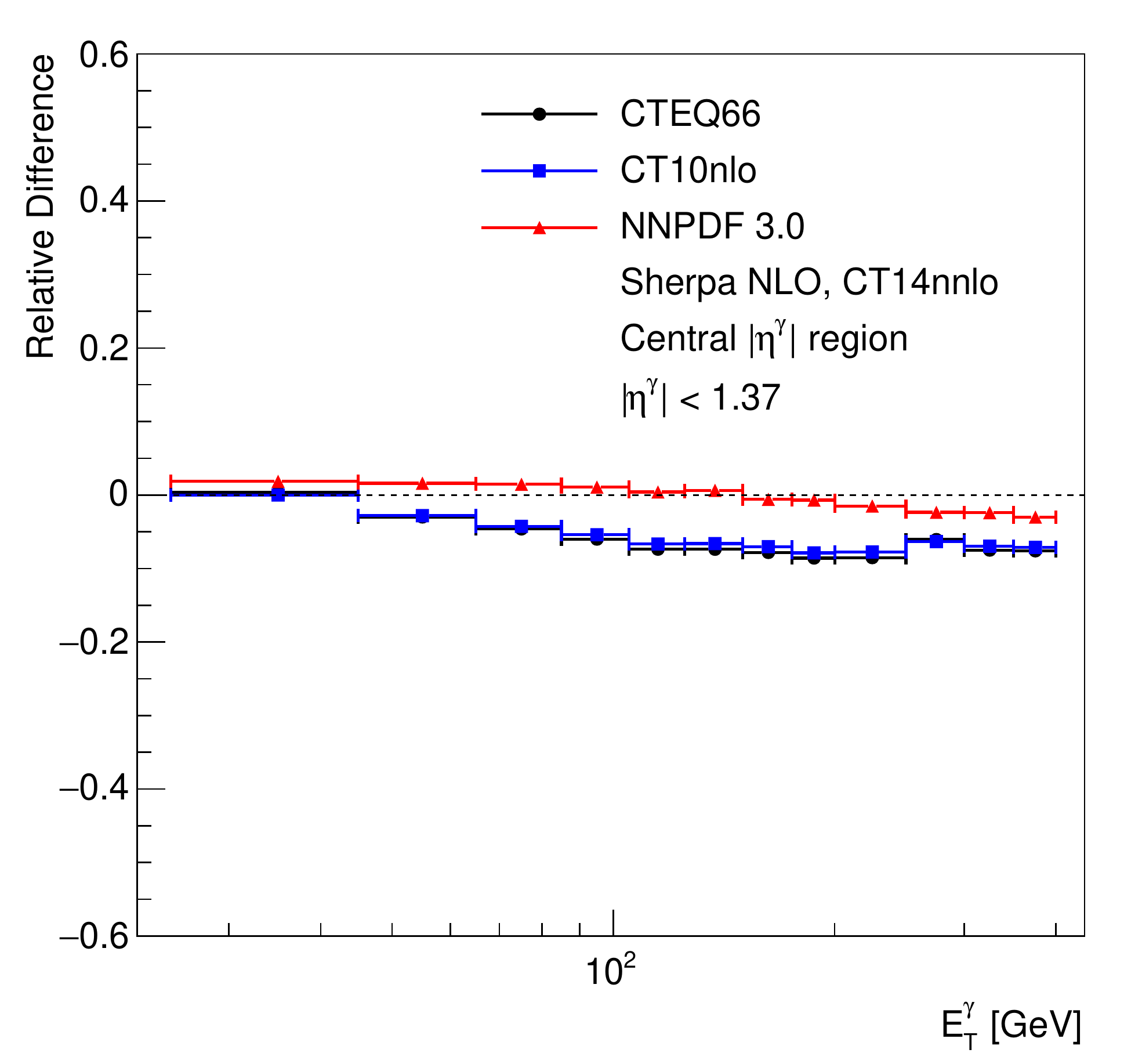}
  \includegraphics[width=\spw]{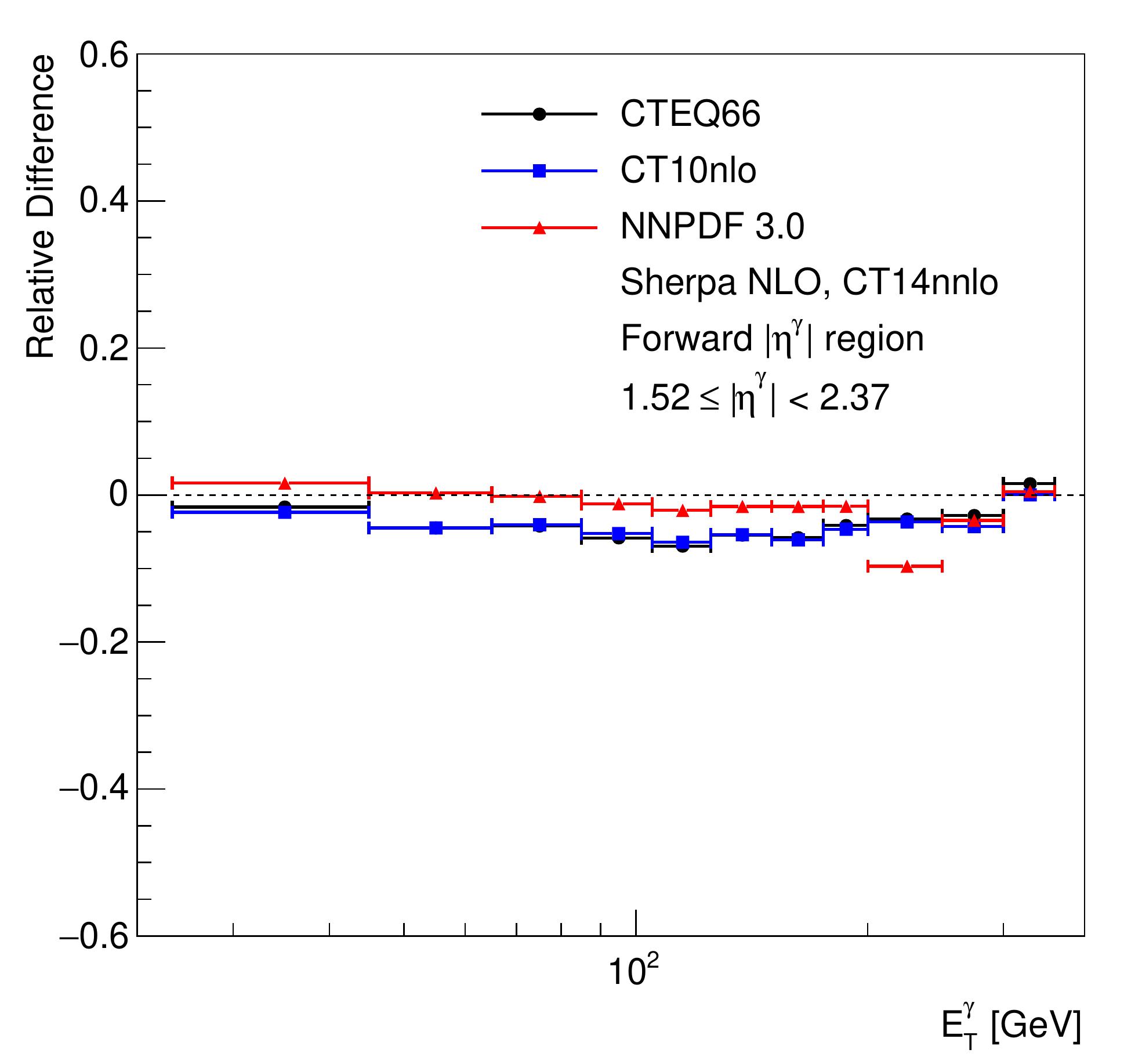}
  \includegraphics[width=\spw]{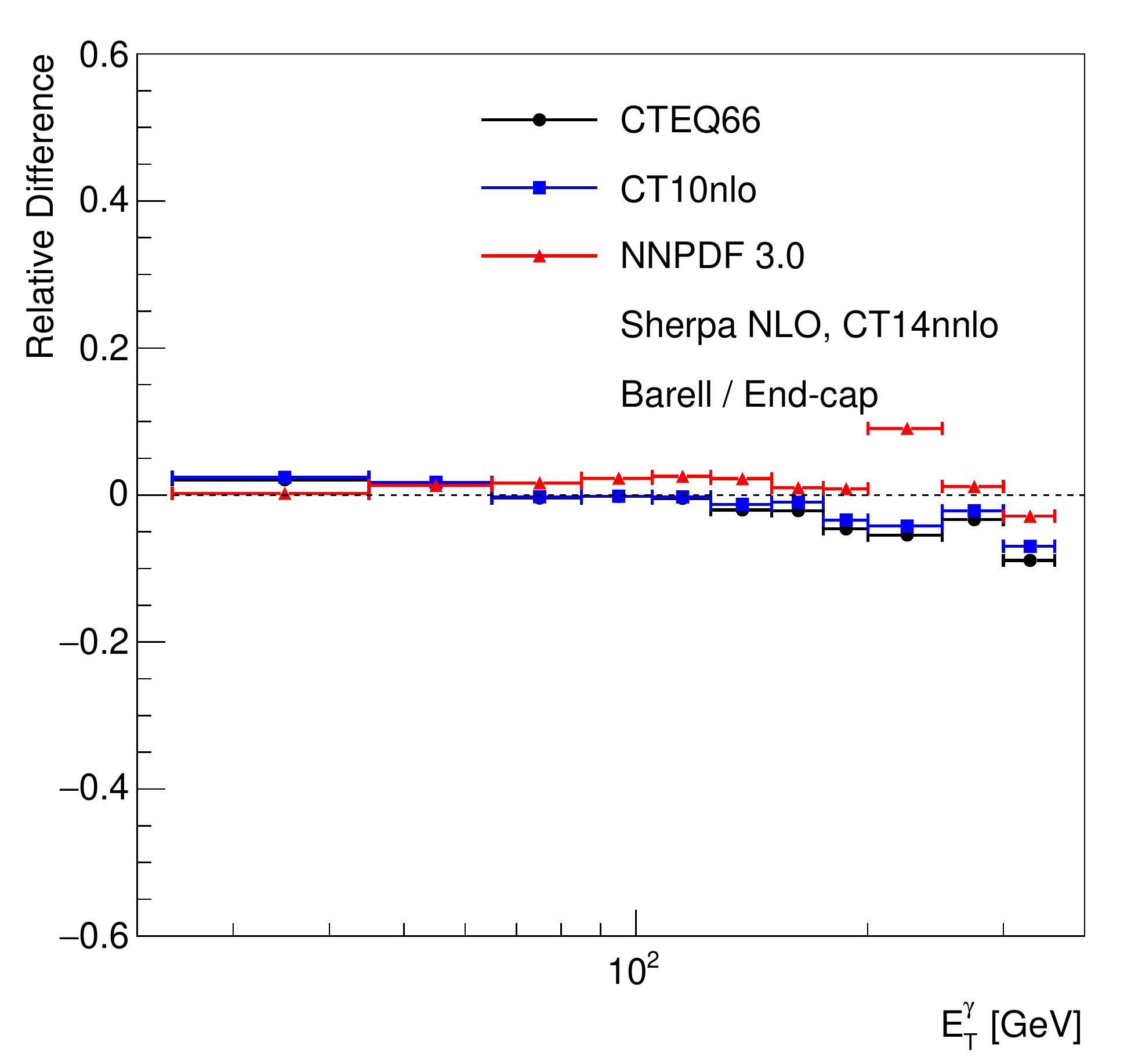}
  \caption{The relative difference between \ety{} spectrum using CTEQ66 (black
           points), CT10nlo (blue points) and NNPDF 3.0 (red points) \pdf{}
           vs. CT14nnlo \pdf{} in two \aetay{} regions (left and middle) and
           their ratio (Barrel/End-cap, right) at zero \ic{} contribution.}%
  \label{fig:results_pdf_var}
\end{figure}

The effects of using different \glspl{pdf} on \ety{} spectra of \ycjet{}
production were investigated. One can see from Fig.~\ref{fig:results_pdf_var},
which shows relative difference of the calculated \ety{} spectra of \ycjet{}
production with different \glspl{pdf}, that the difference is less than 10\% at
the central \aetay{} region and less than 5\% at the forward \aetay{} region. In
the comparison there are four \glspl{pdf}, two calculated with \nlo{} precision
in $\alpha_s$ --- CTEQ66 and CT10nlo, and two calculated with \nnlo{} ---
CT14nnlo and NNPDF 3.0. The size of the difference is comparable to the size of
the uncertainty coming directly from the CT14nnlo eigenvectors. It is also much
smaller than experimental uncertainty of the \atlas{} \ycjet{} measurement and
about half of the scale uncertainties of the simulated samples.

If one looks at ratio between recent \glspl{pdf}, NNPDF 3.0 and CT14nnlo (both
\nnlo{}), the difference is even smaller in the central \aetay{} region.
Unfortunately, the largest difference between the recent \glspl{pdf} is at large
\etay{} in the forward \aetay{} region, the region of largest \ic{}
contribution. Still, one can see, that the uncertainty in \pdf{} has a minor
effect on \ety{} spectra of \ycjet.

\subsection{Intrinsic Charm Fitting Method}%
\label{sec:ic_fitting_method}

The determination of the \w{} value from the ATLAS \ycjet{} measurement
started with validation of the simulated \ety{} spectra against the measurement
in the central \aetay{} region, see Sec.~\ref{sec:results_meas_vs_mc}. One can
see, that the measurement is satisfactorily described in the central \aetay{}
region using the \sherpa{} \nlo{} sample without \ic{}. On the other hand,
the Combined QCD sample description underestimates the measurement, but stays
within the total uncertainties. Since the uncertainties of the measurement and
also simulated samples are comparable in size with the possible \ic{} signal
only an upper limit could be determined instead of precise value of the \ic{}
probability.

To obtain the upper limit \wul{} on possible \ic{} contribution in proton, a
simple method of employing simulated spectra (ratios) containing known values of
\ic{} probability $w$ was employed. In case of the \sherpa{} NLO sample the
quadratic fit was employed and in the case of the Combined QCD sample it was the
quadratic interpolation~\cite{Bednyakov:2017vck}. The templates were generated
for the forward \aetay{} region, where the inclusion of \ic{} has the effect of
increasing the spectrum at high \ety{} and for the Barrel/End-cap ratio where
the inclusion has the effect of decreasing the ratio. To compare a template and
the measured spectra the \chis{} was calculated as follows.

\begin{equation}
  \chi^{2}(w) = \sum_{i = 1}^{n}
                \frac{{\left[ y_{i} - {f(w)}_{i} \right]}^2}{\sigma_i^2}
  \label{eq:results_chi2}
\end{equation}

Here $y_i$ is the measurement, ${f(w)}_{i}$ is the simulated sample with known
\w{} value and $\sigma_i$ is the sum in quadrature of the uncertainties coming
from the measurement and the uncertainties coming from the simulated sample. For
all the templates the \chis{} is calculated creating a curve. The minimum of the
\chis{} curve determines the central value \wc{} and the upper limit \wul{} is
determined as the \w{} value that corresponds to the minimum of
$\chi^{2}_{\min}$ plus one, since there is only one fit parameter.

The employment of this simple fitting method is justified by negligible
correlation between the bins of the \ety{} spectra in the both \aetay{} regions
and in the case of the Barrel/End-cap ratio also by negligible correlations
between bins of the two \aetay{} regions. For the reason of large correlations
between the bins of \ycjet{} and \ybjet{} measurement the fit
using their ratio was not realized.

\subsection{Upper Limit on Intrinsic Charm in Proton}%
\label{sec:results_upper_limit}

The upper limit of the \ic{} contribution in proton obtained within the
\sherpa{} \nlo{} sample is $\wul = 1.97$\% at 68\% \cl{} It is found from the
\ety{} spectrum of the \ycjet{} production measured in the forward \aetay{}
region, see Sec.~\ref{sec:upper_limit_sherpa}. The another
upper limit obtained within the Combined QCD sample is presented in
Sec.~\ref{sec:upper_limit_combined_qcd}.

The simulated samples employed in the \ic{} fitting method are too different in
respect to each other, which prevented one to be used as a cross check for the
other. However, resulting upper limit \wul{} in the case of the \sherpa{}
\nlo{} sample is considered more reliable, since this sample better describes
measured \ety{} spectrum of the \ycjet{} production in the central \aetay{},
this is mainly because of inclusion of all \nlo{} diagrams, parton showers
and hadronization.

\subsection{Sherpa NLO Sample}%
\label{sec:upper_limit_sherpa}

\begin{figure}[h]
  \centering
  \includegraphics[width=\mpw]{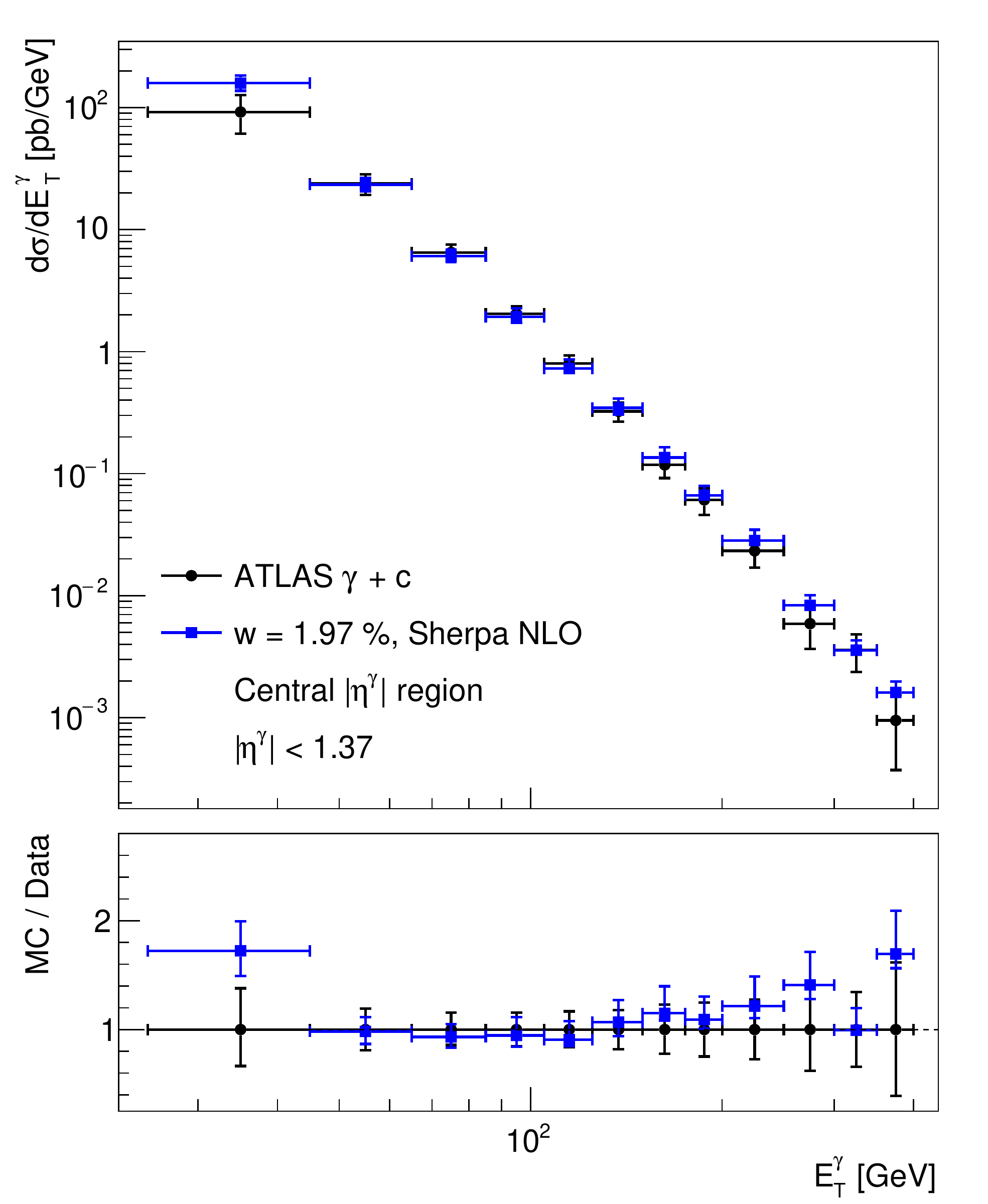}
  \includegraphics[width=\mpw]{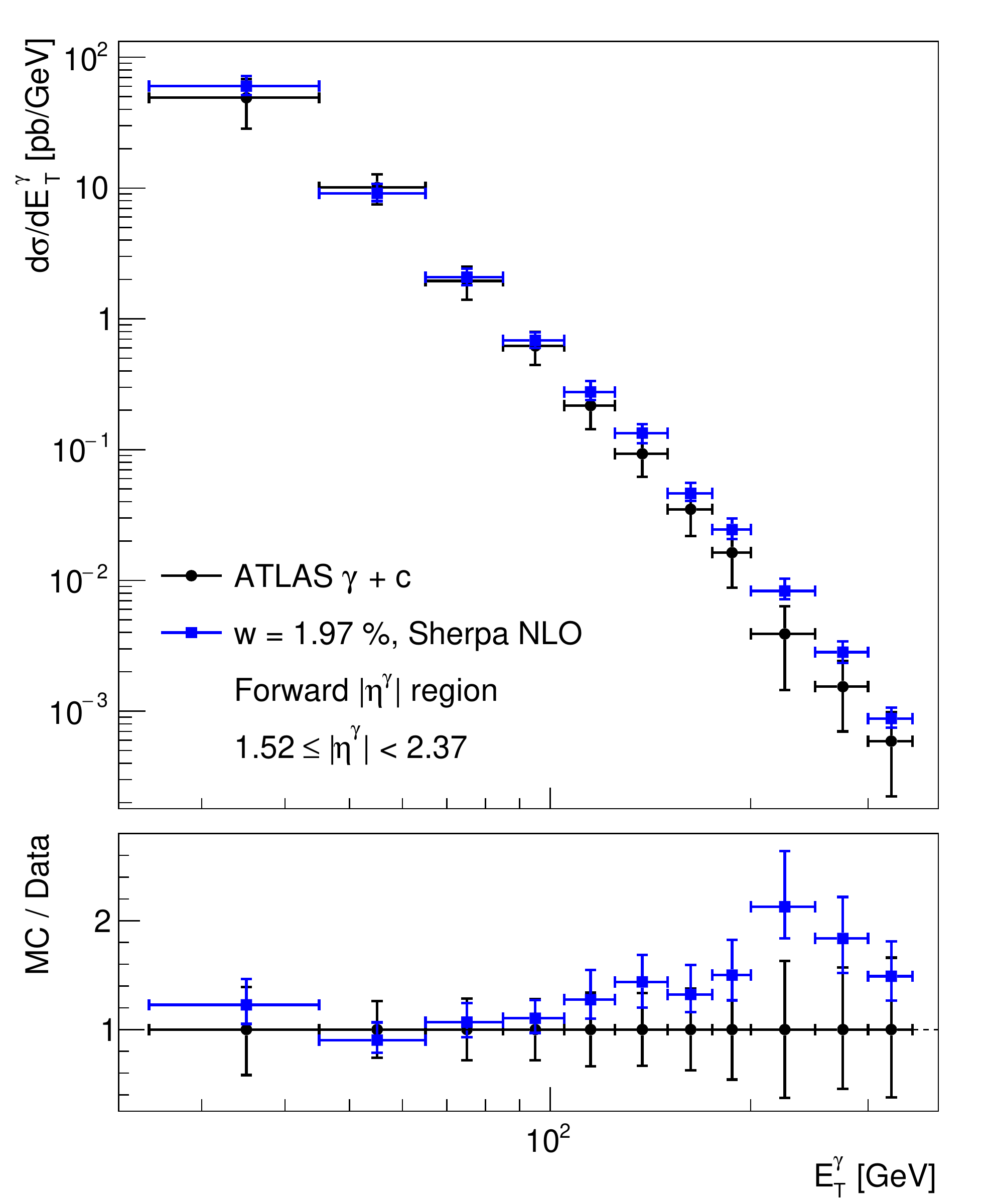}
  \caption{The \ety{} spectrum of \ycjet{} from the \sherpa{} \nlo{} sample
           compared with the \atlas{} measurement in two \aetay{} regions.
           Both panels show the simulated spectrum at the upper limit \ic{}
           contribution $\wul = 1.97$\% at 68\% \cl.}%
  \label{fig:results_ic_wul_sherpa}
\end{figure}

\begin{figure}[h]
  \centering
  \includegraphics[width=\mpw]{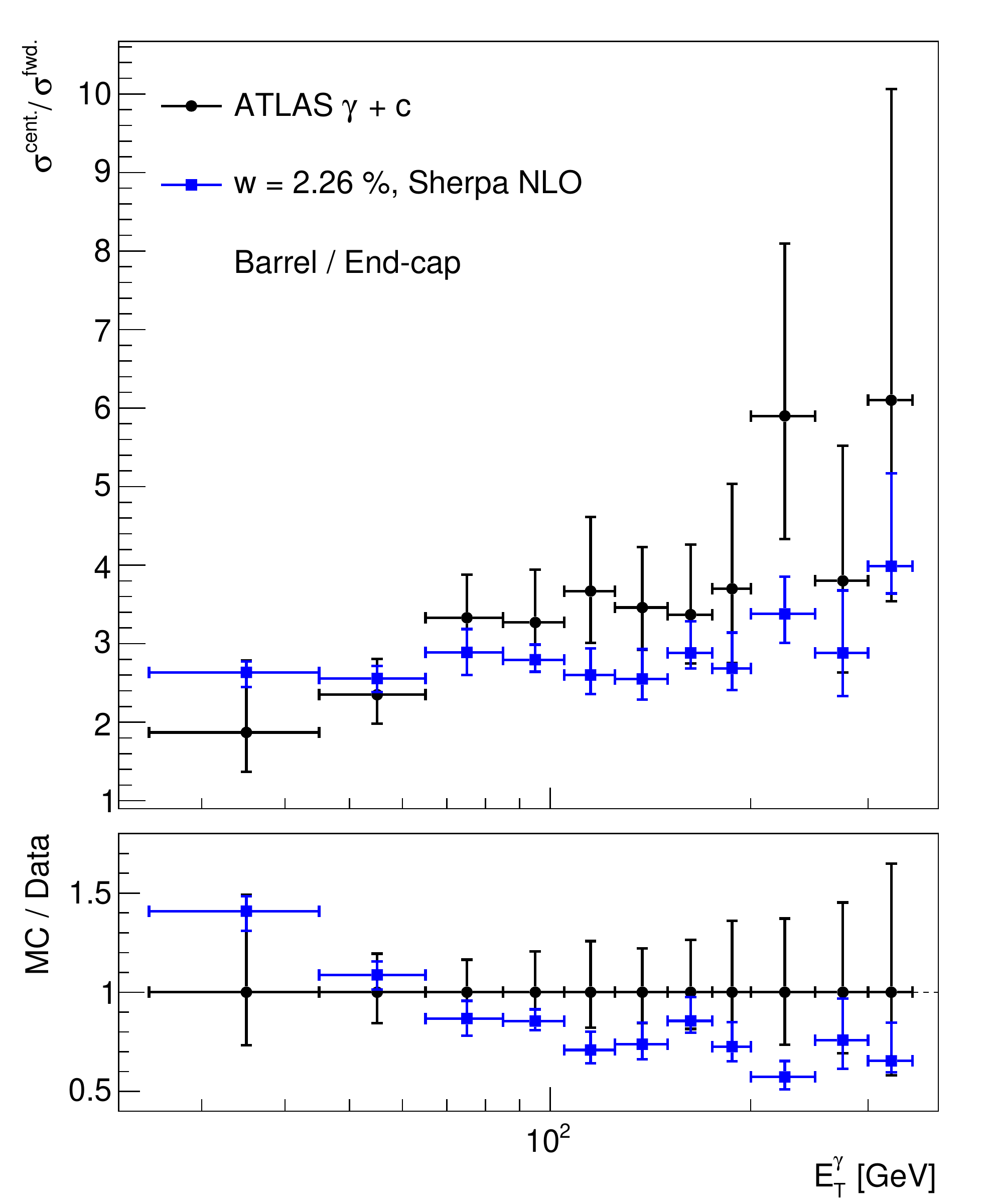}
  \includegraphics[width=\mpw]{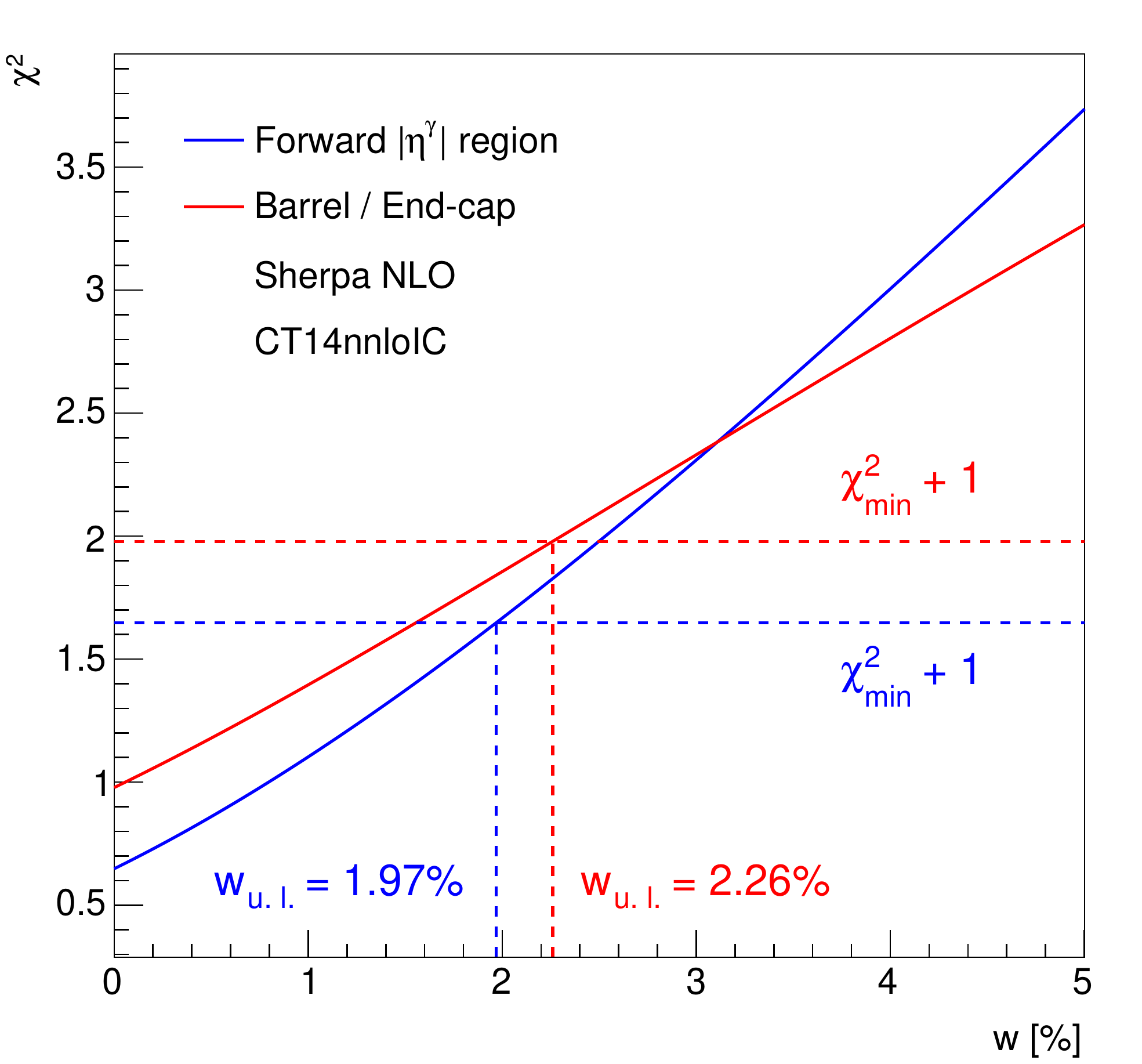}
  \caption{The ratio of the \ety{} spectra in the central \aetay{} region
           to the \ety{} spectra in the forward \aetay{} region (Barrel /
           End-cap ratio) of \ycjet{} process from the \sherpa{} \nlo{}
           sample at the upper limit \ic{} contribution $\wul = 2.26$\% at 68\%
           \cl{} compared with the \atlas{} measurement (left). The \chis{}
           as a function of \w{} for two cases of upper limit determination
           based on the \ety{} spectra in the forward \aetay{} region and the
           Barrel/End-cap ratio (right).}%
  \label{fig:results_ic_wul_be_chi2_sherpa}
\end{figure}

Upper limit on \ic{} in proton determined with the help of the \sherpa{}
\nlo{} simulated sample is presented in Fig.~\ref{fig:results_ic_wul_sherpa}
and Fig.~\ref{fig:results_ic_wul_be_chi2_sherpa}. The
Fig.~\ref{fig:results_ic_wul_sherpa} shows the \ety{} spectra of \ycjet{} with
the contribution from \ic{} at the upper limit value $\wul = 1.97$\% at 68\%
\cl{} in the central and forward \aetay{} regions. The limit was obtained by
fitting the full \ety{} spectra of forward \aetay{} region as described in
Sec.~\ref{sec:ic_fitting_method}. The left panel of
Fig.~\ref{fig:results_ic_wul_be_chi2_sherpa} shows another way of obtaining the
\ic{} limit. In this case the \ic{} upper limit is obtained from the
Barrel/End-cap ratio and resulting value is $\wul = 2.26$\% at 68\% \cl{}. The
limit obtained from employing the forward \aetay{} region is considered more
reliable due to its smaller sensitivity to the incompatibilities between the
simulation and the measurement.
In Fig.~\ref{fig:results_ic_wul_be_chi2_sherpa} (right) the \chis{} dependence on
the \ic{} percentage \w{} for the both cases of upper limit determination is
shown.

Fig.~\ref{fig:results_ic_wul_be_chi2_sherpa} shows a rather weak \chis{}
sensitivity to the \w{} value in the both \ic{} upper limit determination cases,
which is caused by large experimental and theoretical uncertainties. The \chis{}
is slightly more sensitive to the \w{} in the case of employing forward \aetay{}
region, this results in a smaller \ic{} upper limit. The central value of the
fit \wc{} is in the both cases 0. Also in the both cases of upper limit
determination the obtained values are relatively close, the result is around 2\%
at 68\% \cl.

\subsection{Combined QCD Sample}%
\label{sec:upper_limit_combined_qcd}

\begin{figure}[h]
  \centering
  \includegraphics[width=\mpw]{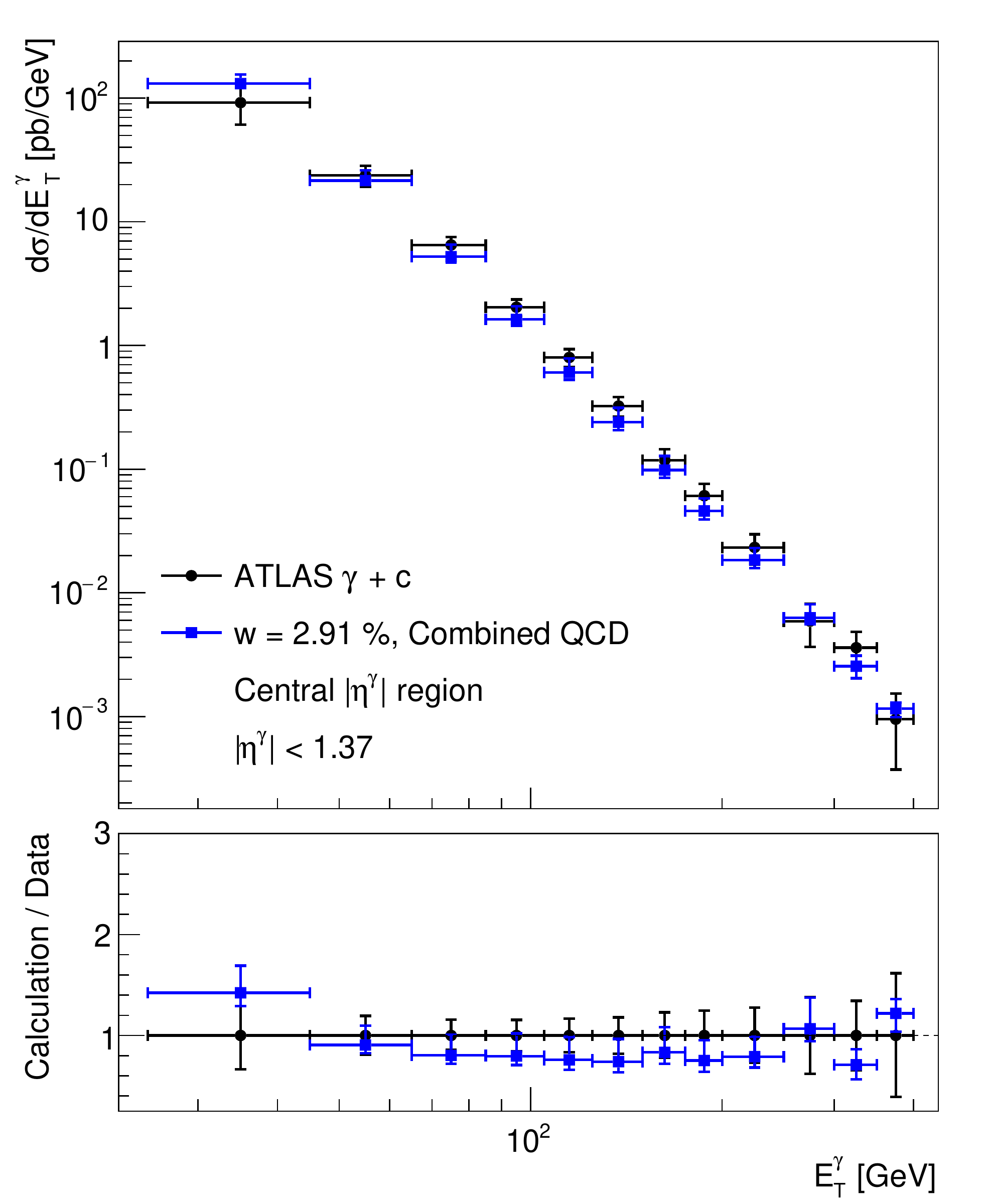}
  \includegraphics[width=\mpw]{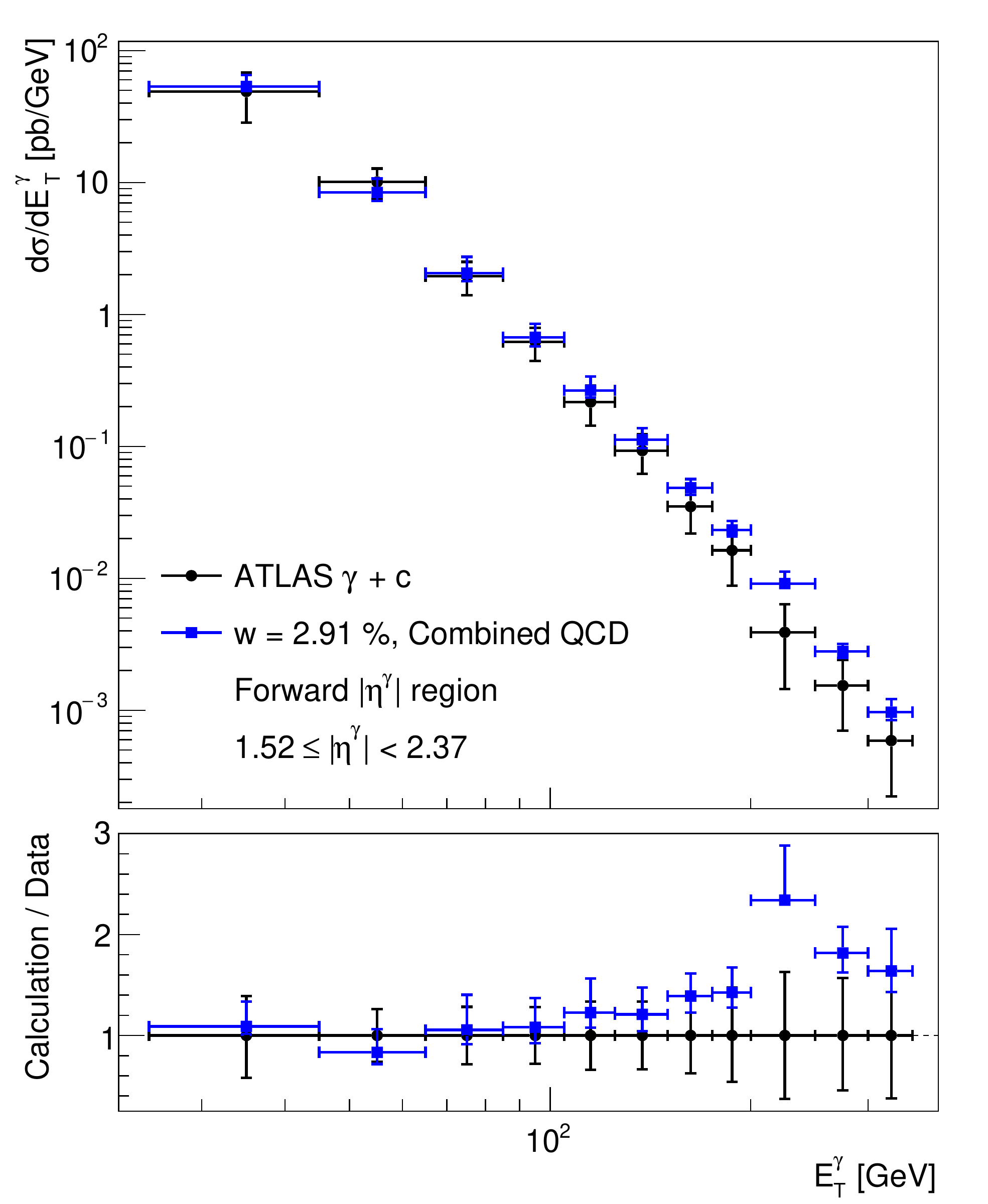}
  \caption{The \ety{} spectrum of \ycjet{} from the Combined QCD sample
           compared with the \atlas{} measurement in two \aetay{} regions.
           Both panels show the simulated spectrum at the upper limit \ic{}
           contribution $\wul = 2.91$\% at 68\% \cl.}%
  \label{fig:results_ic_wul_comb_qcd}
\end{figure}

\begin{figure}[h]
  \centering
  \includegraphics[width=\mpw]{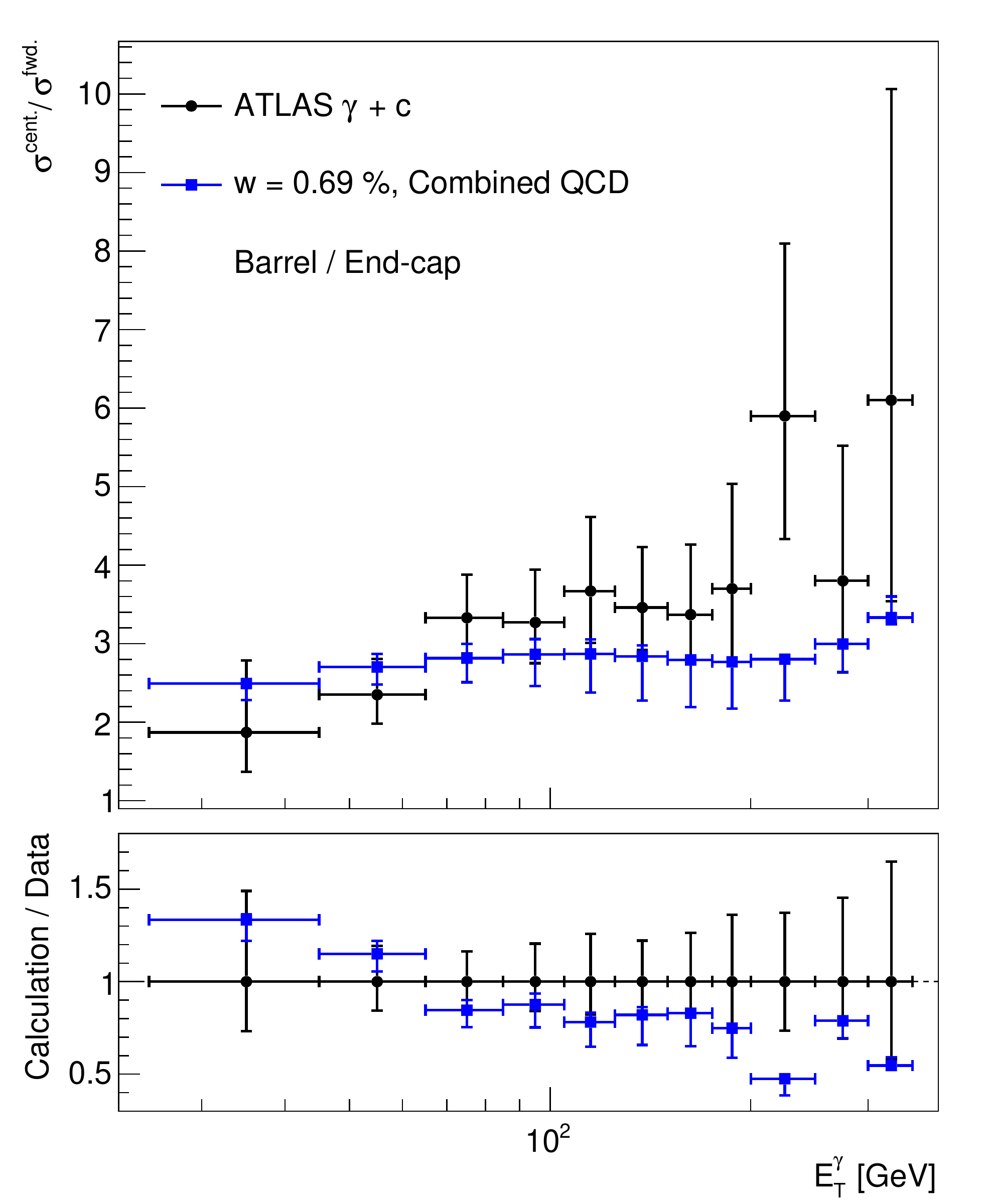}
  \includegraphics[width=\mpw]{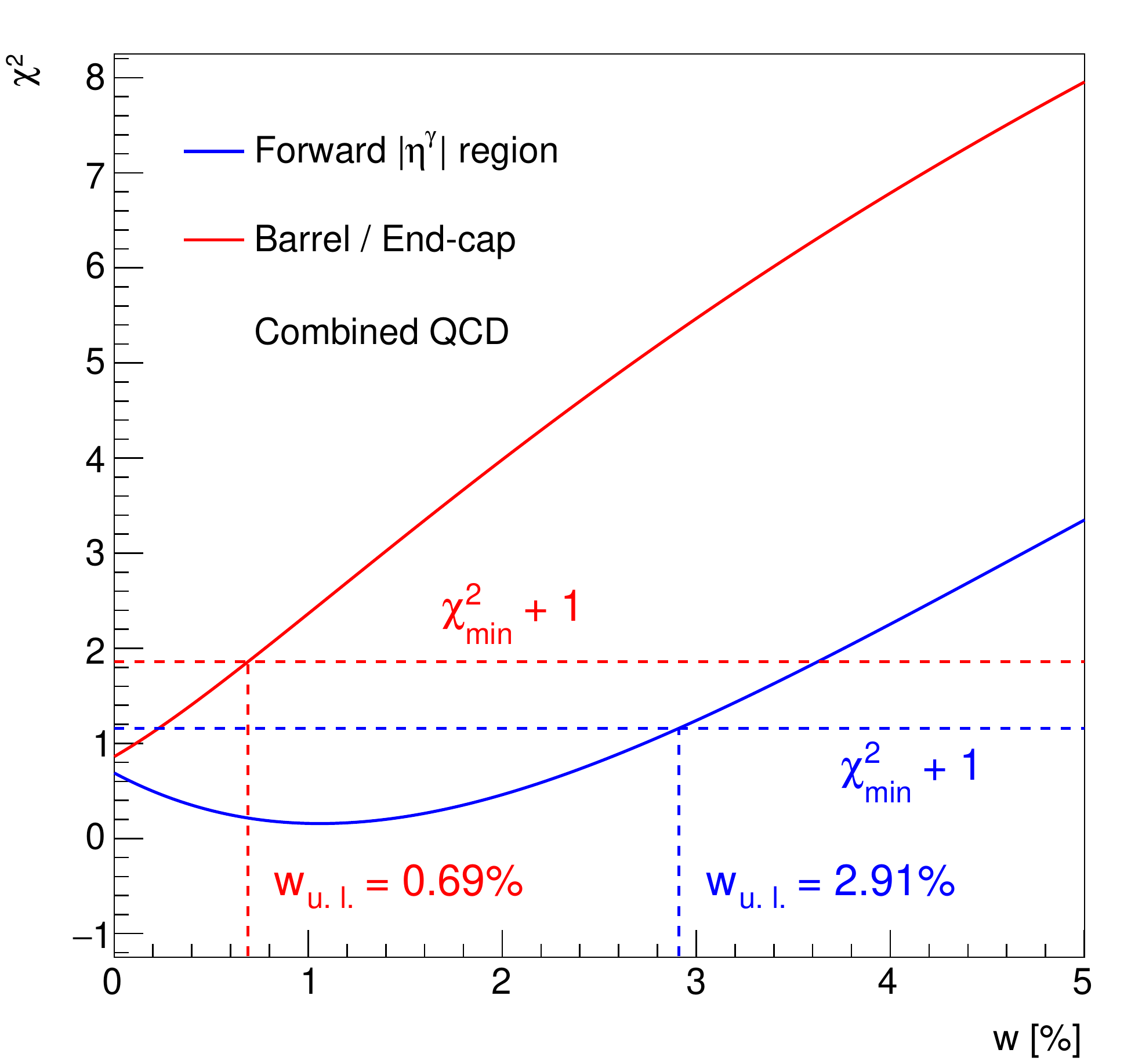}
  \caption{The ratio of the \ety{} spectra in the central \aetay{} region to the
           \ety{} spectra in the forward \aetay{} region (Barrel / End-cap
           ratio) of \ycjet{} process from the Combined QCD sample at the
           upper limit \ic{} contribution $\wul = 0.69$\% at 68\% \cl{} compared
           with the \atlas{} measurement (left). The \chis{} as a function of
           \w{} for two cases of upper limit determination based on the \ety{}
           spectra in the forward \aetay{} region and the Barrel/End-cap ratio
           (right).}%
  \label{fig:results_ic_wul_be_chi2_comb_qcd}
\end{figure}

The \ic{} probability \w{} fitting method was also repeated with the Combined
QCD simulated sample and the \ety{} spectra for both \aetay{} rapidity
regions compared to the \atlas{} measurement are presented in
Fig.~\ref{fig:results_ic_wul_comb_qcd}. The \ety{} spectra contain \ic{} upper
limit value ${\wul = 2.91}$\% at 68\% \cl, which was determined by employing
full \ety{} spectra of forward \aetay{} region as described in
Sec.~\ref{sec:ic_fitting_method}. In the left panel of
Fig.~\ref{fig:results_ic_wul_be_chi2_comb_qcd} the upper limit of ${\wul =
0.69}$\% at 68\% \cl{} obtained by employing the Barrel/End-cap ratio is shown.
The figure also shows in the right panel the \chis{} dependence of both fitting
options on the \ic{} contribution \w.

The upper limit obtained by employing the forward \aetay{} region is not
coinciding with the limit obtained by employing the Barrel/End-cap ratio. Also,
the fits does not produce the same central value, for the forward \aetay{}
region fit it is $\wc = 1.06$ and for the Barrel/End-cap ratio it is $\wc = 0$.
The \chis{} dependence on \w{} in right panel of
Fig.~\ref{fig:results_ic_wul_be_chi2_comb_qcd} shows much different shapes with
the Barrel/End-cap ratio employing fit having a very sharp \chis{} dependency,
resulting in the smallest upper limit \wul{}. The source of this discrepancy can
be attributed to an inability of the Combined QCD model to describe \ety{}
spectra of \ycjet{} production in forward \aetay{} region well enough, see
Fig.~\ref{fig:results_yc_w0}. This just resulted in filling up missing cross
section with the \ic{} contribution, especially at large \ety. The Combined QCD
model does not include parton showers and hadronization, which turns out to be
crucial for the upper limit of \ic{} determination from \ety{} spectra of
\ycjet{} production. Therefore, the results obtained by employing the \sherpa{}
\nlo{} simulated sample, which include these effects, are considered to be
more reliable.

\clearpage

\subsection{Predictions for \sxiii{} Measurement}

\begin{figure}[h]
  \centering
  \includegraphics[width=\mpw]{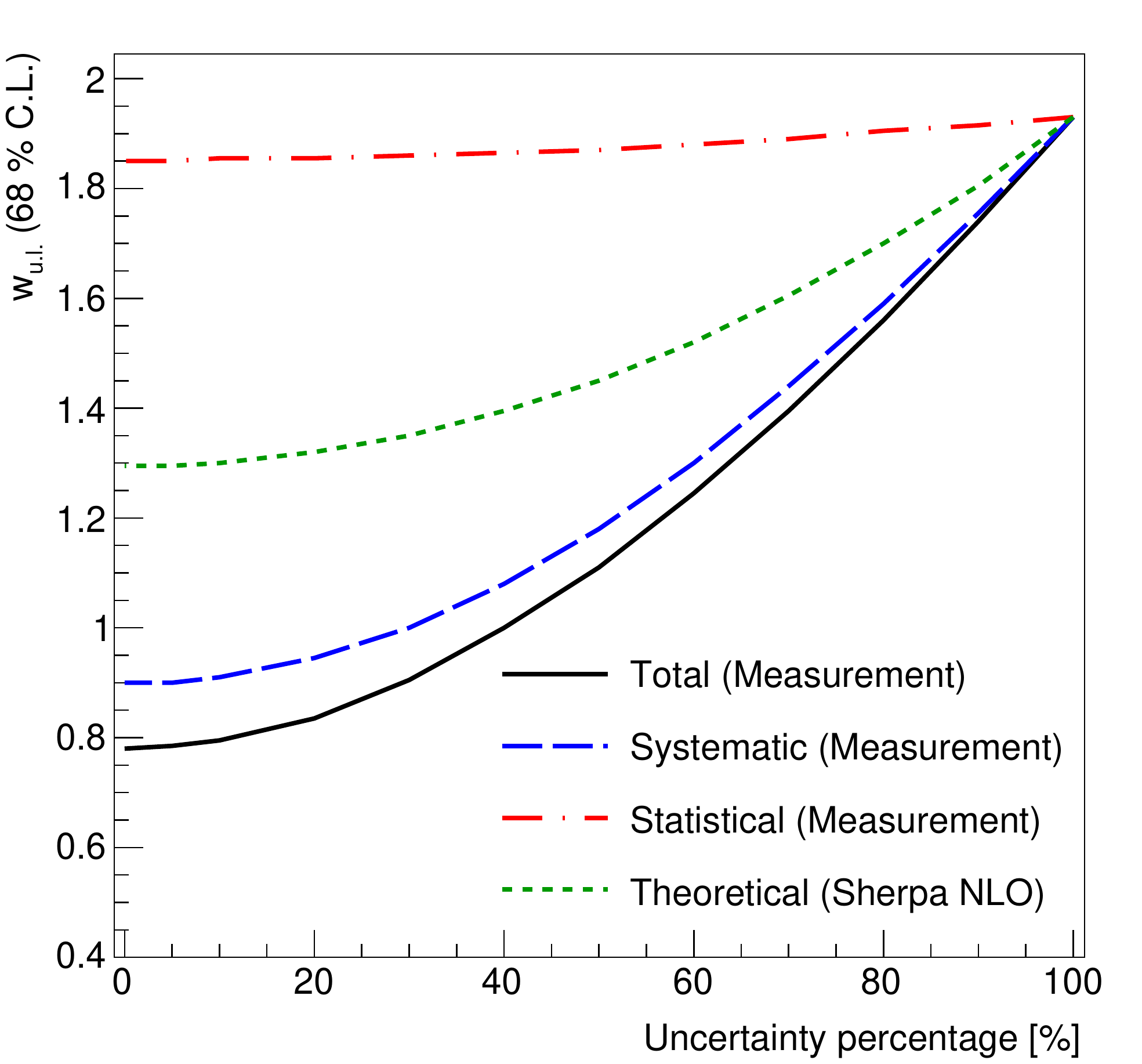}
  \caption{The dependence of the \ic{} upper limit \wul{} at 68\% \cl{} on the
           reduction of a particular uncertainty component.}%
  \label{fig:results_reduce_uncert}
\end{figure}

With the Run 2 of the \lhc{} coming to an end, the final collected luminosity by
the \atlas{} experiment is around 140~fb$^{-1}$. This luminosity is more than
enough to enable more precise measurement of \ycjet{} differential cross-section
in \ety. However, the uncertainty which defines the precision of the measurement
is dominated by the systematic uncertainty connected with the light jet tagging.
In the case of the simulated samples the scale uncertainties are the most
significant. The effects of the uncertainties of the measurement or the
simulated samples on the upper limit of the \ic{} contribution to proton \pdf{}
can be seen in Fig~\ref{fig:results_reduce_uncert}, which shows the possible
upper limit when one reduces a particular uncertainty in the \ic{} fit described
in Sec.~\ref{sec:results_upper_limit}. Three sources of uncertainty are reduced
separately from full size (100\% on the right) down to none (0\% on the left).
One can see, that the reduction of statistical uncertainty of the measurement
has no particular effect on the reduction of the \ic{} upper limit \wul{}. The
reduction of the theoretical uncertainty of the simulated samples (\sherpa{}
\nlo) has significant effect down to around 50\% of the original uncertainty.
The largest effect has the reduction of the systematic uncertainty of the
measurement down to around 20\%.

\begin{figure}[h]
  \centering
  \includegraphics[width=\mpw]{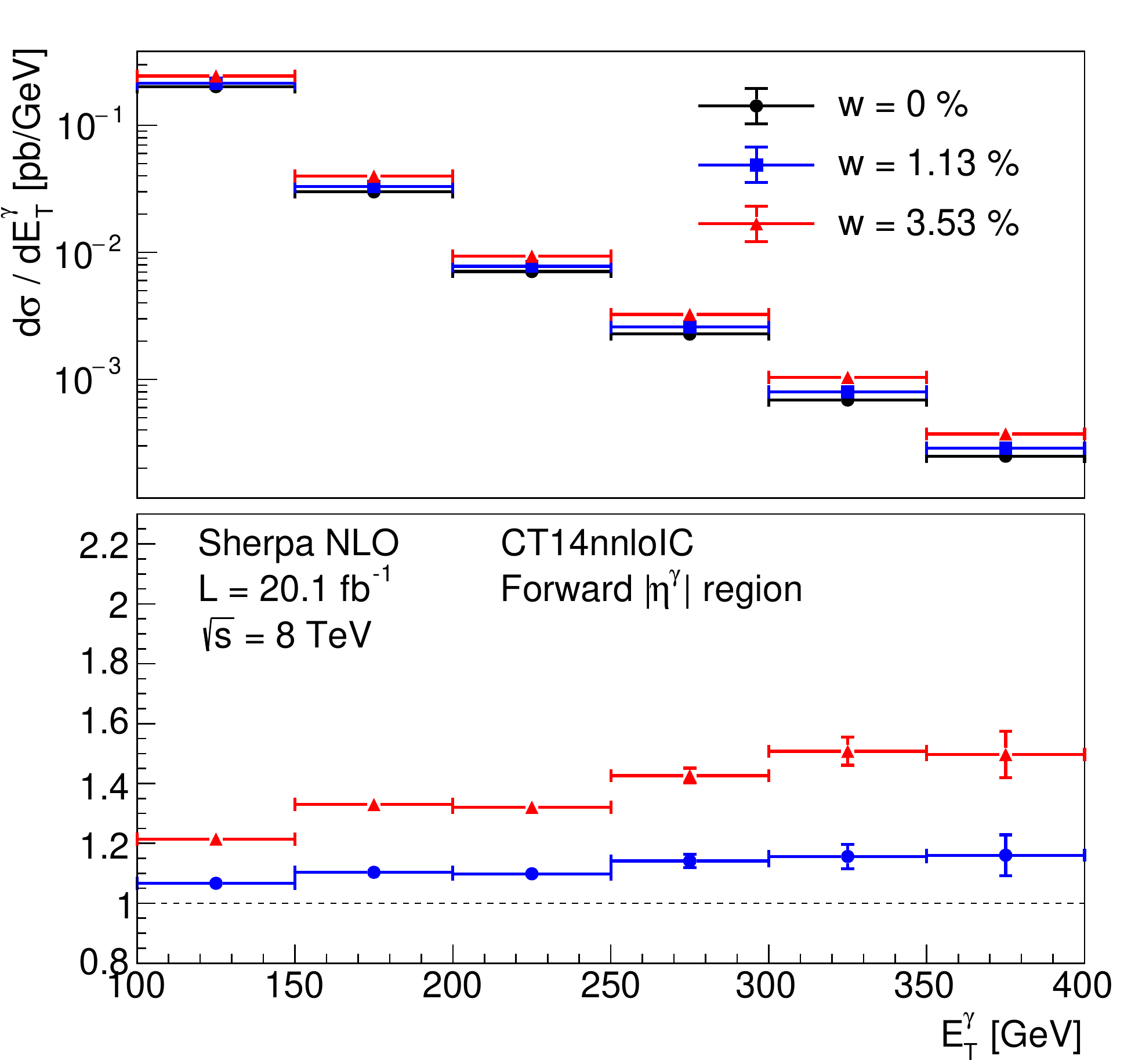}
  \includegraphics[width=\mpw]{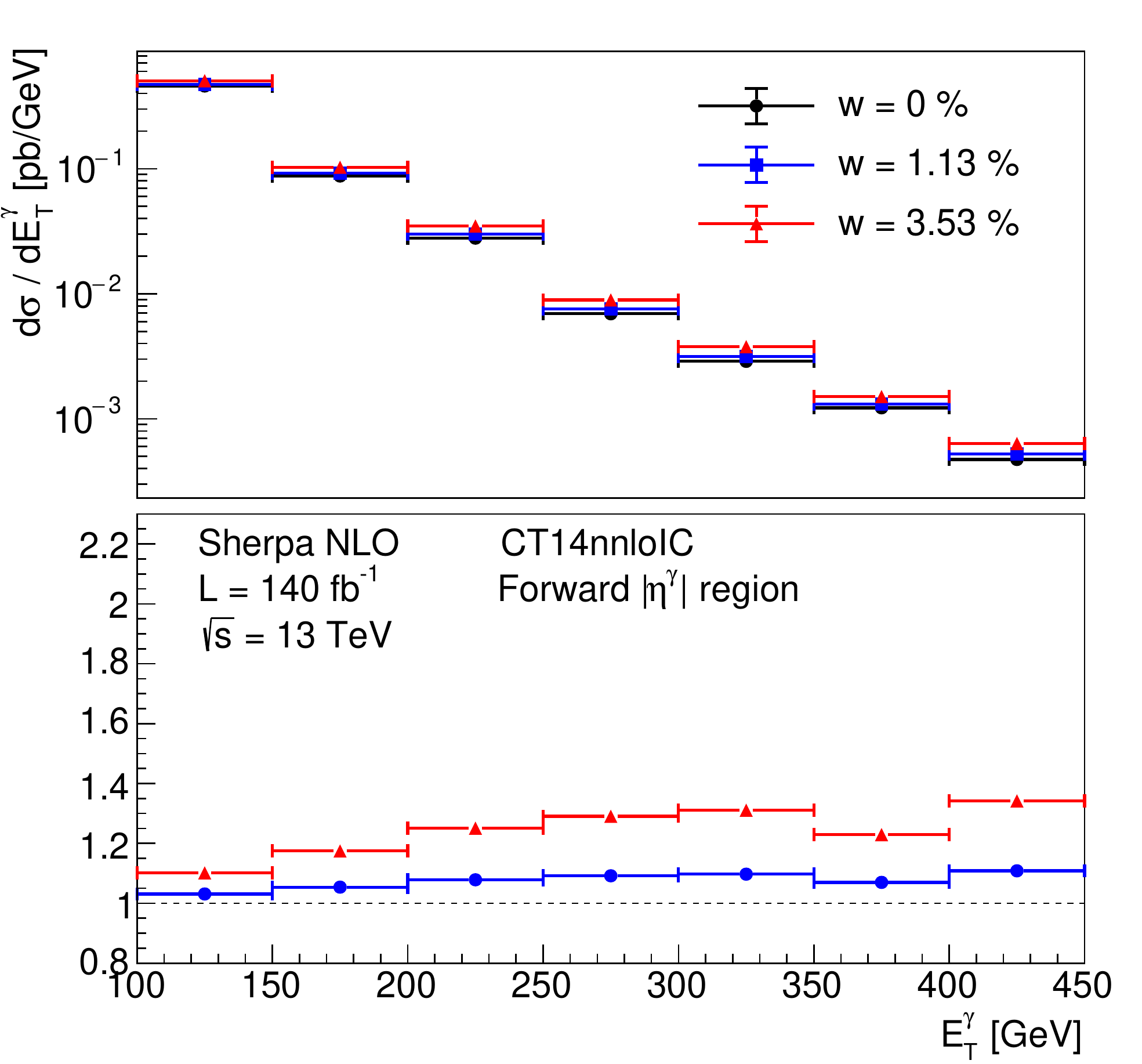}
  \caption{Differential cross-sections in \ety{} of the \ycjet{}
           production at two center-of-mass energies \sviii{} (left) and
           \sxiii{} (right). The error bars show the statistical uncertainties
           only.}%
  \label{fig:results_pt_ic_13tev}
\end{figure}

\begin{figure}[h]
  \centering
  \includegraphics[width=\mpw]{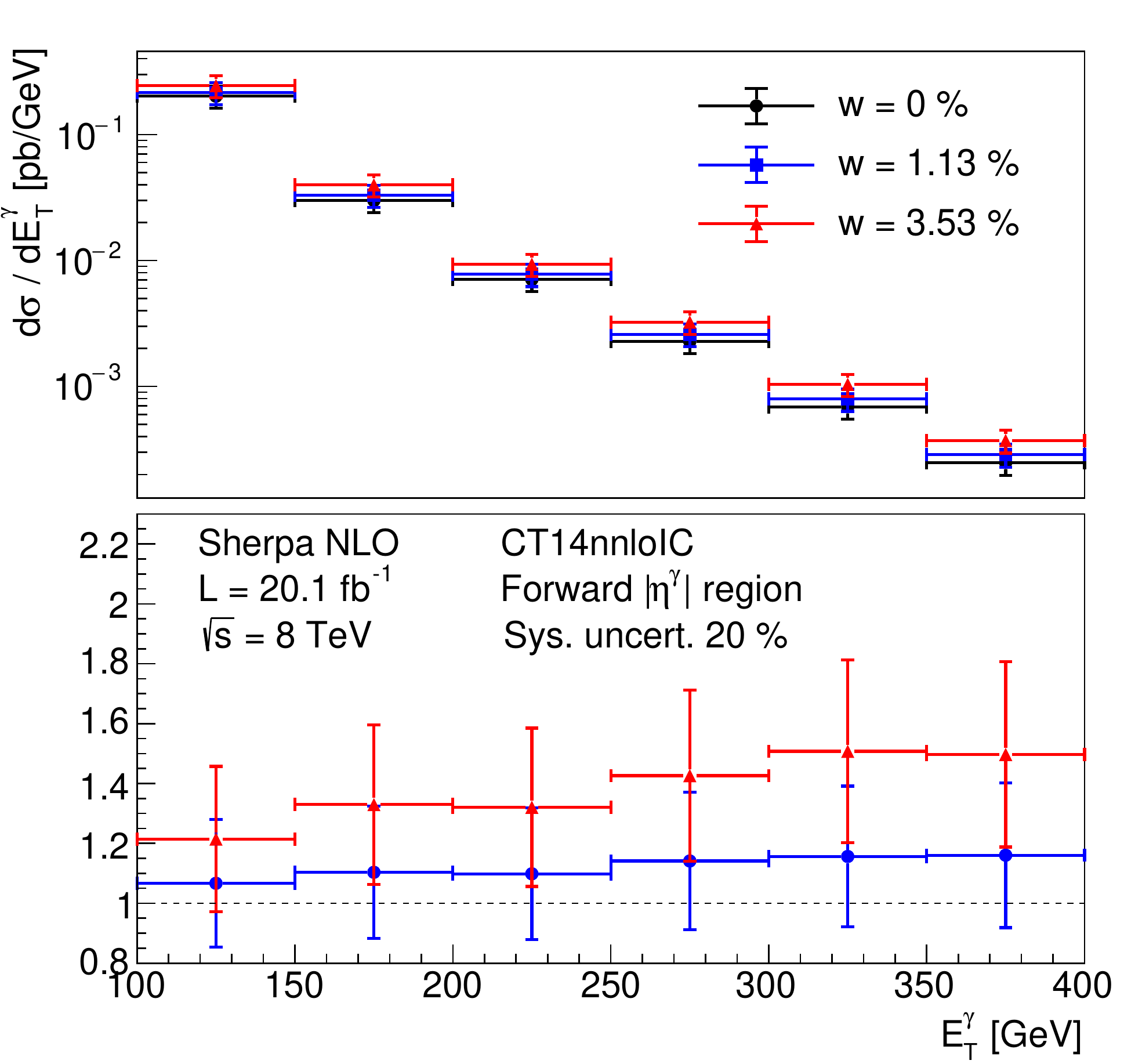}
  \includegraphics[width=\mpw]{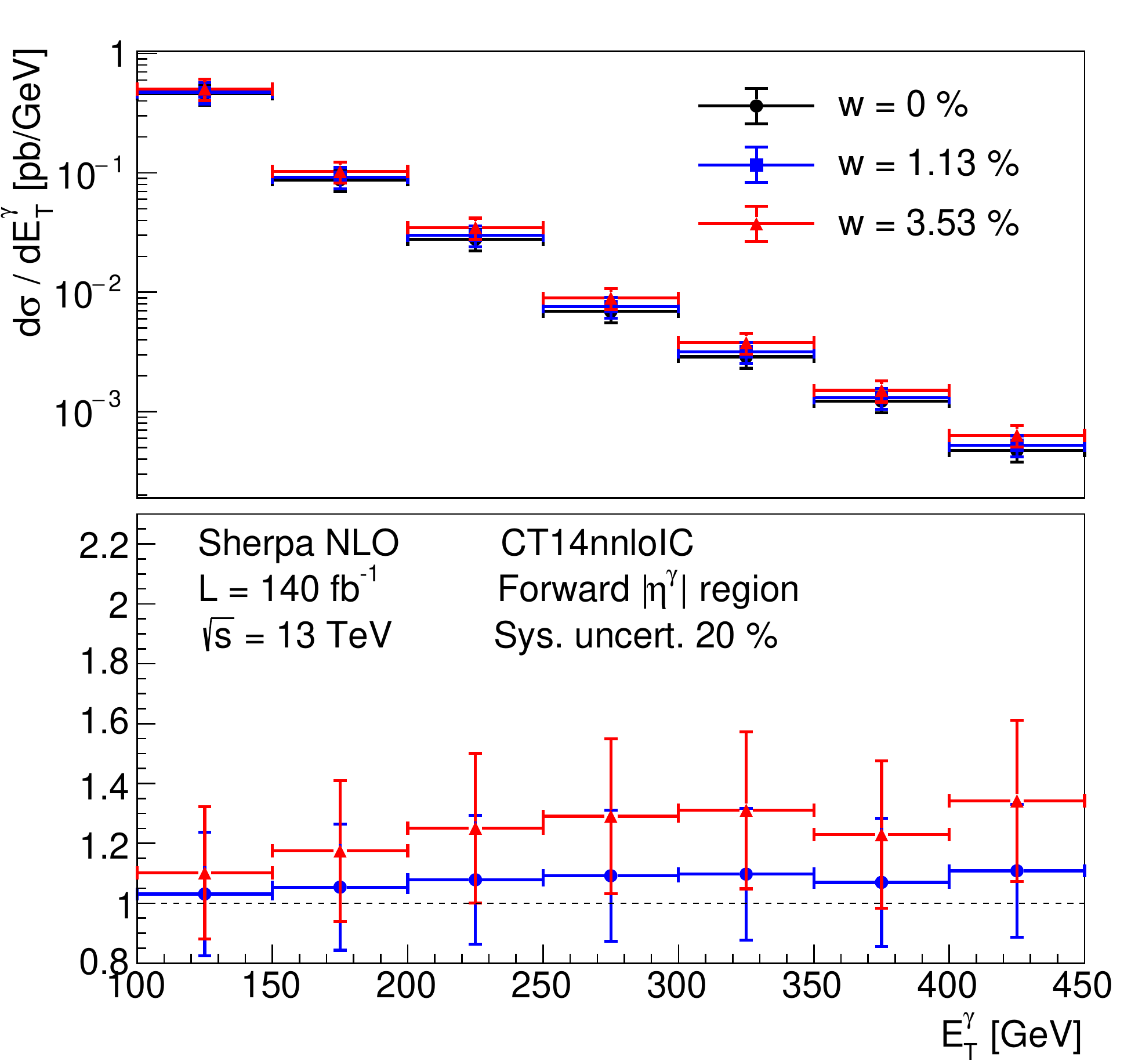}
  \caption{Differential cross-sections in \ety{} of the \ycjet{}
           production at two center-of-mass energies \sviii{} (left) and
           \sxiii{} (right). The error bars show the statistical and optimistic
           expected systematic uncertainties summed in quadrature.}%
  \label{fig:results_pt_ic_13tev_sys}
\end{figure}

The effect of \ic{} on the \ety{} spectra of \ycjet{} at \sxiii{} is shown in
Fig.~\ref{fig:results_pt_ic_13tev}. The figure shows a comparison of the \ic{}
effect at \sviii{} versus \sxiii{}. One can see that in the case of \sxiii{} the
\ic{} effect is smaller, this is due to two reasons. First, the peak of the
\ic{} contribution will be shifted towards higher \ety{}, this comes from the
relationship 
between \ety{} and \xf{}. Second, it is
suspected, that the contribution to the prompt photon spectra from diagrams
which do not propagate \ic{} grows quicker than from the ones which propagate
it.
Fig.~\ref{fig:results_pt_ic_13tev_sys} shows the effect of \ic{} in comparison
to an optimistic systematic uncertainty prediction of a possible future
measurement at \sxiii.

\begin{figure}[h]
  \centering
  \includegraphics[width=\mpw]{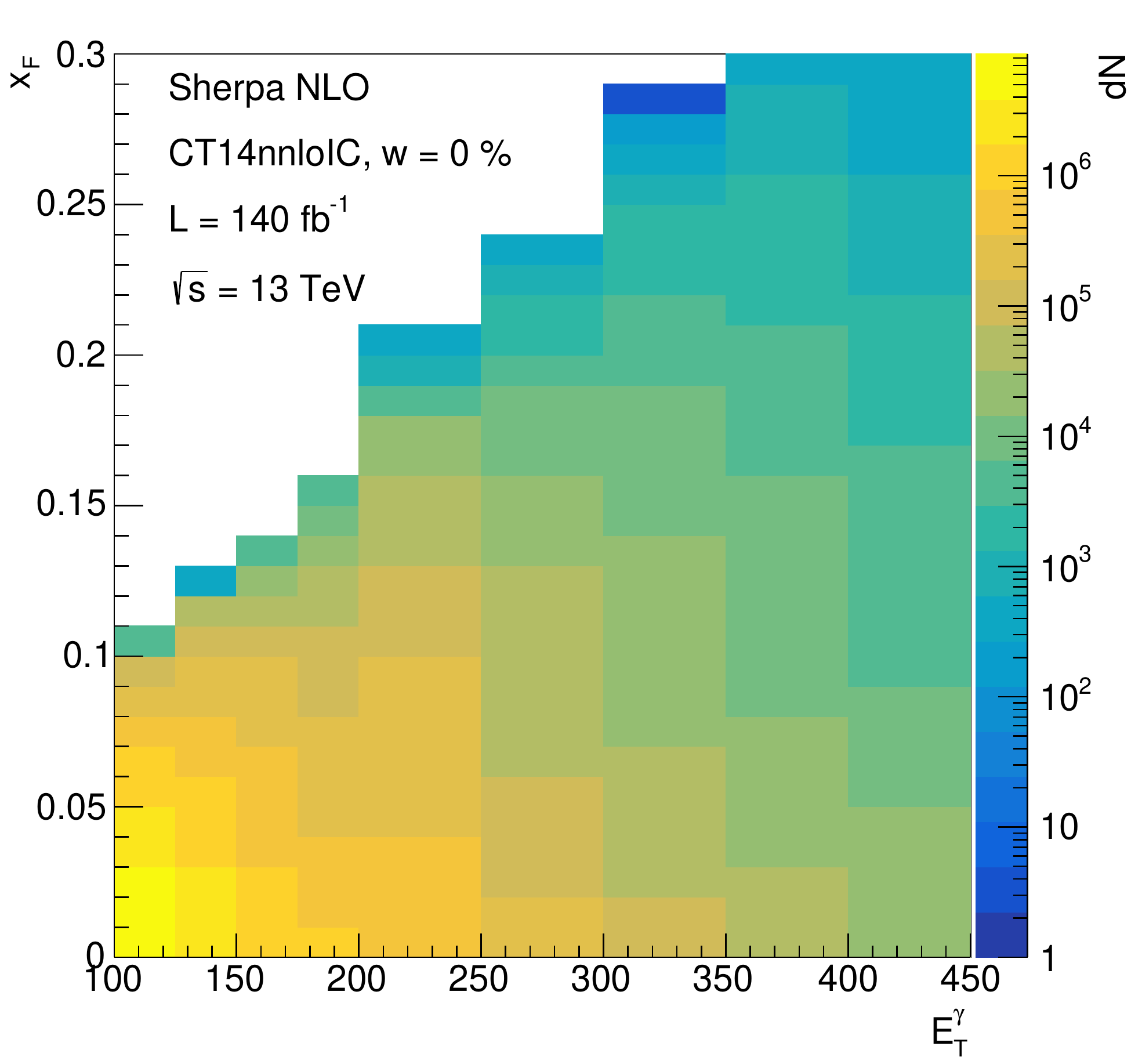}
  \includegraphics[width=\mpw]{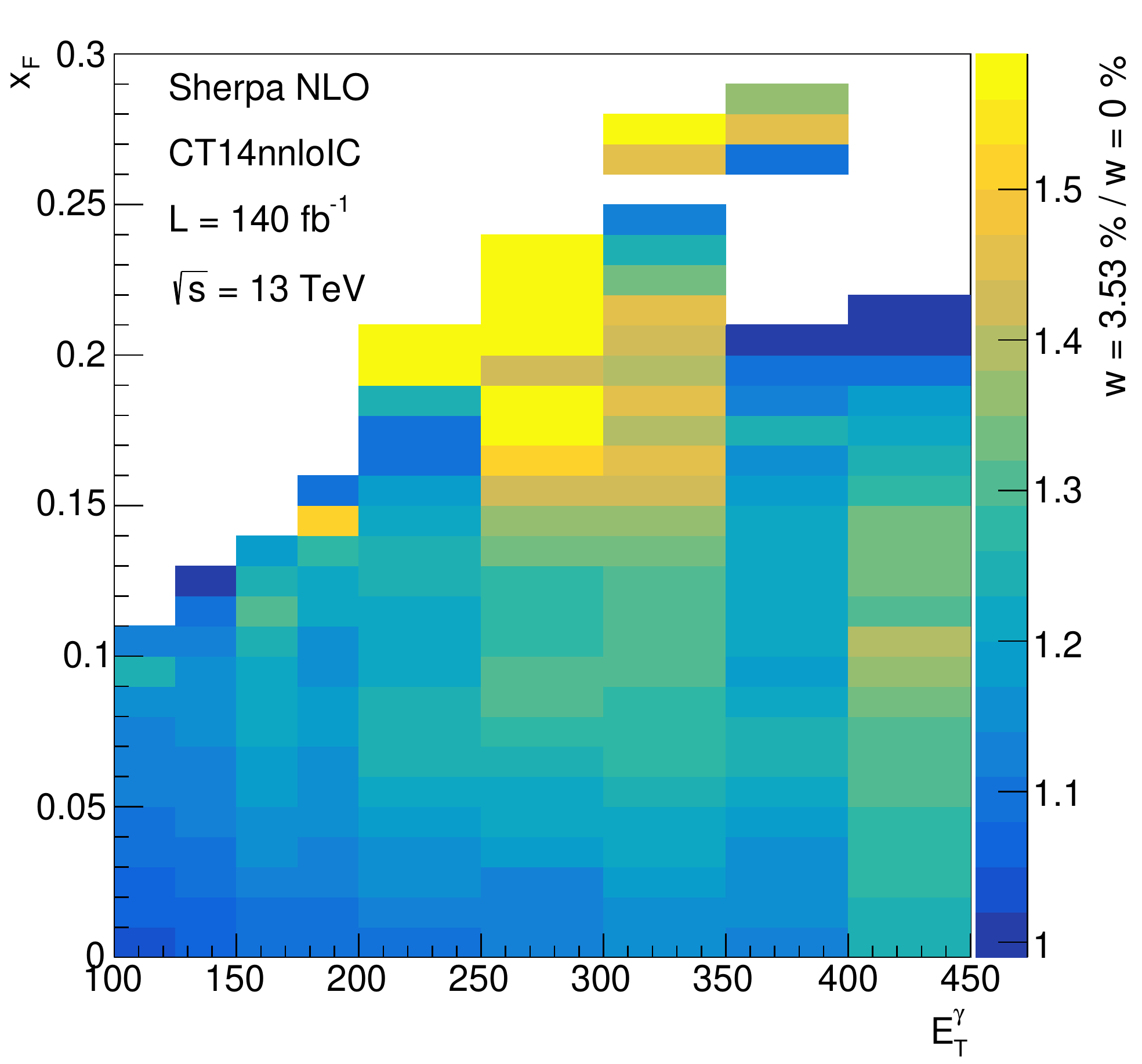}
  \caption{Number of events in \ety{} spectra of the \ycjet{}
           production at center-of-mass energy \sxiii{} versus the cut on photon
           momentum fraction \xf{} (left). Ratio of the spectra with \ic{}
           contribution $\w = 3.5$\% to no \ic{} contribution (right).}%
  \label{fig:results_pt_xf_13tev}
\end{figure}

The effect of possible \ic{} presence can be enhanced by introducing a cut on
photon momentum fraction \xf{}, since the peak of \ic{} contribution is located
at around 0.2--0.3. The Fig.~\ref{fig:results_pt_xf_13tev} shows the expected
number of events in \ety{} spectra at \sxiii{} with integrated luminosity
140~fb$^{-1}$ and the ratio of \ic{} ($\w = 3.53$\%) to no \ic{} spectrum, when
the cut on \xf{} is gradually introduced starting at 0. One can see, that the
ratio has comparable size to the \sviii{} case in
Fig.~\ref{fig:results_pt_ic_13tev} at around the cut of $\xf = 0.15$. The empty
cells in upper left corner are the result of the gradual cut on \xf{} and the
detector cut on $\etay < 2.37$. The expected \ic{} upper limit from a possible
future \sxiii{} measurement could be improved only if a substantial reduction of
systematic uncertainties of the measurement takes place.

\clearpage

\subsection{Summary}

A first estimate of the intrinsic charm probability in the proton has been
carried out utilizing recent ATLAS data on the prompt photon production
accompanied by the $c$-jet at $\sqrt{s} = 8$~\tev~\cite{Aaboud:2017skj}.  We
estimate the upper limit of the \ic{} probability in proton about 1.93\%.  In order
to obtain more precise results on the intrinsic charm contribution one needs
additional data and at the same time reduced systematic uncertainties  which
come primarily from $c$-jet tagging. In particular, measurements of cross
sections of $\gamma + c$ and $\gamma + b$ production in $pp$-collisions at
$\sqrt{s} = 13$~\tev{} at high transverse momentum with high
statistics~\cite{Lipatov:2016feu} will be very useful since the ratio of photon
+ charm to photon + bottom cross-sections is very sensitive to the \ic{}
signal~\cite{Lipatov:2016feu, Brodsky:2016fyh}. The ratio, when \ety{} grows,
decreases in the absence of the \ic{} contribution and stays flat or increases when
the \ic{} contribution is included. Furthermore, measurements of $Z/W+c/b$
production in $pp$ collision at 13~\tev{} could also give additional significant
information on the intrinsic charm contribution~\cite{Beauchemin:2014rya,
Lipatov:2016feu,Brodsky:2016fyh,Lipatov:2018oxm}. Our study shows that the most
important source of theoretical uncertainty on $w_{cc}$, from the theory point of
view, is the dependence on the renormalization and factorization scales. This
can be reduced by the application of the Principle of Conformality (PMC), which
produces scheme-independent results, as well the calculation of the NNLO pQCD
contributions. Data at different energies at the LHC which checks scaling
predictions and future improvements in the accuracy of flavor tagging will be
important. These advances, together with a larger data sample (more than
100~fb$^{-1}$) at 13~\tev, should provide definitive information from the LHC on
the contribution of the non-perturbative intrinsic heavy quark contributions to
the fundamental structure of the proton.

%
%
\section{Hard processes of vector bosons accompanied by heavy flavor jets}%
\label{sec:z_c_jet}

\subsection{Theoretical approaches to associated \ZHF{} production}

To calculate the total and differential cross-sections of associated \ZHF{}
production within the combined QCD approach, we strictly follow the 
scheme described earlier in Ref.~\cite{Baranov:2017tig}. In this scheme, the 
leading contribution comes from the ${\cal O}(\alpha \alpha_s^2)$
off-shell gluon-gluon fusion subprocess
$g^{*} + g^{*} \to Z + Q + \bar{Q}$ (where $Q$ denotes the heavy 
quark), calculated in the \kt-factorization approach.
The latter has certain technical advantages in 
the ease of including higher-order radiative corrections in the
form of the TMD parton distributions (see~\cite{Andersson:2002cf,
Andersen:2003xj, Andersen:2006pg} 
for more information). To extend the consideration 
to the whole kinematic range, several subprocesses involving 
initial state quarks, namely flavor excitation 
$q + Q\to Z + Q + q$, quark-antiquark annihilation $q + \bar{q}\to Z + Q + 
\bar Q$ and quark-gluon scattering $q + g\to Z + q + Q \bar{Q}$, are taken into 
account using the collinear QCD factorization (in the tree-level 
approximation). 
The \ic{} contribution is estimated using the 
${\cal O}(\alpha \alpha_s)$ QCD Compton scattering $c + g^* \to Z + c$,
where the gluons are kept off-shell but the incoming non-perturbative 
intrinsic charm quarks are treated 
as on-shell ones\footnote{The perturbative charm contribution is already taken 
into account in the off-shell gluon-gluon fusion subprocess.}.
Thus we rely on a combination of two 
techniques, with each of them being used for the kinematics
where it is more 
suitable\footnote{An essential point of 
consideration~\cite{Baranov:2017tig} is using a numerical solution of the CCFM 
evolution equation~\cite{Ciafaloni:1987ur,Catani:1989yc,Catani:1989sg} to
derive the  TMD  gluon density in a proton. The 
latter smoothly interpolates between the small-$x$ BFKL gluon dynamics and 
high-\x{} DGLAP dynamics. Following~\cite{Baranov:2017tig}, below 
we adopt the latest JH'2013 parametrization~\cite{Hautmann:2013tba}, 
adopting the JH’2013 set 2 gluon as the default
choice.}
(off-shell gluon-gluon fusion subprocesses at small $x$
and quark-induced subprocesses at large $x$ values).
More details of the above calculations can be found 
in Ref.~\cite{Baranov:2017tig}.

In contrast to earlier studies~\cite{Beauchemin:2014rya,Lipatov:2016feu} of
\ZHF{} production within the MCFM routine (that performs calculation in the
fixed order of \pqcd), in the present paper the \sherpa{}
2.2.1~\cite{Gleisberg:2008ta} MC generator is applied. It uses matrix elements
that are provided by the built-in generators Amegic++~\cite{Krauss:2001iv} and
COMIX~\cite{Gleisberg:2008fv}; OPENLOOPS~\cite{Cascioli:2011va} is used to
introduce additional loop contributions into the \nlo{} calculations. We use
matrix elements calculated at the next-to-leading order (NLO) for up to 2 final
partons and at the leading-order (LO) for up to 4 partons. They 
are merged with the \sherpa~parton showering~\cite{Schumann:2007mg} following 
the 
ME+PS@NLO prescription~\cite{Hoeche:2012yf}. This is different from the study of 
$Z+c$ production carried out in Ref.~\cite{Hou:2017khm} where the matrix element was 
calculated in the LO and merged following the ME+PS@LO 
method~\cite{Hoeche:2009rj}. The 
latter approach was also used in this study as a cross-check, with the LO 
matrix 
element allowing for up to 4 final partons.
In both approaches, the five-flavor scheme (5FS) is used where $c$ and $b$ 
quarks 
are considered as massless particles in the matrix element and massive in both 
the initial and final state parton showers. 
\sherpa~can also model the full
chain of hadronization and unstable particle decays for an accurate comparison 
with experimental measurements of
\HF~jets.

\subsection{Comparison with the LHC data at $\sqrt s = 7$ and 8~TeV}

In this section we present comparisons of our calculations for \ZHF{} production
made with the \sherpa{} generator and within the combined QCD approach to the
LHC Run~1 data, in order to verify the applicability of these approaches for
further predictions. Following~\cite{Bednyakov:2013zta,Beauchemin:2014rya,
Lipatov:2016feu,Brodsky:2016fyh,Ball:2016neh}, we mainly concentrate on the
transverse momentum distributions of $Z$ bosons and/or \HF~jets, where the \ic{}
effects are expected to appear\footnote{Recent ATLAS and CMS experimental data
on associated $Z + b$ production taken at $\sqrt{s} = 7$~TeV as functions of
other kinematical variables within the framework of the combined QCD approach
are considered in Ref.~\cite{Baranov:2017tig}.}.

The first comparison is performed for associated $Z + b$
production measured by the ATLAS Collaboration~\cite{Aad:2014dvb} at
$\sqrt{s}=7$~TeV. According to~\cite{Aad:2014dvb}, the
following selection criteria were applied to generated events. Two leptons
originating from the $Z$ boson decay are 
required to have an invariant mass 
$76~\mathrm{GeV} < m_{\ell\ell} <106$~GeV with a minimum transverse momentum of 
each lepton $\pt[^\ell] > 20$~GeV and rapidity $|y^{\ell}|<2.5$.
In \sherpa~generated events, jets are built using all stable particles
excluding 
the lepton pair from the $Z$ boson decay with the anti-$k_t$ algorithm with a
size parameter $R = 0.4$. They are required to have 
a rapidity $|y^{\jet}| < 2.4$ 
and minimum transverse momentum $\pt[^\jet] > 20$~GeV. Each jet is also required 
to be separated from any of the two leptons by $\Delta{R}_{\jet,\ell} > 0.5$. 
Jets are identified as $b$-jets, if there is a weakly decaying 
$b$ hadron with a transverse momentum $\pt[^b] > 5$~GeV within a cone $\Delta{R} 
= 0.3$ around the jet direction.
The same kinematic requirements are applied to final state
$b$ quarks 
(treated as $b$-jets at a parton level) when using the
combined QCD approach. \sherpa~results were obtained 
within the ME+PS@NLO model. In both approaches
the CTEQ66 PDF 
set~\cite{Nadolsky:2008zw} was used.

\begin{figure}
  \centering
  \includegraphics[width=\mpw]{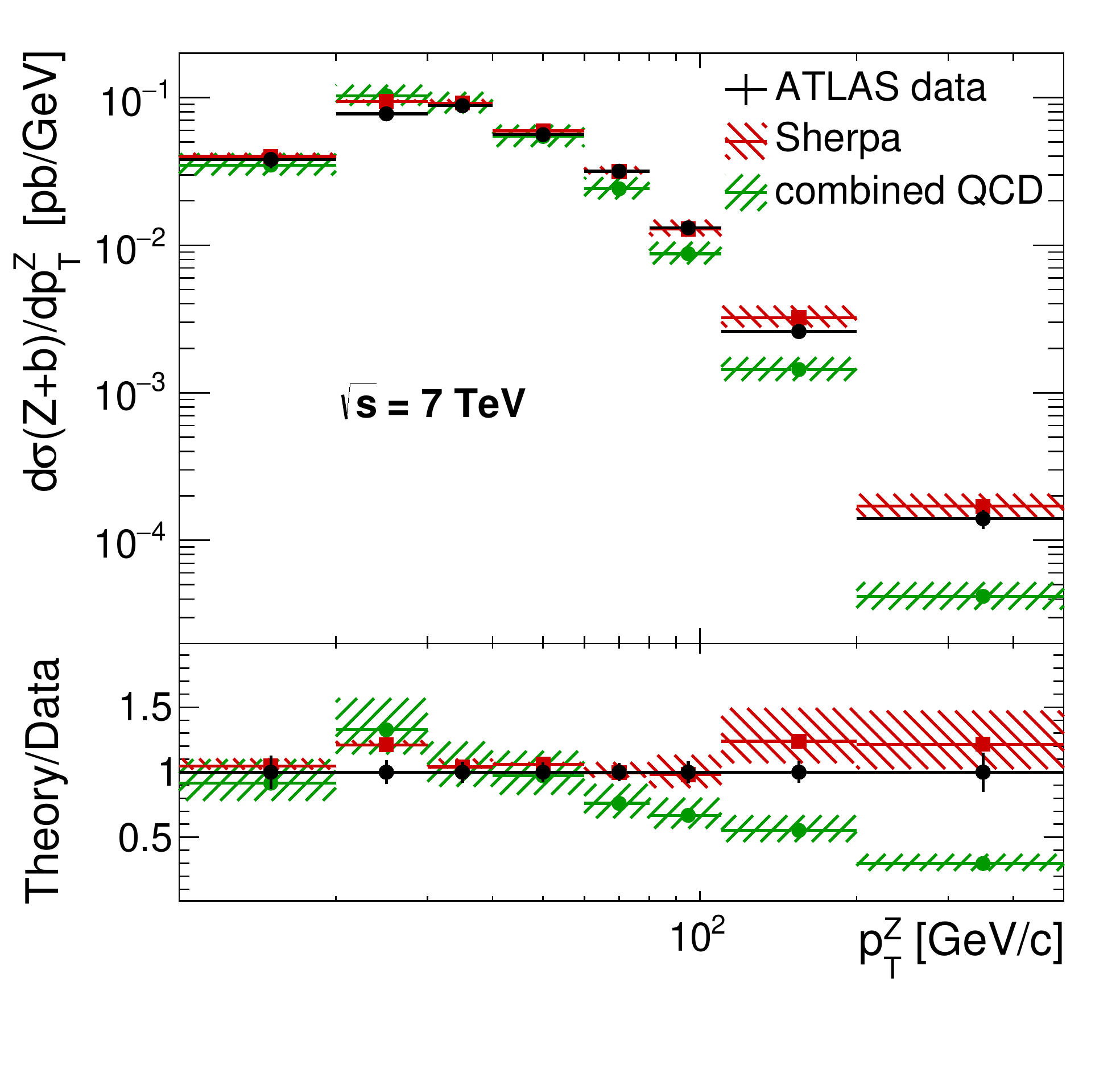}
  \caption{Cross section of $Z+b$-jet production as a function of the $Z$ boson
           transverse momentum in the full rapidity region $|y^{Z}| < 2.5$ at
           $\sqrt{s} = 7$~TeV. The main panel shows the ATLAS measurement
           result~\cite{Aad:2014dvb} compared to \sherpa{} calculations and to
           combined QCD calculations. The uncertainty bands represent
           uncertainties in the QCD scale. The bottom panel shows the ratio of
           calculations to data.}%
  \label{fig:Zb-7TeV}
\end{figure}

In Fig.~\ref{fig:Zb-7TeV} the associated $Z + b$-jet production cross 
section (for events with at least one $b$-jet) calculated as a function of
the $Z$ boson transverse momentum $\pt[^Z]$ is presented in
comparison with the ATLAS data~\cite{Aad:2014dvb}.
Here and below central values, marked by horizontal lines, correspond to 
the default choice of factorization and renormalization
scales $\mu_R = \mu_F = m_\mathrm{T}$, where $m_\mathrm{T}$ is the $Z$ boson
transverse mass. Theoretical uncertainties of our calculations correspond to
the maximum deviation between the nominal spectrum and those obtained by
the usual factor 2 variations of renormalization and
factorization scales. 

One can see that the \sherpa~results are
in perfect agreement with the ATLAS data within the scale
uncertainties in the whole $\pt[^Z]$ range. In the combined QCD approach, we
observe some underestimation of the data at high $\pt[^Z]$ and a slight
overestimation at low transverse momenta. The latter can be attributed to the
TMD gluon density used in the calculations, because the region
$\pt[^Z] < 100$~GeV is fully dominated by the off-shell
gluon-gluon fusion subprocess~\cite{Baranov:2017tig}.
However, the results obtained within both approaches under consideration in 
this region are rather close to each other.
A noticeable deviation of the combined QCD calculations from the data at 
large $\pt[^Z]$ is explained by the absence of
the effects of parton showers, hadronization and additional contributions of 
NLO diagrams, including loop ones, in these calculations.
Such contributions, which are taken into account by \sherpa, considerably
improve the description of data. The influence of the parton showers and of
higher-order \pqcd{} corrections is investigated in detail in the next Section.
It is important to note that our results obtained with \sherpa{} are
in good agreement with the results obtained within a similar
approach~\cite{Krauss:2016orf}.

Now, we turn to the associated $Z + c$-jet production measured by the CMS
Collaboration at $\sqrt{s} = 8$~TeV~\cite{Sirunyan:2017pob}. The following
selection criteria are applied to generated events for this comparison. Two
leptons originating from a $Z$ boson decay must have an invariant mass
$71~\mathrm{GeV} < m_{\ell\ell} <111$~GeV, a minimum transverse momentum of
$\pt[^\ell] > 20$~GeV and rapidity $|y^{\ell}|<2.1$. Jets built with the
anti-$k_t$ algorithm with a size parameter $R=0.5$ are required to have
$\pt[^\jet] > 25$~GeV and $|y^{\jet}| < 2.5$ and to be separated from the
leptons by $\Delta{R}_{\jet,\ell} > 0.5$. Similar $b$ and $c$ flavor
identification criteria to those described above are used.

\begin{figure}
  \centering
  \includegraphics[width=\mpw]{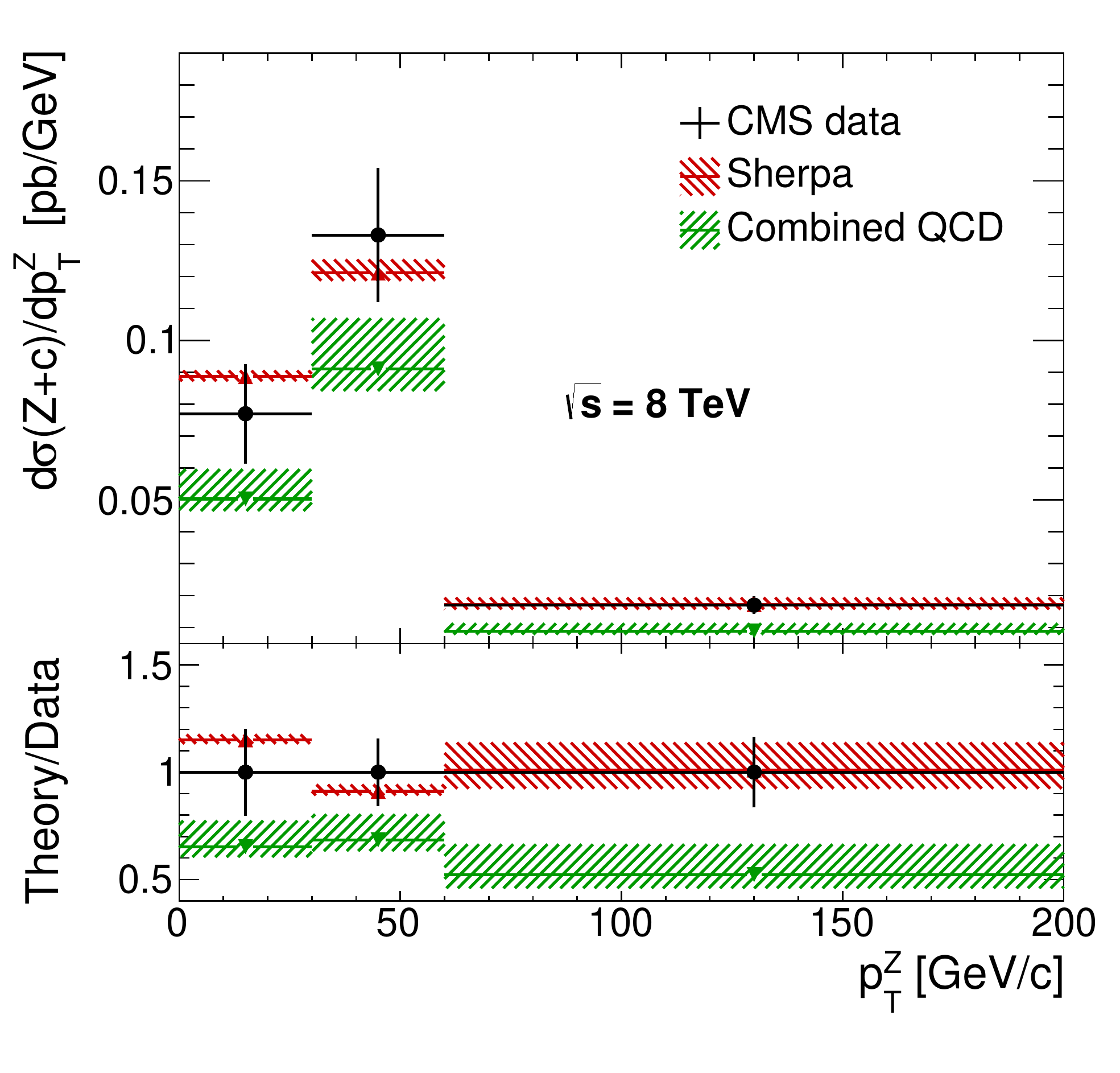}
  \includegraphics[width=\mpw]{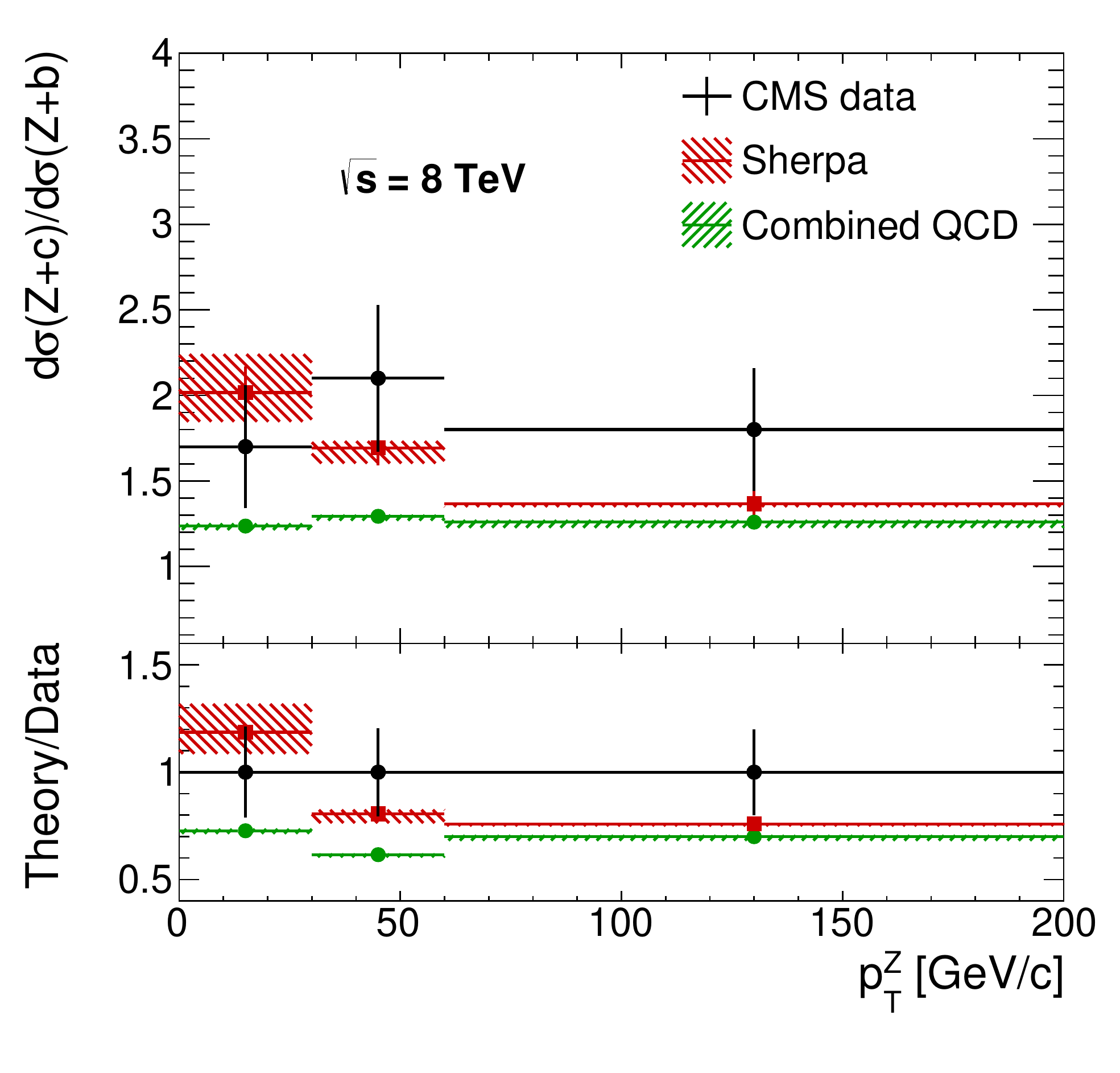}
  \includegraphics[width=\mpw]{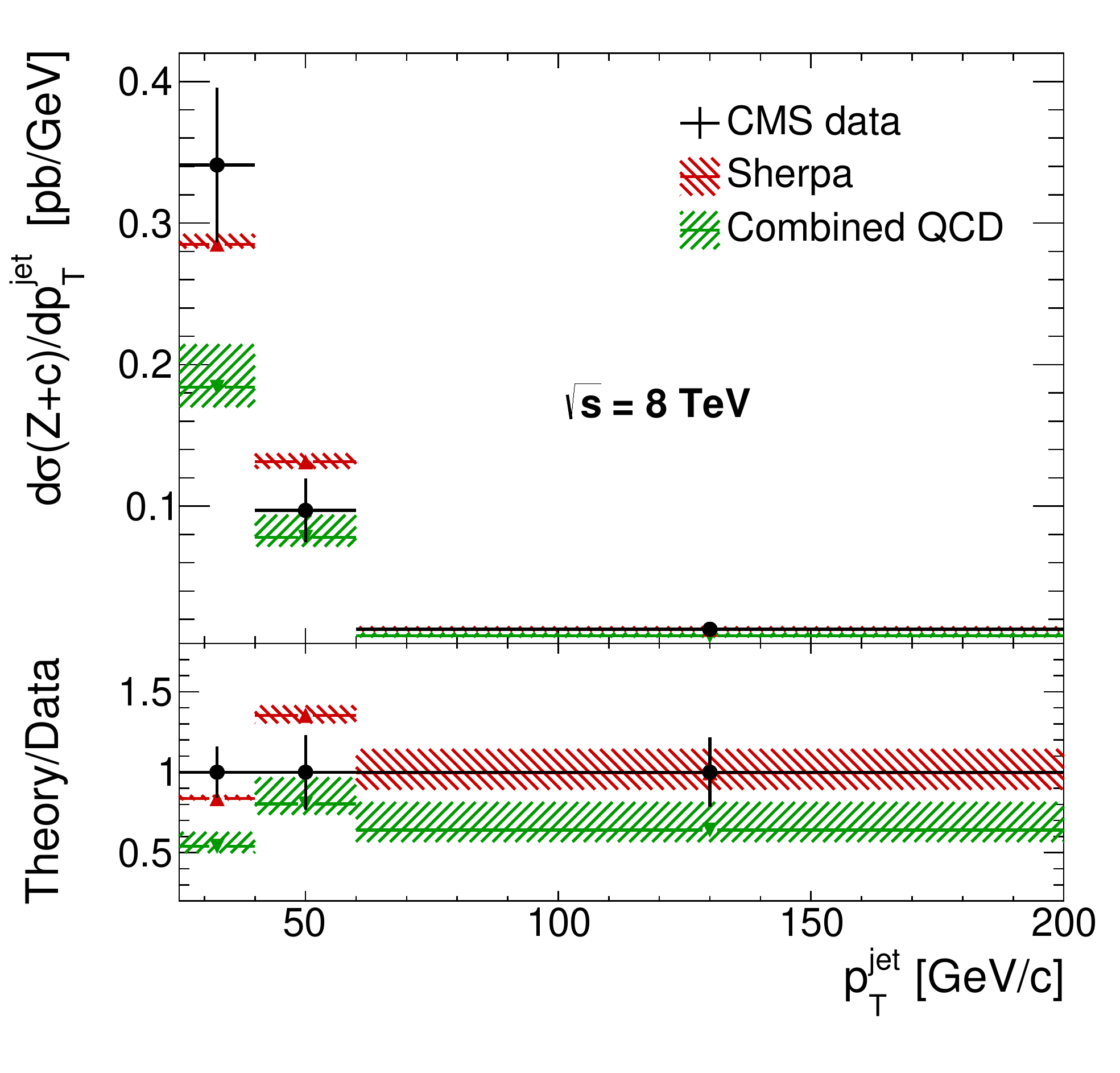}
  \includegraphics[width=\mpw]{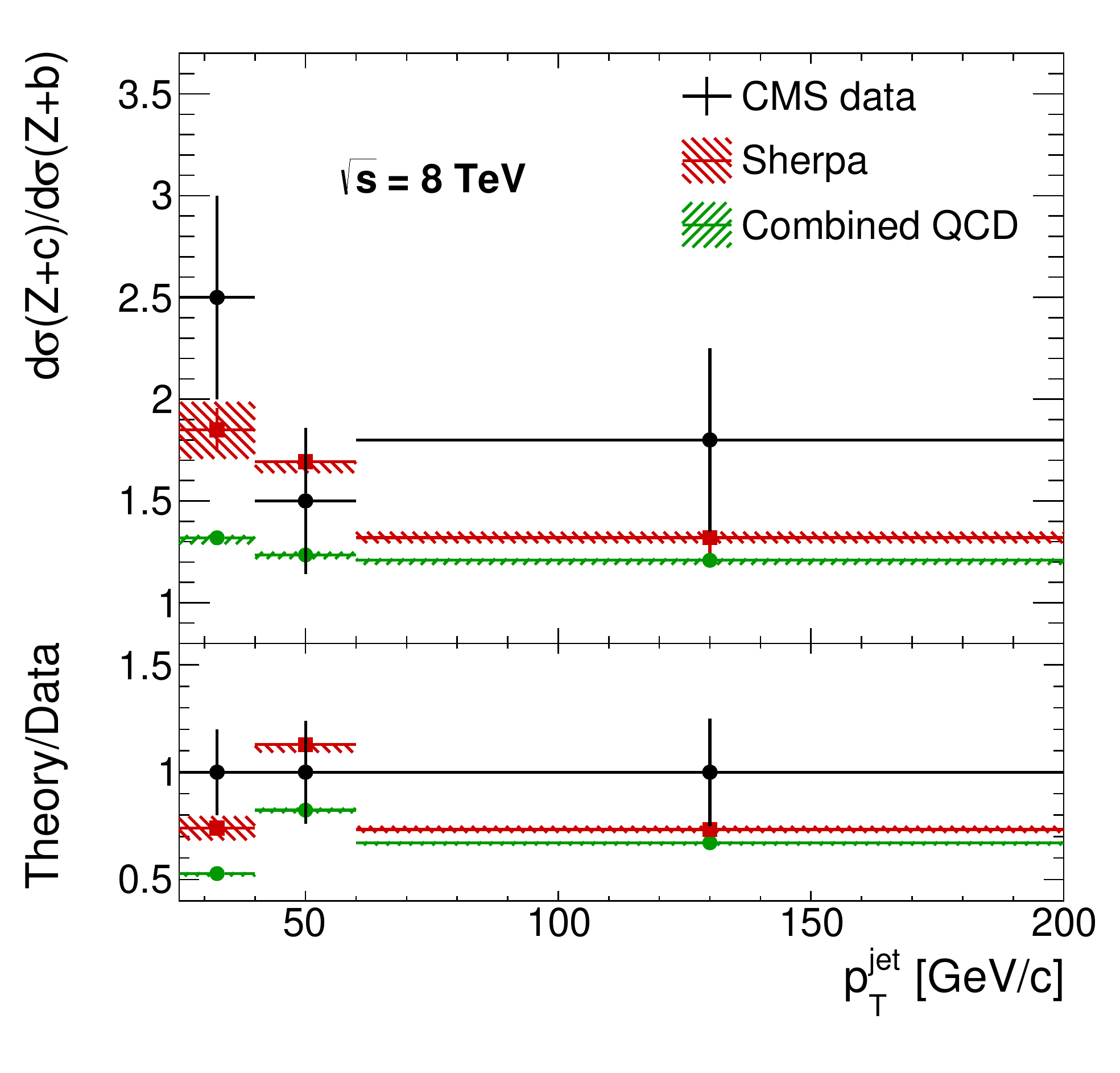}
  \caption{Cross section of $Z+c$-jet production (left) and the ratio of cross
           sections of $Z+c$-jet and $Z+b$-jet production (right) as a function
           of the $Z$ boson (top) and HF jet (bottom) transverse momenta in the
           full rapidity region $|y^{Z}| < 2.5$ at $\sqrt{s} = 8$~TeV. The main
           panels show the CMS measurement result~\cite{Sirunyan:2017pob}
           compared to results of \sherpa{} and the combined QCD calculations.
           The uncertainty bands represent the uncertainties in the QCD scale.
           The bottom panels show the ratio of calculations to data.}%
  \label{fig:Zc-8TeV}
\end{figure}

In Fig.~\ref{fig:Zc-8TeV} our results for the differential cross 
sections of associated $Z + c$-jet production calculated as functions of the $Z$ boson and $c$-jet 
transverse momenta are shown in comparison 
with the CMS data~\cite{Sirunyan:2017pob}. 
A comparison with the measured ratio of the cross-sections $\sigma{(Z + 
c)}/\sigma{(Z + b)}$ is also presented.
We find that the particle-level \sherpa{} calculations agree well with the data.
The parton-level combined QCD calculations also
describe the CMS data within the theoretical and experimental uncertainties 
(except at low $\pt[^\mathrm{c}] < 40$~GeV),
although they tend to underestimate the \sherpa{} results.
As in the case of associated $Z + b$-jet production, we
attribute the latter to the parton showering effects and
additional NLO contributions, missing in the combined QCD calculations (to be
precise, mainly in the tree-level quark-induced subprocesses, since the
off-shell gluon-gluon fusion only gives a negligible
contribution at large transverse momenta). Note that the scale uncertainties
of our calculations partially cancel out when considering the
$\sigma{(Z + c)}/\sigma{(Z + b)}$ ratio\footnote{This ratio, being considered
in the forward rapidity region $1.5 < y^Z < 2.5$, is sensitive to the \ic{}
content of a proton~\cite{Lipatov:2016feu}. However, we checked that in the
kinematical region probed by the CMS experiment~\cite{Sirunyan:2017pob} this
\ic{} dependence is negligible.} (see Fig.~\ref{fig:Zc-8TeV}, right plots).

One can see that a better description of the CMS data is achieved by employing
the \sherpa{} generator, therefore we consider \sherpa{} calculations to be more
reliable. Thus, we mainly concentrate on them when investigating the possible
effects from \ic{} in the LHC experiments below.

\subsection{$Z$+HF spectra for $\sqrt{s} = 13$~TeV and prediction for the IC
            contribution}
The purpose of the calculation of \ZHF~differential cross
sections in this paper is to investigate the
effect of an \ic{} signal on the observables, which can be
measured at the LHC by general purpose detectors at
$\sqrt{s} = 13$~TeV. As it was mentioned above, a sensitivity
to the \ic{} at ATLAS and CMS experiments on $Z+c$-jet production
can be achieved in the forward rapidity region $1.5 < |y^{Z}| < 2.5$ and
$\pt[^Z] > 50$~GeV~\cite{Beauchemin:2014rya, Lipatov:2016feu}. In this
kinematical region the shape of the 
$\sigma{(Z + c)}/\sigma{(Z + b)}$ ratio is sensitive to  
effects of \ic{} and is less affected by scale
uncertainties than those of the transverse momentum spectra.
This fact provides an opportunity to measure
the \ic{} contribution.

In \sherpa~, predictions for \ZHF~production are calculated
within the ME+PS@NLO model using the CT14nnlo \pdf{} set~\cite{Hou:2017khm}
containing PDFs with \ic{} probabilities $w_\mathrm{IC} = 0$,~$1$ and
$2$\%~\cite{Hou:2017khm}.
The following selection criteria are used in this analysis.
Two leptons from the $Z$ boson decay are required to have
a mass
$76~\mathrm{GeV} < m_{\ell\ell} <106$~GeV, transverse momentum 
$\pt[^\ell] > 28$~GeV and rapidity $|y^{\ell}|<2.5$. Jets 
are reconstructed from  all stable particles, excluding the  
leptons,  with the anti-$k_t$ algorithm with parameters $R = 0.4$ and
are required to have $|y^{\jet}| < 2.5$ and
$\pt[^{\jet}] > 20$~GeV, $\Delta{R}_{\jet,\ell} > 0.4$. The
identification of heavy flavor jets is performed as
follows. If there is a weakly decaying $b$ hadron 
with $\pt[^\mathrm{b}] >5$~GeV within a cone of $\Delta{R} = 0.5$ around the jet 
direction, the jet is identified as a $b$-jet. If it is not
identified as such, the same criteria are applied for $c$-hadrons, and the jet
is identified as a $c$-jet, if one is found.

\begin{figure}
  \centering
  \includegraphics[width=\mpw]{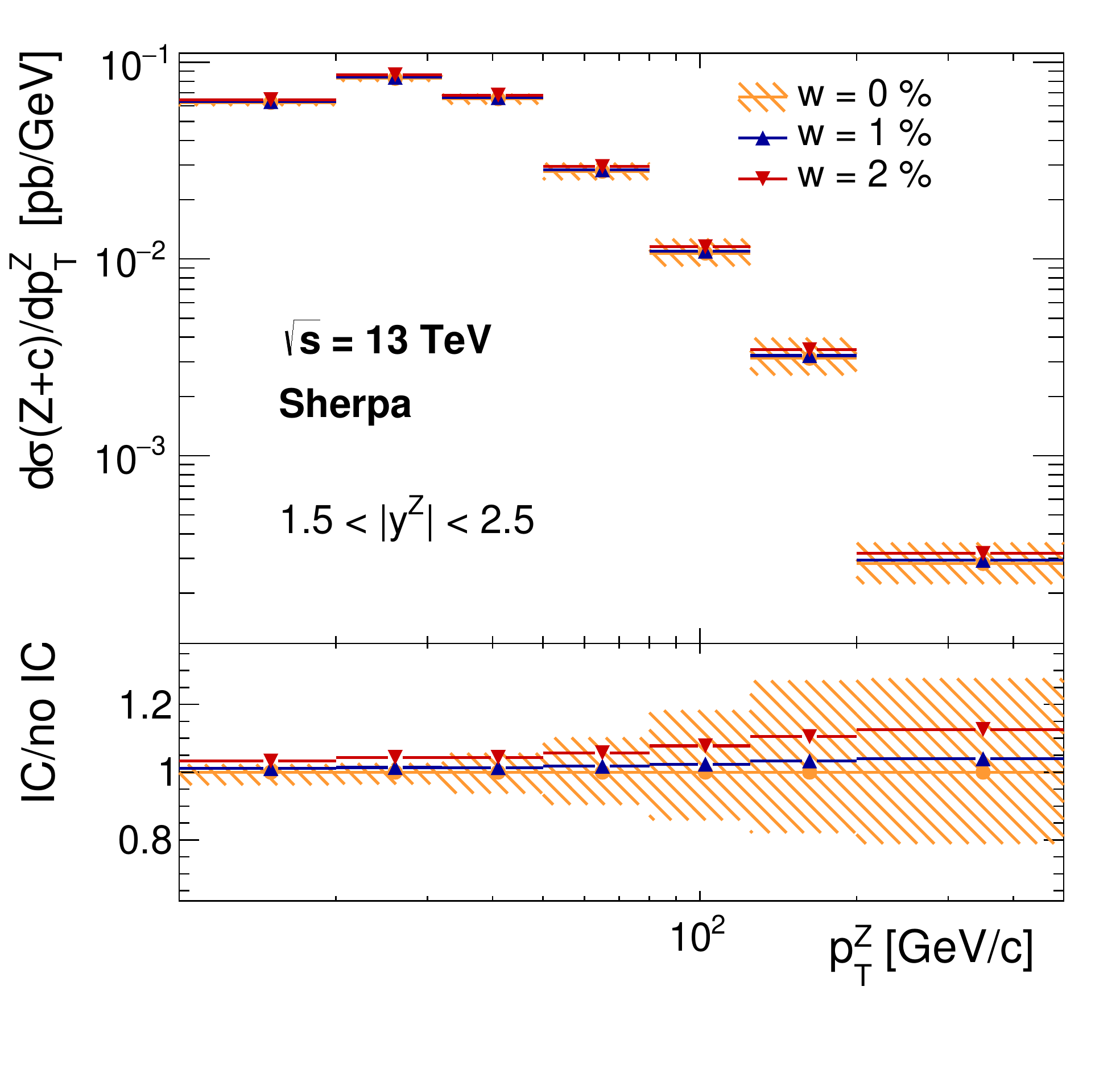}
  \includegraphics[width=\mpw]{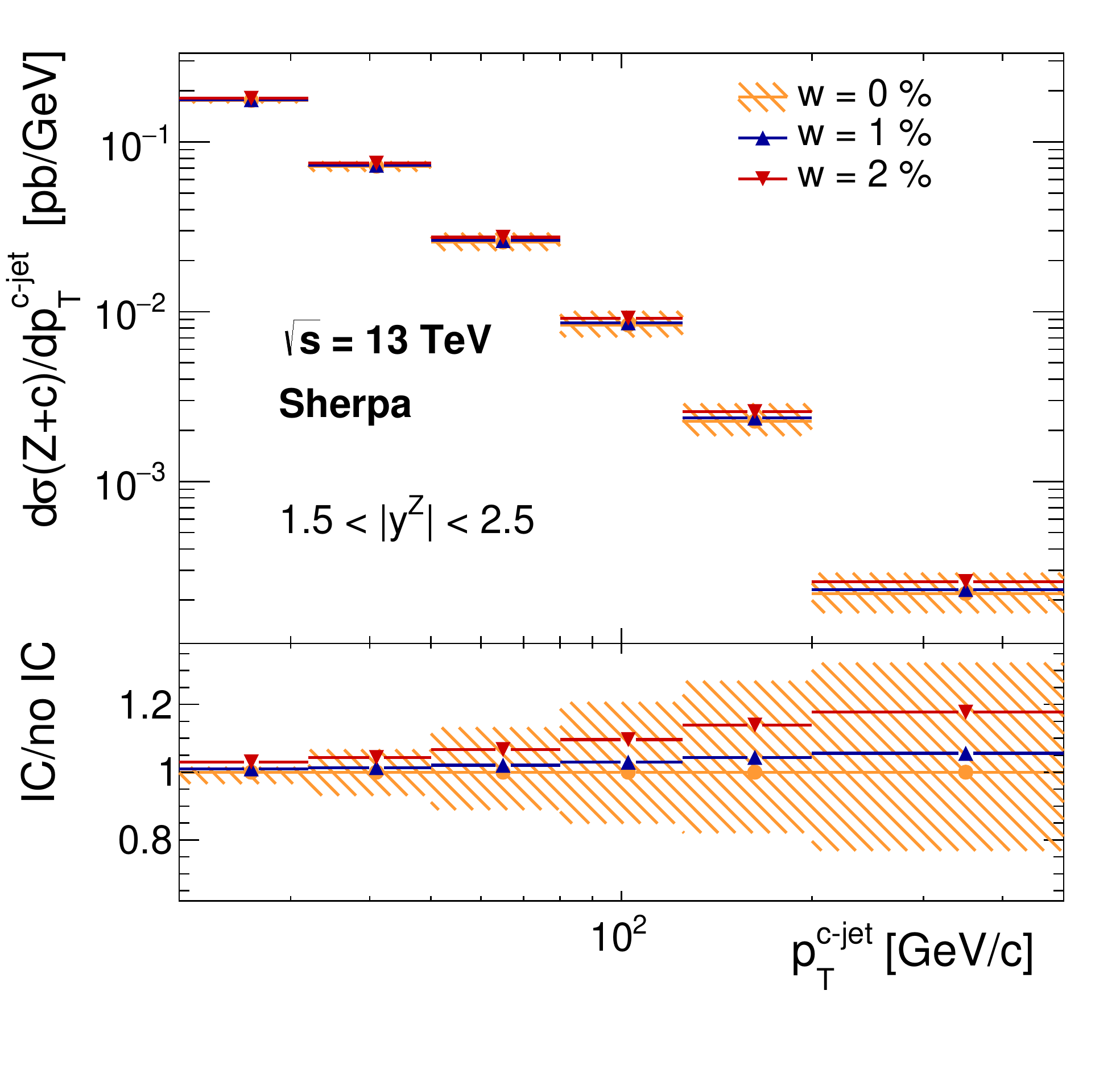}
  \caption{Predictions for the cross-section of $Z+c$-jet production as a
           function of the $Z$ boson (left) and $c$-jet (right) transverse
           momentum in the forward rapidity region $1.5 < |y^{Z}| < 2.5$ at
           $\sqrt{s} = 13$~TeV. The predictions are made with the \sherpa{}
           generator using the CT14nnlo PDF with different values for the IC
           contribution $w$. The bottom panels show the ratio of predictions for
           non-zero values of $w$ to those for $w=0$\%. The uncertainty bands
           represent the uncertainties in the QCD scale (shown only for $w=0$\%
           predictions).}%
  \label{fig:Zc_Sherpa_predictions}
\end{figure}

\begin{figure}
  \centering
  \includegraphics[width=\mpw]{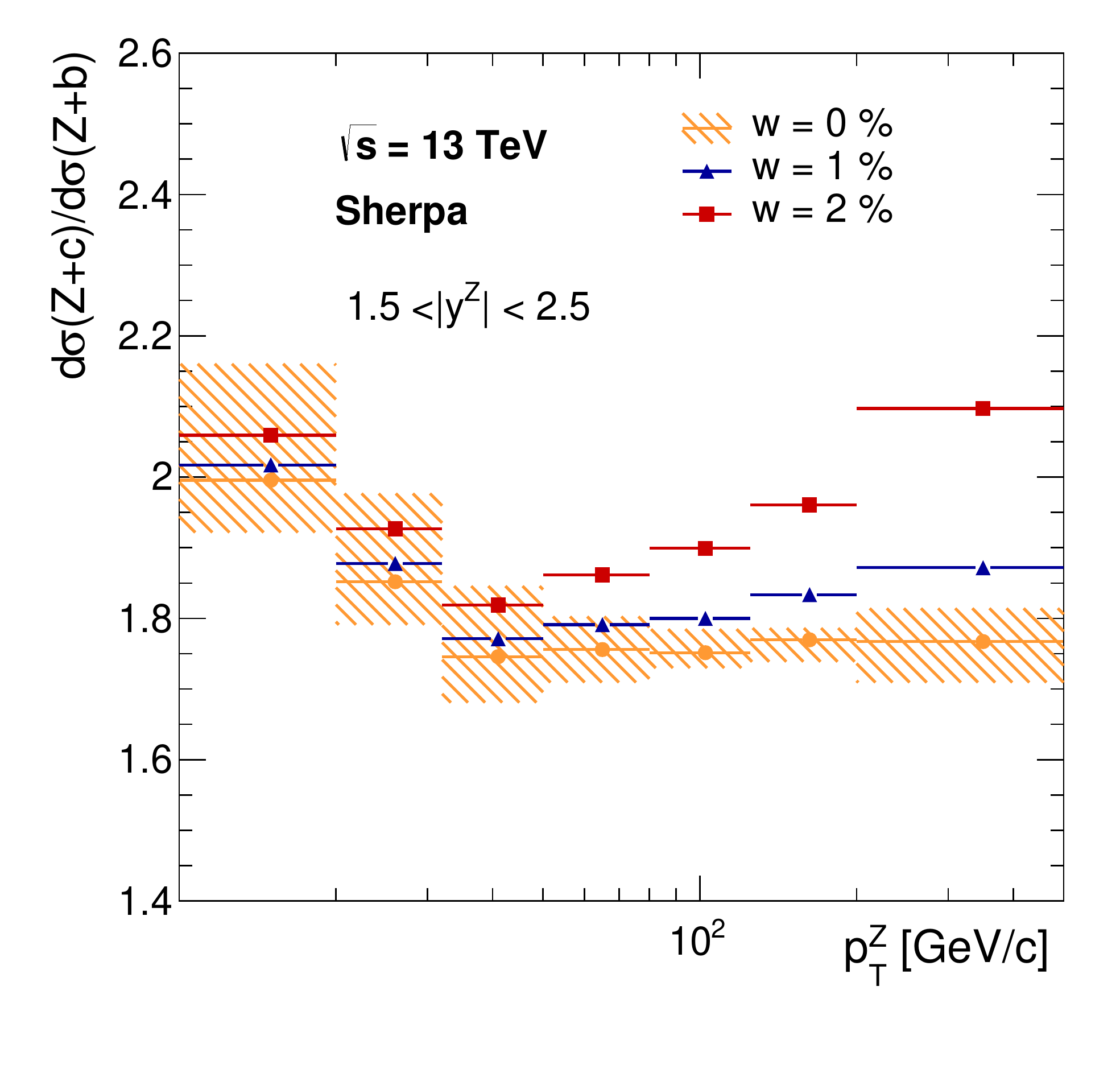}
  \includegraphics[width=\mpw]{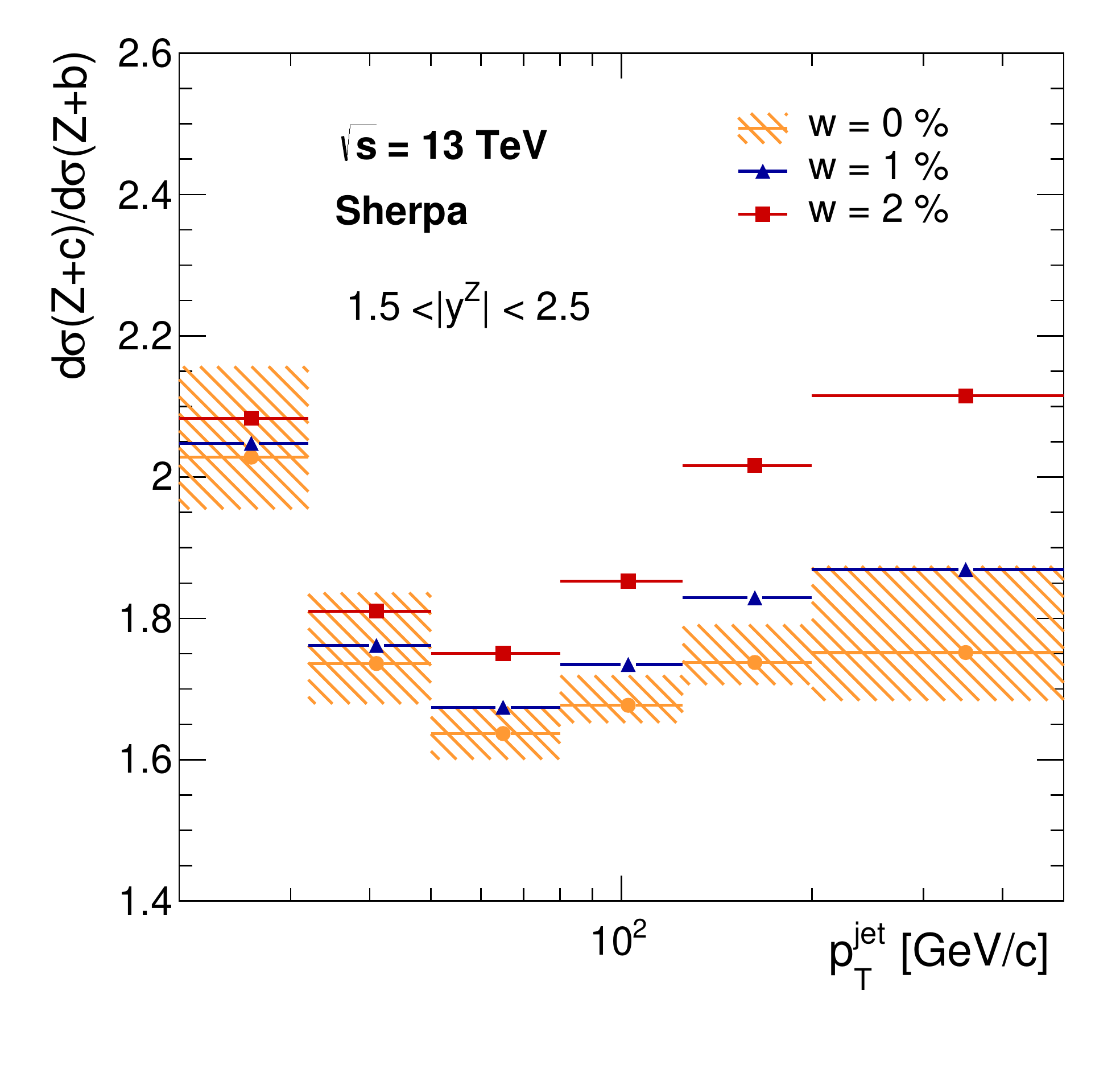}
  \caption{Predictions for the ratio of $Z+c$-jet and $Z+b$-jet production
           cross-sections as a function of the $Z$ boson (left) and $c$-jet
           (right) transverse momenta in the forward rapidity region
           $1.5 < |y^{Z}| < 2.5$ at $\sqrt{s} = 13$~TeV. The predictions are
           made with the \sherpa{} generator using the CT14nnlo PDF with
           different values of the \ic{} contribution $w$. The uncertainty bands
           represent the uncertainties in the QCD scale (shown only for
           $w = 0$\% predictions).}%
  \label{fig:Sherpa_ratios}
\end{figure}

In Fig.~\ref{fig:Zc_Sherpa_predictions} differential cross-sections
of associated $Z + c$-jet production calculated in the 
forward rapidity region $1.5 < |y^{Z}| < 2.5$ at $\sqrt{s} = 13$~TeV as  
functions of the $c$-jet and $Z$ boson transverse momenta
are shown. The effect of \ic{} becomes visible at $\pt \gtrsim 200$~GeV in both
distributions, but the theoretical uncertainties are still higher than the
size of this effect in the whole transverse momentum region 
studied. However, in the ratios of differential cross
sections $\sigma{(Z + c)}/\sigma{(Z + b)}$ the effect of \ic{} can be visible at
significantly lower $Z$ boson or \HF~jet
transverse momenta than in the differential cross-sections
themselves. Predictions for these ratios are shown in
Fig.~\ref{fig:Sherpa_ratios}.

\begin{figure}[h]
  \centering
  \includegraphics[width=\mpw]{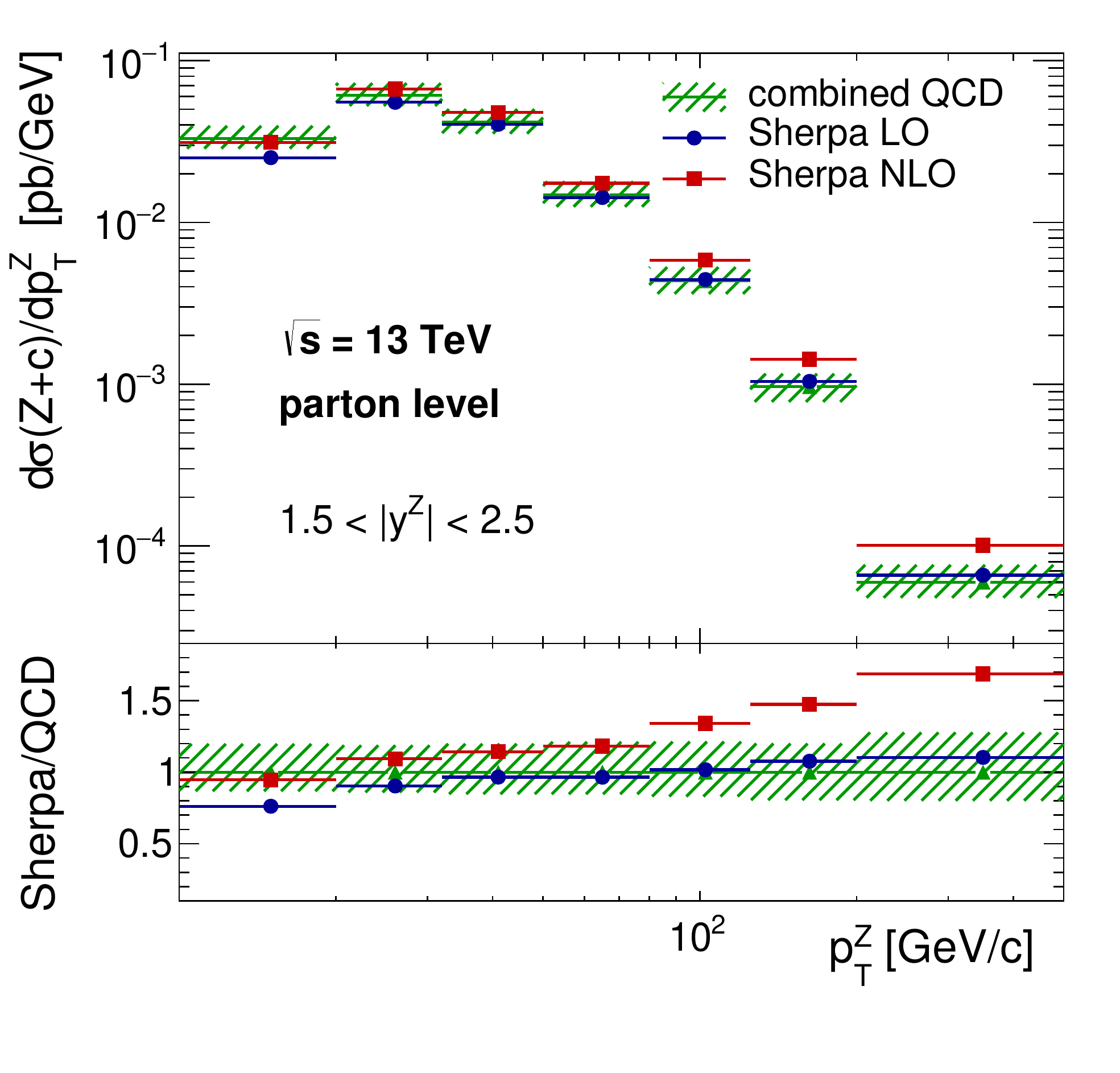}
  \includegraphics[width=\mpw]{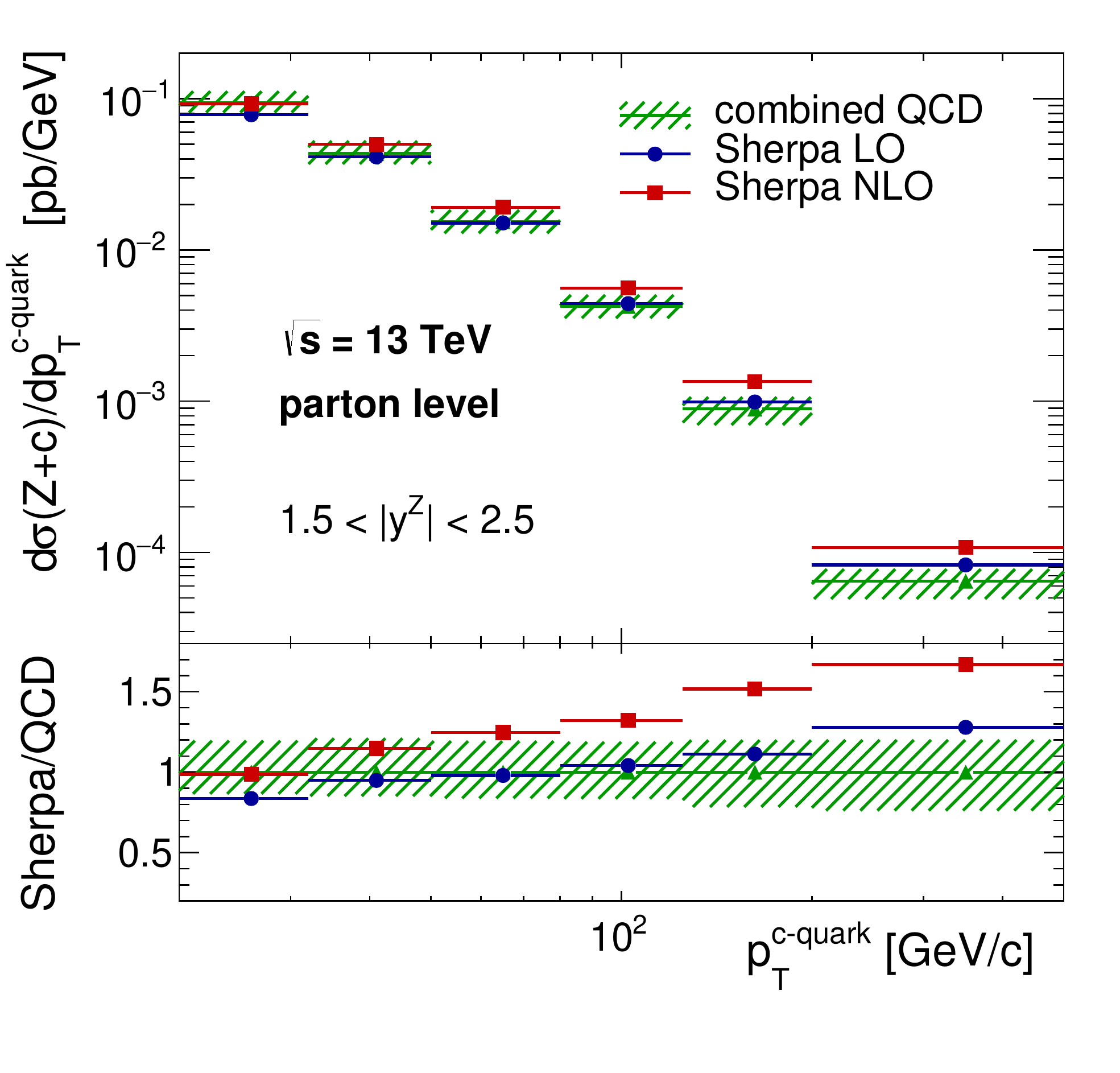}
  \caption{Parton level predictions for the production cross-section of a $Z$
           boson with a $c$ quark as a function of the $Z$ boson (left) and
           $c$-quark (right) transverse momenta in the forward rapidity region
           $1.5 < |y^{Z}| < 2.5$ at $\sqrt{s} = 13$~TeV. The predictions are
           made by combined QCD calculations and the \sherpa{} generator using
           LO and NLO matrix elements. The CTEQ66 PDF set without any intrinsic
           charm contribution is used. The uncertainty bands represent the
           uncertainties in the QCD scale (shown only for combined QCD
           predictions).}%
  \label{fig:Comb_QCD_Sherpa_comp_13TeV}
\end{figure}

To investigate the influence of parton showers and higher-order 
pQCD corrections on the predictions, we repeated the above
\sherpa~calculations at a parton level using LO and NLO
matrix elements. The results of these calculations are 
shown in Fig.~\ref{fig:Comb_QCD_Sherpa_comp_13TeV}
in comparison with the combined QCD predictions. First, one
can see that the best agreement with the combined QCD approach 
at large transverse momenta is given by the \sherpa~calculations using the LO
matrix element. This is not surprising
because the combined QCD predictions are represented in this kinematical region
by the quark-induced subprocesses calculated in the usual collinear QCD
factorization with the same accuracy. At low and moderate transverse momenta 
the results of the combined QCD approach are consistently
close to parton-level \sherpa~predictions obtained at the NLO level,
that demonstrates it is effective to take into account
higher-order pQCD corrections in the off-shell gluon-gluon fusion subprocess
supplemented with the CCFM gluon dynamics. Therefore, we can conclude that 
there are no large contradictions 
between our two theoretical approaches at the parton level. 
The combined QCD approach can be used to predict \ZHF{}
production cross-sections at the  
parton level at moderate transverse momenta, but 
such approximation becomes worse towards high transverse momenta where the 
effects described above are quite large.

\begin{figure}[h]
  \centering
  \includegraphics[width=\mpw]{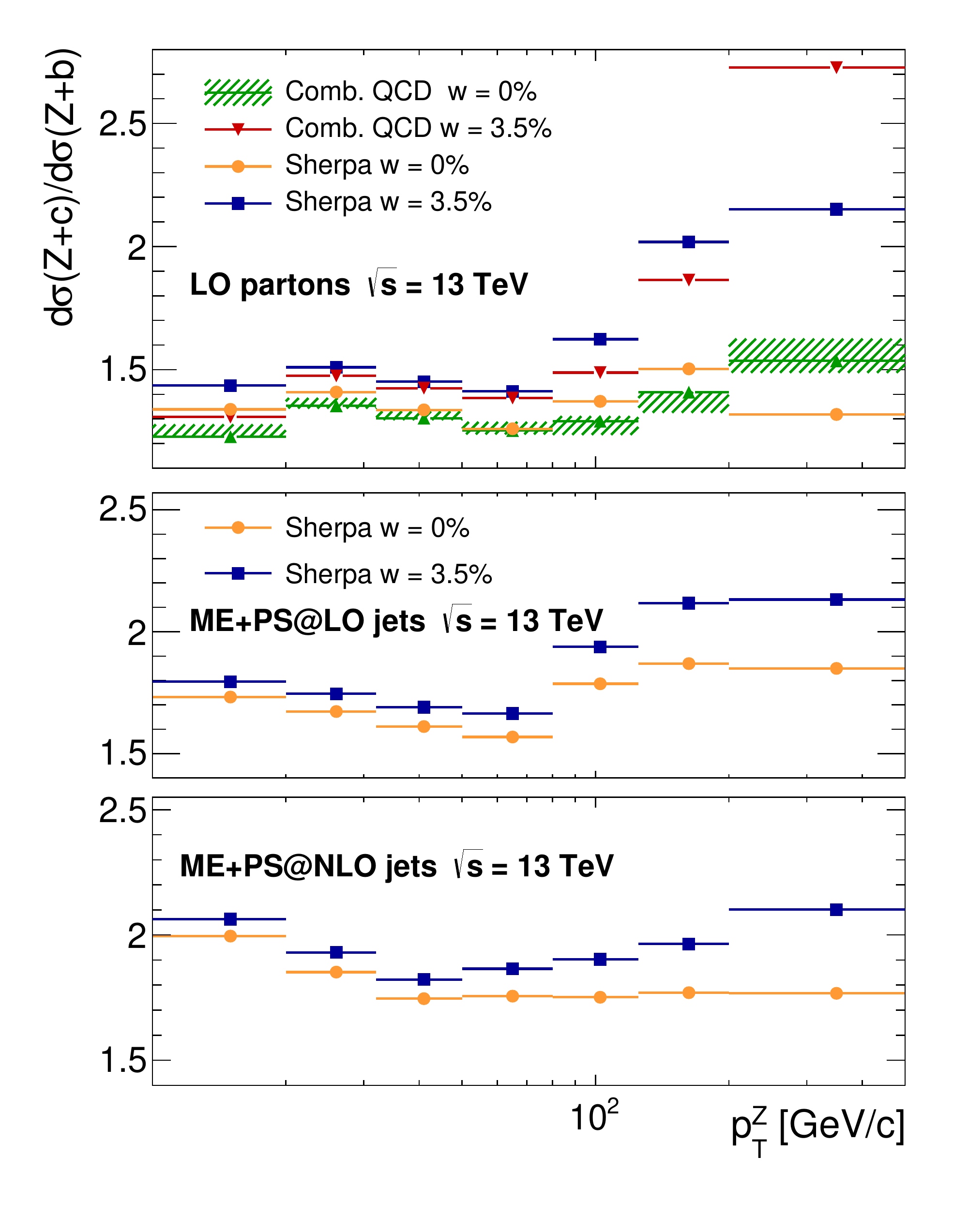}
  \includegraphics[width=\mpw]{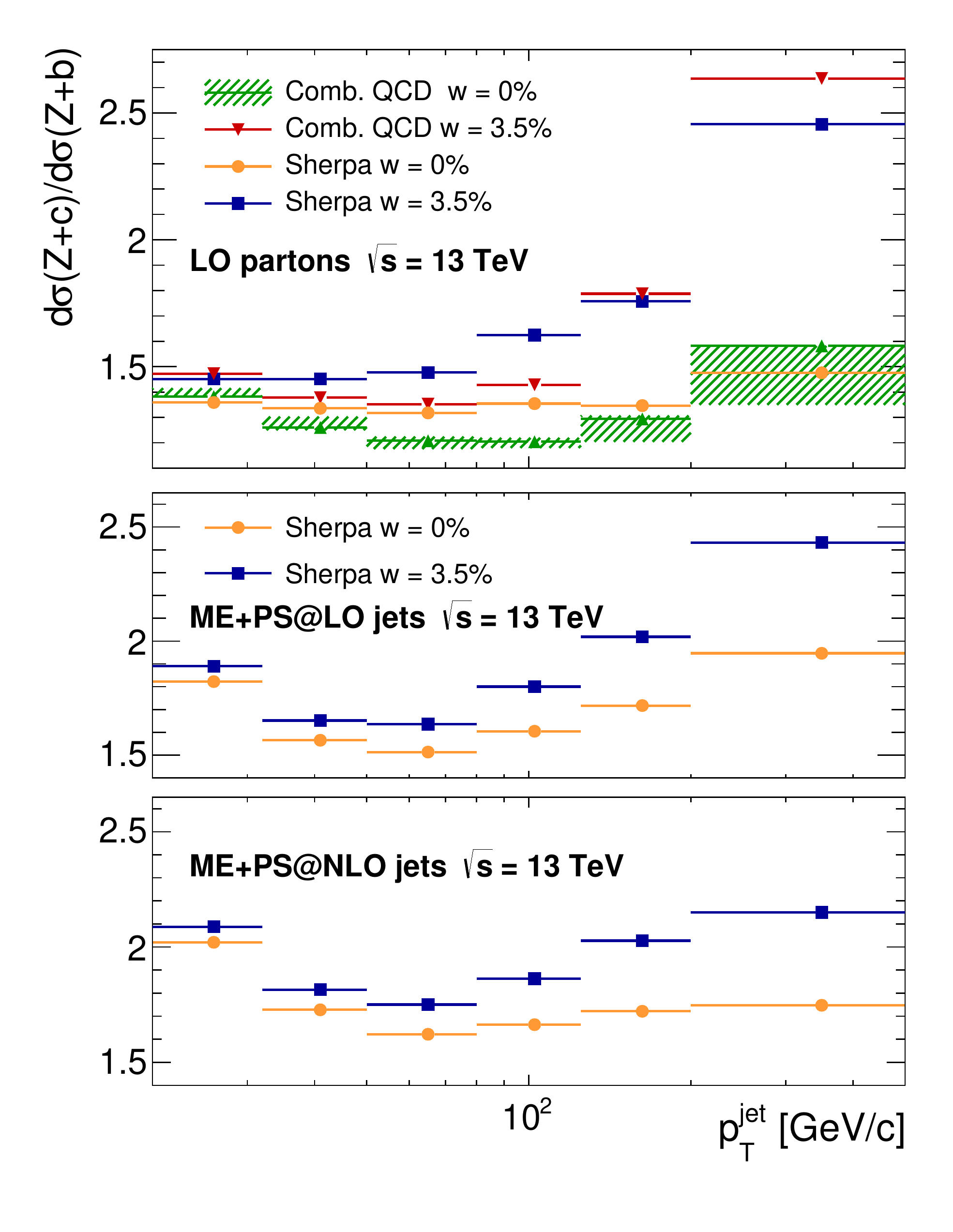}
  \caption{Predictions for the ratio of the production cross-sections of
           $Z+c$-jet and of $Z+b$-jet as a function of the $Z$ boson (left) and
           \HF{} jet (right) transverse momenta in the forward rapidity region
           $1.5 < |y^{Z}| < 2.5$ at $\sqrt{s} = 13$~TeV. The predictions are
           made with the \sherpa{} generator at the parton level using the LO
           matrix element (top panels) and at the particle level using the
           ME+PS@LO (middle panels) and ME+PS@NLO (bottom panels) models.
           Predictions of the combined QCD are also shown in the top panel.
           CTEQ66(c) PDF sets are used with IC contribution values $w = 0$ and
           3.5\%.}%
  \label{fig:explanation_IC_sup}
\end{figure}

Next, the effects of adding 
parton showers and NLO corrections to the parton 
level \sherpa{} \lo{} predictions for differential cross-section
ratios $\sigma{(Z + c)}/\sigma{(Z + b)}$ are illustrated in 
Fig.~\ref{fig:explanation_IC_sup}.
These ratios are calculated using CTEQ66(c) PDF sets with $w_\mathrm{IC} = 0$\%
and $3.5$\%.
One can see that including parton showers does 
significantly decrease the excess in the spectrum caused by
the non-zero \ic{} component, while adopting the
ME+PS@NLO instead of the ME+PS@LO approach makes little
difference. 
Thus, both \sherpa{} predictions made at a particle level give the \ic{} effect
in the forward region at $200 < \pt < 500$~GeV (irrespectively of whether \pt{}
of the jet or of the $Z$ boson is considered) of the order of 10--20\%, compared
to the much larger effect predicted by the parton-level calculations (\sherpa{}
at \lo{} or the combined QCD approach) to be at the level
of a factor of about 2. This observation is
in qualitative agreement with that 
made in Ref.~\cite{Hou:2017khm} when comparing the predictions for integral cross 
sections of $Z+c$-jet production from fixed order MCFM calculations and those 
from \sherpa~within the ME+PS@LO approach.

\begin{figure}[h]
  \centering
  \includegraphics[width=\mpw]{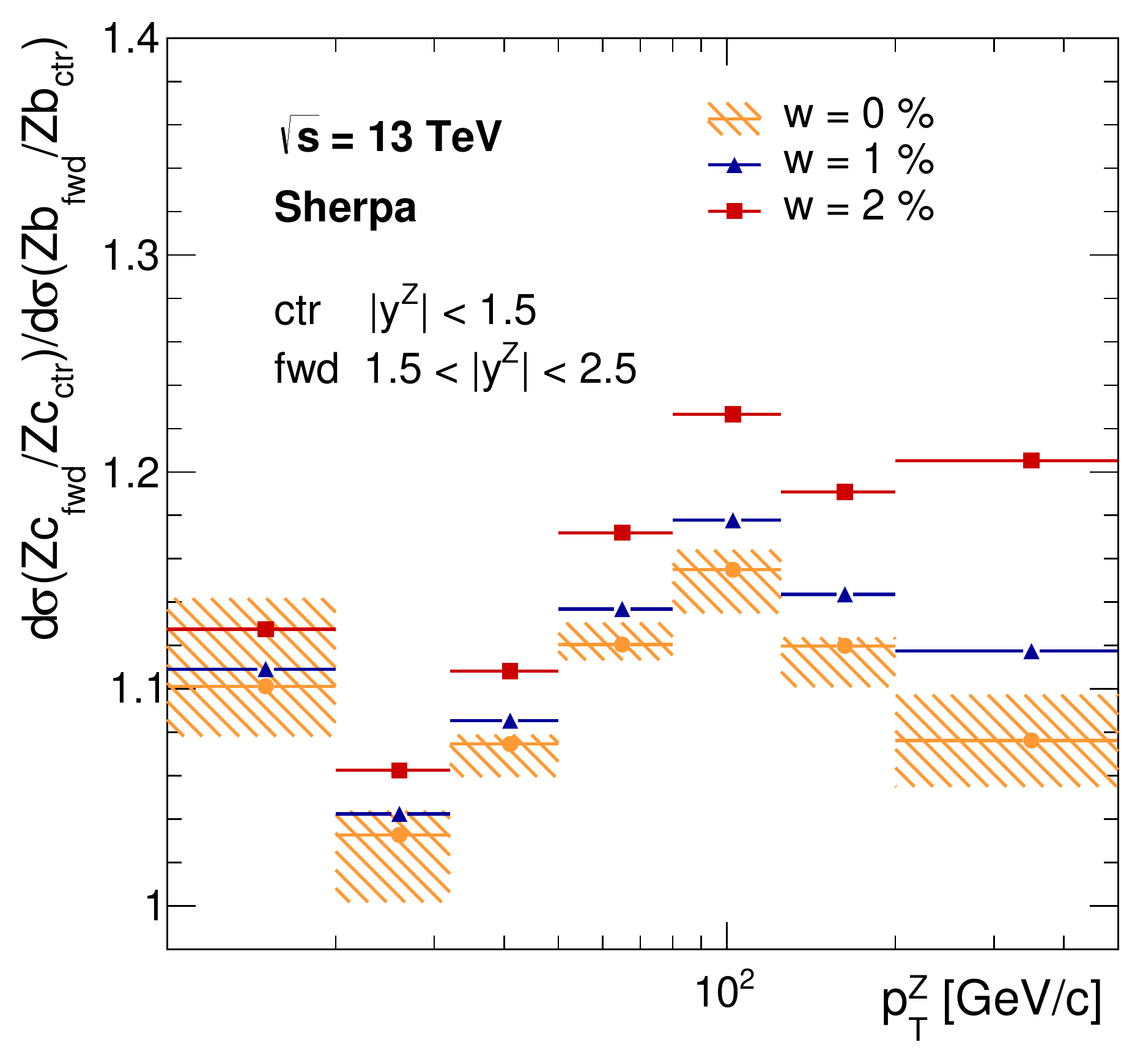}
  \includegraphics[width=\mpw]{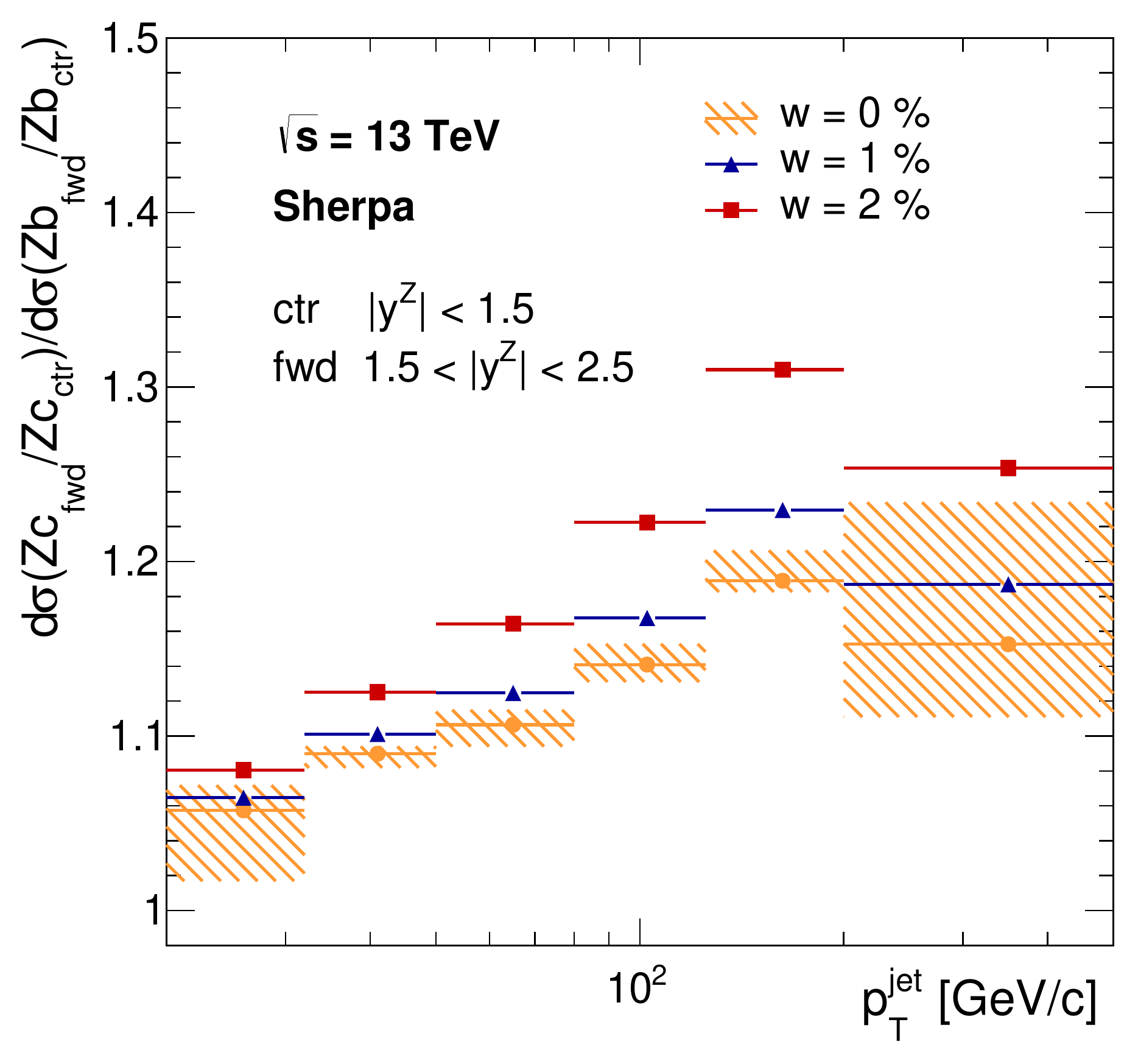}
  \caption{Predictions for the double ratio as a function of the $Z$ boson
           (left) and jet (right) transverse momenta at $\sqrt{s} = 13$~TeV.
           The double ratio is the ratio of the $Z+c$-jet production
           cross-section in the forward region $|y^{Z}| < 1.5$ to the
           cross-section in the central region $1.5 < |y^{Z}| < 2.5$, divided by
           the same ratio for $Z+b$-jet production. The predictions are made
           with the \sherpa{} generator using CT14nnlo PDF with different \ic{}
           contribution values $w$. The uncertainty bands represent the
           uncertainties in the QCD scale (shown only for $w = 0$\%
           predictions).}%
  \label{fig:double_ratio}
\end{figure}

Now we turn to the discussion of our theoretical uncertainties and uncertainties
of the \lhc{} measurements. These uncertainties have been
shown~\cite{Bednyakov:2017vck} to impose a strong restriction on the precision
of the \ic{} probability estimation from the experimental data. So new
observables which may be less affected by such uncertainties are of high
interest. A new variable satisfying this criterion can be defined as follows.
The ATLAS and CMS rapidity range is divided into a central region
$|y^{Z}| < 1.5$ and a forward region $1.5<|y^{Z}|<2.5$. Then, the ratio of the
$Z + c$ production cross-sections in the forward region and in the central
region is divided by the same ratio for $Z+b$ production. This so-called double
ratio $\sigma(Zc_{fwd}/Zc_{ctr})/\sigma(Zb_{fwd}/Zb_{ctr})$ is shown in 
Fig.~\ref{fig:double_ratio} as a function of the transverse 
momentum of the $Z$ boson $\pt[^{Z}]$ at the left and
of the leading jet $\pt[^{\jet}]$ at the right. 
One can see that in those ratios the \ic{} effect
is already visible at the transverse momentum
$\pt \gtrsim 50$~GeV. 
This value is much less than if one studies the differential cross 
sections of $Z+c$ production.
Moreover, the uncertainties related to the QCD scale in
theoretical calculations are significantly suppressed in this
double ratio (see Fig.~\ref{fig:double_ratio}).
Therefore, the latter could be a more promising variable in the 
search for intrinsic  
charm at LHC as compared to other observables considered
previously.

Moreover, to obtain more reliable information on the probability 
of \ic{} being present in the proton from future LHC data
at $\sqrt{s}= 13$~TeV one can perform a better estimation of
theoretical scale uncertainties and reduce systematic
uncertainties. This problem can be addressed by employing the
``principle of maximum conformality'' (PMC)~\cite{Brodsky:2013vpa} which sets
renormalization scales by shifting the $\beta$ terms in the pQCD series into
the running coupling. The PMC predictions are independent of
the choice of renormalization scheme --- a key
requirement of the renormalization group. 
However, up to now there is no direct
application of the PMC to the hard processes discussed in this paper.
One can expect forthcoming ATLAS and CMS 
experimental results on associated \ZHF~production
to be sensitive to the effect of \ic{} in a proton.

\subsection{Summary}

Associated production of the $Z$ boson and heavy flavor
jets in $pp$ collisions at LHC energies has
been considered applying the \sherpa~Monte Carlo 
generator and the combined QCD factorization approach using PDF sets with
different intrinsic charm components. The combined QCD
approach employs both the \kt-factorization and the collinear QCD
factorization with each of them used in the kinematical conditions of its
reliability. The best description of the ATLAS and CMS data on the 
$Z + b$ and $Z + c$ production at $\sqrt{s} = 7$ and $8$~TeV was obtained
within the \sherpa~5FS ME+PS@NLO model. Effects arising from
parton showers and higher-order pQCD corrections have been
investigated. We found these effects to strongly suppress
the sensitivity of our predictions to the intrinsic charm content of a proton.
However, despite this suppression, one can expect forthcoming
ATLAS and CMS measurements of \ZHF~production at $\sqrt{s} = 13$~TeV 
to be very important to search for the \ic{} contribution in
the proton. We suggest to measure a new
observable, namely, the double ratio of cross 
sections $\sigma(Zc_{fwd}/Zc_{ctr})/\sigma(Zb_{fwd}/Zb_{ctr})$, which is extremely 
sensitive to the \ic{} signal. This observable can be very promising
for precision estimation of the \ic{}
probability, since it is less affected
by QCD scale uncertainties, as compared to
the observables considered previously.

%
%
\section{Future Experiments}%
\label{sec:future}

A primary objective of the proposed fixed-target experiment AFTER@LHC will be to
study heavy hadron production at high \xf{} in $pA$ collisions at far forward
rapidities~\cite{Brodsky:2015fna}. These measurements will also have direct
impact for astrophysics since intrinsic charm is important for charm production
in cosmic ray experiments that measure charm production from high energy
experiments interacting in the earth's atmosphere. It is also important for
estimating the high energy flux of neutrinos observed in the IceCube experiment.
In fact, one finds~\cite{Laha:2016dri} that the prompt neutrino flux arising
from charm hadroproduction by protons interacting in the earth's atmosphere
which is due to  intrinsic charm is comparable to the extrinsic contribution if
one normalizes the intrinsic charm differential cross-sections to the ISR and
the LEBC-MPS collaboration data.

The intrinsic heavy quark Fock states  in the nuclear target itself will  also be excited  in a high energy LHC proton-nucleus collision.   The resulting heavy quarks will be produced at small rapidities relative to to the target rapidity; i.e.,
{\it nearly at rest  in the laboratory}.  For example, the coalescence of the produced 
heavy quarks with comoving light quarks will lead  to the production of a
heavy hadron such as a $\Lambda_\mathrm{b}(udb)$ at \emph{small rapidity}
$y_{\Lambda_\mathrm{b}} \simeq \ln{x_b}$, relative to the rapidity of the
nucleon in the target.
In addition, heavy-quark hadrons such as 
double-charm baryons, and exotic multiquark hadrons such as  $|[u\bar u][Q\bar Q]$, tetraquarks, pentaquarks, and even octoquarks
containing heavy quarks will be produced nearly
at rest in the nuclear target rest frame  in the $p A$ collision in a fixed
target experiment where they can be easily observed.  
One can also study the hadro-production of exotic hadrons such a heavy hexa-diquarks~\cite{West:2020rlk} (the color singlet bound state of six diquarks) containing a heavy quark.
The \ic{} signal can  also be studied in hard processes such as the 
production of prompt photons or
$Z^0$- or $W$  bosons accompanied by heavy quark jets.  Typical  underlying
subprocesses  are $g c \to \gamma  c$ or $g c \to Z^0  c$.

Measurements of $Z^0$ production accompanied by $c$-jet in $pp$ collisions at
$\sqrt{s}= 13$~TeV in the mid-rapidity range is currently being worked on at the
ATLAS experiment, LHC\@. The main goal of this measurement is the search for the
\ic{} signal in the \pt-spectra of $Z^0$ or $c$-jet. In the
Section~\ref{sec:z_c_jet} of this review and in Ref.~\cite{Lipatov:2018oxm} the
corresponding predictions are presented.

The inclusive production of $D$ mesons in $pp$ collisions at LHC energies and
their large rapidities $y$ and transverse momenta \pt{} can give also the
information on the \ic{} contribution to the proton PDF\@. The corresponding
predictions for such experiment at LHCb were presented in
Ref.~\cite{Lykasov:2012hf}. It was shown that in the \pt-spectrum of
$D^0$-mesons the enhancement at ${2.5 < y < 4.5}$ and $\pt \geq 10$~GeV can be
observed due to the \ic{} contribution. The similar \ic{} signal could be
searched for at future NA61/SHINE experiment on the $D$-meson production in $AA$
collisions at the fixed target.

The measurement of $D^0 \rightarrow K^\pm \pi^\mp$ and
$\Lambda_\mathrm{c} \rightarrow p K^{-} \pi^{+}$ at $\xf > 0.7$ at the LHC at
$\sqrt{s} = 14$~TeV would be possible in the forward multiparticle spectrometer
(FMS) being proposed as a new sub-detector for CMS\@. The FMS
measurements will also be sensitive to the large asymmetries in
$\x [c(\x, Q) - \bar{c}(\x, Q)]$ predicted for intrinsic charm, see
Fig.~\ref{fig:cbarc_asymm}.

Recently the AnDY experiment at RHIC~\cite{Bland:2019aha} has observed both
single- and double-$\Upsilon$ production in the forward  direction in Cu+Au
collisions. It will be important to test whether the observed production rates
are compatible with the hadronization of single and double intrinsic bottom Fock
states, which are predicted to have probabilities suppressed by
${m_\mathrm{c}^{2} / m_\mathrm{b}^2} \sim 1/10$ relative to the single and
double intrinsic charm probabilities.
%
%
\section{Acknowledgments}

We thank V.A.~Bednyakov, M. Ga\'{z}dzicki, A.A.~Glasov, R.~Keys, E.V.~Khramov,
V.V. Lyubushkin, S.~Prince, F.~Sforza, Yu.Yu.~Stepanenko, S.~Tokar, S.M.~Turchikhin for helpful
discussions. The authors are grateful to S.P.~Baranov, H.~Jung, M.A.~Malyshev,
N.A.~Abdulov, A.A.~Prokhorov for very useful
collaboration. A.V.L.\ is grateful to DESY Directorate for the support in the
framework of Cooperation Agreement between MSU and DESY on phenomenology of the
LHC processes and TMD parton densities.
This work has been partially supported by U.S. D.O.E. Grant  No. DE–
AC02–76SF00515, SLAC-PUB-17538.

\bibliography{bibliography}





\end{document}